\documentclass{elsart}
\usepackage{graphicx,amssymb}

\newcommand{\be}{\begin{equation}}
\newcommand{\ee}{\end{equation}}
\newcommand{\bea}{\vspace{0.25cm}\begin{eqnarray}}
\newcommand{\eea}{\end{eqnarray}}

\def\PLA{{Phys. Lett.}  A }
\def\PLB{{Phys. Lett.}  B }
\def\PRL{{Phys. Rev. Lett.} }
\def\PRA{{Phys. Rev.} A }

\def\PRD{{Phys. Rev.} D }

\begin{document}

{\bf Research on Hidden Variable Theories: a review of recent
progresses. } \vskip 1cm

{ Marco Genovese} \vskip 0.5cm {\it Istituto Elettrotecnico
Nazionale Galileo Ferraris, Strada delle Cacce 91, 10135 Torino,
Italy \\} \vskip 1cm

\vskip 1cm
 {\bf Abstract}

Quantum Mechanics (QM) is one of the pillars of modern physics: an
impressive amount of experiments have confirmed this theory and
many technological applications are based on it.  Nevertheless, at
one century since its development, various aspects concerning its
very foundations still remain to be clarified. Among them, the
transition from a microscopic probabilistic world into a
macroscopic deterministic one and quantum non-locality. A possible
way out from these problems would be if QM represents a
statistical approximation of an unknown deterministic theory.

This review is addressed to present the most recent progresses on
the studies related to  Hidden Variable Theories (HVT), both from
an experimental  and a theoretical   point of view, giving a
larger emphasis to results with a direct experimental application.

More in details, the first part of the review is a historical
introduction to this problem. The Einstein-Podolsky-Rosen
argument and the first discussions about HVT
are introduced, describing the fundamental Bell's proposal for a
general experimental test of every Local HVT
and the first attempts to realise it.

The second part of the review is devoted to elucidate the recent
progresses toward a conclusive Bell inequalities experiment obtained with
entangled photons and other physical systems.

Finally, the last sections are targeted to  shortly discuss
Non-Local HVT.

\newpage

{\bf \large Table of contents}

{\bf 1 - Introduction}

{\bf 2 - Historical introduction}

2.1) The EPR paradox.

2.2) Hints on the macro-objectivation problem.

2.3) The von Neumann theorem.

2.4) The Bell inequalities.

2.5) Quantum states violating Bell inequalities.

2.6) Theorem about no faster than light transmission by using EPR correlations.

2.7) Toward an experimental test of Local Realism: introductory remarks.

2.8) First experimental tests of Bell inequalities.

2.9) The Orsay experiment.

{\bf 3 - PDC experiments on Local Realism.}

3.1) The Parametric Down Conversion.

3.2) PDC experiments with interferometers.

3.3) First tests of local realism by using PDC polarisation entanglement.

3.4) Bright sources of polarisation entangled states.

3.5) Applications of bright sources of entangled photon pairs.

3.6) Tests of local realism by equalities.

3.7) LHVT built for surviving PDC experiments.

3.8) Test of local realism in Hilbert spaces with dimension larger than 2.

3.9) Tests of non-contextuality.

3.10) Other Quantum Optics experiments connected with local realism and quantum non-locality.

{\bf 4 - Tests of local realism with other physical systems than photons}

4.1) Tests of Bell inequalities with mesons.

4.2) Other tests of local realism with mesons.

4.3) Bell inequalities experiments with ions.

{\bf 5 -Some conclusions about Local Hidden Variable Theories}

{\bf 6 - Non-Local Hidden Variable Models}

6.1) The de Broglie - Bohm theory.

6.2) Nelson Stochastic model.

6.3) Experimental tests of NLHVT against SQM.

{\bf 7 - Determinism at Planck scale?}

{\bf 8 - Conclusions}

\newpage

\section{Introduction}

Quantum Mechanics (QM) represents nowadays one of the pillars of
modern physics: so far a huge amount of theoretical predictions
deriving from this theory has been confirmed by very accurate
experimental data. No doubts can be raised on the validity of this
theory. Nevertheless, even at one century since its birth, many
problems related to the interpretation of this theory persist:
non-local effects of entangled states; wave function reduction and
the concept of measurement in Quantum Mechanics; the transition
from a microscopic probabilistic world to a macroscopic
deterministic world\footnote{Leaving out of consideration
classical chaotic system, where, non the less, non determinism is
due to an imperfect knowledge of initial conditions (whilst a
perfect one is, at least in principle, possible in classical
mechanics).}
 perfectly described by classical mechanics (macro-objectivation)
and so on. A possible way out from these problems would be if QM
represents a statistical approximation of an unknown deterministic
theory, where all observables have defined values fixed by unknown
variables, the so called Hidden Variable Theories (HVT), a
suggestion dating since the celebrated Einstein-Podolsky-Rosen
paper of 1935.

Even though, in the past decades, many interesting theoretical
works were focused on the solution of this problem originating
a large debate, nevertheless they could not lead to any conclusive
solution \cite{books3,books4,books1,books2}.

However, in the last years the experimental know-how concerning
creation and manipulation of quantum systems has hugely increased,
permitting the realisation of several experiments originally
thought as Gedanken Experiment and  the conceiving of new ones.

The dream of testing the theoretical ideas proposed in connection
with foundation of quantum mechanics is now becoming reality.

Furthermore, the improved knowledge about quantum systems  is now
leading to the growth of new areas of physics as quantum
information theory, which could lead to the realisation of many
different interesting technological applications as quantum
cryptography \cite{Qcrip},  quantum computers \cite{Qinf,Qinf2,Qinf3,Qinf4},
quantum metrology \cite{Qmet1,Qmet2,Qmet3,Qmet4,Qmet5}, ...

Considering the large conceptual relevance of the problem of
completeness of QM and the fundamental role of entanglement in
quantum information and metrology, the purpose of this work is to
give the basic concepts related to this sector of research
(without the impossible pretension of being exhaustive in a field
where hundreds of papers appear every year) and a review of recent
experiments that have permitted to throw new light on it, with hints
to possible future developments.

In detail, the outline of the paper is the following: after
describing how this research field was born with the celebrated
Einstein-Podolsky-Rosen paper \cite{EPR} and how possible
experimental investigations on every Local Hidden Variable Theory
(LHV) became possible after the Bell work in 1964 \cite{bell}, we
will review many interesting experiments (largely based on
entangled photon pairs) devoted to  test Bell inequalities
\cite{pos1,pos2,pos3,pos4,ac1,ac2,ac3,ac4,pp,lam,asp,typei1,typei2,typei3,typei4,kw,kw2,HSZ,strek,rarlong,gislong1,gislong2,weihs,hugo,gisbb,gisbb2,tbent,tblong,type2,kurt,shapII,dig,aspel,whi,nos},
leading to a substantial agreement with standard quantum mechanics
(SQM) and strongly disfavouring LHV theories. We will then
consider why, nevertheless,  none of these experiments has yet
been conclusive since, so far,  one has always been forced to
introduce a further additional hypothesis \cite{santos,decaro},
due to the low total detection efficiency, stating that the
observed sample of particle pairs is a faithful subsample of the
whole: a problem known as { \it detection  loophole}.

After  discussing the pioneer  experiments in 70's, we will
delineate in some detail the celebrated experiment of Aspect et
al. \cite{asp}, which represented a miliary stone in the field
since it allowed for the first time to realise well space-like
separated measurements, albeit with a very low collection
efficiency.

In the following section are described
 the big progresses in the direction of a conclusive
test of HVT that have been obtained in 90's by using parametric down conversion
(PDC) process for generating entangled photon pairs with high
angular correlation.  The first experiments of this kind had, by
construction, a limited total efficiency
\cite{typei1,typei2,typei3,typei4} and were far from
eliminating the detection loophole    \cite{santos}. Anyway, more
recent experiment have drawn nearer to a conclusive experiment
\cite{kw,kw2,HSZ,strek,rarlong,gislong1,gislong2,weihs,hugo,gisbb,gisbb2,tbent,tblong,type2,kurt,shapII,dig,aspel,whi,nos}
(approaching an efficiency about $20 \%$).

In section 4, after a general discussion about alternative systems
for testing local realism as, in specific, entangled mesons, a
recent experiment \cite{nist} performed using Be ions will be
presented. Here very high efficiencies (around 98 \%) were
reached, but the two subsystems (the two ions) were not really
separated systems during measurement.

Finally, we will outline non-local hidden variable models (NLHVT),
which are not excluded by Bell inequalities tests, and on the
possibility of testing them against SQM.

On the whole in this review we have tried to give a general panorama on the research about local realism
(with particular emphasis to recent experimental progresses), endeavouring to be as
complete as possible with the hope that this work could be useful both to the neophytes who want to approach
this field and to experts who can find a useful summary and a rich source of references.


\section{Historical introduction}

\subsection{ The EPR paradox }

The debate whether Quantum Mechanics is a complete theory and probabilities
have a non-epistemic character (i.e. nature is intrinsecally probabilistic)
or whether it is a statistical approximation of a deterministic theory and
probabilities are due to our ignorance of some parameters (i.e. they are epistemic)
dates to the beginning of the theory itself.

The fundamental paper where this problem clearly emerged appeared in 1935
when Einstein, Podolsky and Rosen risen this question by considering an explicit example
\cite{EPR}.

For this purpose, they introduced the concept of { \it element of
reality} according to the following definition: if, without
disturbing in any way a system, one can predict without any
uncertainty the value of a physical quantity, then there is an
element of physical reality corresponding to this quantity (i.e. introducing the hypothesis of realism).
They formulated also the reasonable hypothesis (at the light of special
relativity) that any non-local action was forbidden. A theory is
complete when it describes any element of reality.

They concluded  that either one of their premises was wrong or
Quantum \ Mechanics was not a complete theory, in the sense that
not every element of physical reality had a counterpart in the
theory.

More in details,  they considered a system consisting of two
particles prepared in a state such that the sum of their momenta
$(p_1+p_2)$ and the difference of their positions $(x_1-x_2)$ were
both defined at the same time (situation possible in QM since they
correspond to commuting operators). The case where $(x_1-x_2)$ has
eigenvalue $a$ and $(p_1+p_2)$ has eigenvalue 0 is described in QM
by a Dirac delta function $\delta (x_1-x_2-a)$.
This is an example of an
entangled state: a state of two or more particles which
cannot be factorised in single-particle states, namely (as stated
in original Schr\"odinger definition \cite{scent}) a compound state
whose subsystems are not probabilistically independent\footnote{See
\cite{GhirMar} for an in deep discussion on the definition of entanglement.}.

By measuring the position of particle 1 (2) one can predict with
certainty the position of particle 2 (1), without disturbing in
any way this second particle (which at the moment of the
measurement can be very far from the other). Position of particle
2 (1) is therefore an element of reality according to the previous
definition. On the other hand, a measurement of momentum of particle
1 (2) allows one to infer momentum of particle 2 (1) without acting on
it, thus also momentum of particle 2 (1) is an element or reality.
But according to QM position and momentum are conjugate variables
and perfect knowledge of one of them implies complete ignorance on
the value of the other. Therefore, in this case QM cannot predict
all the element of realities and thus QM is an incomplete theory.

The same argument has then been presented by Bohm for a spin system,
allowing a very clear physical understanding and avoiding the use
of a delta of Dirac wave function.

Let us thus consider a singlet state of two spin 1/2 particles

\bigskip
\begin{equation}
\vert \psi_{0} \rangle = {\frac{ \vert \uparrow \rangle \vert \downarrow \rangle
- \vert \downarrow \rangle \vert \uparrow \rangle }{\sqrt {2}} }
 \label{Singlet}
\end{equation}

where $\vert \uparrow \rangle $ and $\vert \downarrow \rangle $
represent a single particle of spin up and down respectively (in
the following we will often use the general quantum information
notation for two-level systems, qubits,  $|0 \rangle, | 1
\rangle$). This state is manifestly entangled.  Let us notice that
the complete wave function for the system is obtained by
multiplying the spin wave functions by appropriate space wave
functions,  depending on the space co-ordinates of both particles.
When the two space wave functions, e.g. gaussian,   are widely
separated at the moment of measurement, one can  therefore realise
two wave packets entangled in spin, but well separated in space.

Repeating the EPR argument, let us suppose that the two particles
separate (i.e. the spatial wave functions of the two particles do
not overlap anymore), being addressed to two remote measurement
apparatuses. Usually the two experimenters are called Alice and
Bob, a convention that we will sometimes use in the following. Let
us then suppose that Alice measures the z component of the spin of
the first particle; this permits to immediately know the z spin
component of the particle 2 (which is opposite) without disturbing
in any way the second particle. Thus the z component of the spin
of the second particle is an element of reality according to the
previous definition.

But, since the singlet state is invariant under rotations, we
could refer to any other axis (as x or y etc.): thus we can argue
that any other spin component of particle 2 is an element of
reality. However, spin components on different axes are
incompatible variables in Quantum Mechanics, for which one cannot
assign definite values at the same time: thus this implies that
Quantum Mechanics cannot be a complete theory (according to the
EPR definition), since it does not allow a prediction of all
elements of reality.

The interesting property of an entangled system in QM is that
before measurement the values of some specific observables of the
two (or more) subsystems are not defined.  The measurement of the
observable for one of the subsystems (e.g. the z component of spin
of particle 1 in the previous example) instantaneously fixes the
value of the correlated observable of the second subsystem (e.g.
the z component of spin of particle 2 in the previous example),
independently of the distance of the second particle. An entangled
system must be considered as a whole, independently of separation
of its parts. This seems to introduce a sort of instantaneous
effect which is often called "quantum non-locality" and which in
the past caused a strong discussion about the compatibility of QM
and special relativity. A subsequent paragraph will be devoted to
this point.

Following the EPR presentation, this property of entangled systems
and its implications suscitated a large debate
\cite{books3,books4,books1,books2,stat,kara}. The position of many
authors can be summarised by quoting Ballentine  \cite{stat},
saying that "[to deny] the reality of both spin values until the
measurement has actually been performed" represents "an extreme
positivist philosophy" and  "entails the unreasonable, essentially
solipsist position that the reality of particle 2 depends upon
some measurement which is not connected to it by any physical
interaction".

Within this debate emerged the development of the so-called Local
Hidden Variable Theories\footnote{often denoted with the term
Local Realistic Theories (LRT) as well.}: namely of the proposal
that it exists a deterministic (and local) theory describing
nature, where the precise value of all observables of a physical
system are fixed by some unknown variables (the hidden variables).
Quantum Mechanics would only be a statistical approximation of
this theory.  This situation resembles statistical termodinamics,
which describes in a probabilistic way systems composed of many
particles, each behaving in a perfectly deterministic way
according to the classical equations of motion \cite{belinf}. Let
us emphasize the deep philosophical difference between the two
cases: in Standard Quantum Mechanics nature is intrinsically
probabilistic, whilst in a hidden variable theory quantum
probabilities become epistemic, they are due to our ignorance
about hidden variables whose knowledge would give us precise
information about every physical quantity.

As we will discuss later, also non-local HVT (where the action on
a subsystem can have immediate effect on the other subsystem
independently of the distance) or contextual HVT (where the
verity value of a statement about a physical property depends not
only on hidden variables but on the whole experimental condition
as well) have been considered later.

 In general, in any hidden variable
theory every particle has an assigned value for each
observable, determined by a hidden variable (or a set of hidden
variables) $x$. A statistical ensemble of particles has a certain
distribution $\rho (x)$ of the hidden variable and thus the
average value of an observable A is given by:

\bigskip

\begin{equation}
<A>=\int dx\rho (x)A(x)  \label{aa}
\end{equation}

Of course, considering the great success of QM in predicting many
different experimental data, a judicious HVT should be built such
that the average $<A>$ given by (\ref{aa}) reproduces the QM
predictions.

\subsection{Hints on the macro-objectivation problem}

Before proceeding further in discussing the history of studies on
Hidden Variable models, let us just give a rapid hint to another
largely debated problem of the very foundations of QM, which would
find a natural solution in the HVT framework: the
macro-objectivation problem.

The problem of macro-objectivation derives from the fact that
Schr\"odinger equation is linear and thus requires that a
macroscopic system interacting with an entangled state  gets
entangled as well.

For example let us consider a macroscopic measurement apparatus
described by the state $| \chi _0 >$ (i.e. a wave function as
complicated as necessary) which interacts with the (microscopic)
states $ \vert \phi_1 \rangle $ and $
 \vert \phi_2 \rangle $. The interaction, representing the measurement and lasting a time
interval $\Delta t$, can be described by a linear evolution
operator $U(\Delta t)$. The results of the measurement are then

\be \vert \chi_0 \rangle \vert \phi_1 \rangle \rightarrow U(\Delta
t) [\vert \chi_0 \rangle \vert \phi_1 \rangle ] = \vert \chi_1
\rangle  \vert \phi_1 \rangle \ee

and \be \vert \chi_0 \rangle \vert \phi_2 \rangle \rightarrow
\vert \chi_2 \rangle  \vert \phi_2 \rangle \ee
where the states $\vert \chi_1
\rangle$ and $\vert \chi_2
\rangle$ represent the state of macroscopic apparatus after measurement
corresponding, for example, to two different directions of a pointer.

If $| \chi _0 >$ interacts with the superposition state \be a
\vert \phi_1 \rangle + b \vert \phi_2 \rangle \ee because of
linearity of the evolution equation, one would have \be \vert
\chi_0 \rangle [a \vert \phi_1 \rangle + b \vert \phi_2 \rangle ]
\rightarrow[a \vert \chi_1 \rangle  \vert \phi_1 \rangle +b \vert
\chi_2 \rangle  \vert \phi_2 \rangle ] \label{vns} \ee which is an
entangled state involving the macroscopic apparatus as well. If
one considers intermediate physical systems, as subsystems of the
measurement one, they get entangled in the same way. The chain of
entangled systems starting with the microscopic state in a
superposition and ending with the detection apparatus is often
called von Neumann chain.

Of course at a macroscopic level we do not perceive anything which
can be thought as a superposition of two macroscopic situations,
for example if the measuring apparatus has a pointer which will be
up or down according to if it has measured a property 1 or 2, we
always observe the pointer in one well defined position and never
in an undefined superposition of pointer up and down at the same
time.

A very illuminating example of this problem was proposed by
Schr\"odinger. Let us consider a box with a cat inside. In the box
there is also a measurement apparatus gauging a property of
a quantum system, which is in a superposition state for the
measured observable. According to the obtained result the
apparatus opens or not a poison bottle. Thus, in this case the von
Neumann chain includes the quantum system, the measuring apparatus
and the poison bottle. But at the end also the cat is involved: if
the poison has been diffused the cat dies, otherwise it survives.
The result of this analysis is therefore that we have a
superposition of cat alive and dead, which looks rather a
paradoxical situation. From this example in the literature a
superposition of two macroscopic states is usually dubbed a
"Schr\"odinger cat".

Therefore,  measurements in quantum mechanics would seem to require
some process breaking the entanglement: among the possible
outcomes only one will be realised and observed in the measurement
process. Only one state in the superposition survives the
measurement process, i.e. in the previous example the measuring
apparatus will be found or in the situation described by $\vert
\chi_1 \rangle$, with probability $|a|^2$, or in the one described
by the state $\vert \chi_2 \rangle $, with probability $|b|^2$ and
the measured state (if the measurement is non-destructive) will
be, correspondingly,  in the state $\vert \phi_1 \rangle$ or
$\vert \phi_2 \rangle$ respectively, after the measuring process.
This is called the wave function collapse. However, this request
must be justified more precisely. We have to understand at which
point of the measurement process the  collapse occurs and how this
collapse happens.

A first answer \cite{books4} is to split the world into a
macroscopic one following classical mechanics and a microscopic
one following QM (substantially the one adopted by the Copenaghen
school). However this solution, even if perfectly useful for
practical calculation of quantum processes, is weak from a
conceptual point of view since it does not permit to identify the
border between quantum and classical worlds. How many particles
should a body have for being macroscopic? What about "macroscopic"
systems as superconductors which exhibit quantum properties? This
answer looks to be too over-simplified.

 Various different ideas have been considered for
 explaining/understanding decoherence at macroscopic level, without reaching for any of them
a general consensus in the physicists community. Among them
(without any purpose to be exhaustive): the many universes models
\cite{everett}, modal interpretations \cite{modal1,modal2,modal3},
decoherence and quantum histories schemes
\cite{books2,gr,zurek,GMH}, transactional interpretation
\cite{cram}, 'informational' interpretation \cite{zeiinf,fuchs},
dynamical reduction models (where a non-linear modification of
Schr\"odinger equation is introduced) \cite{GRW,pearle,mil}, reduction by
consciousness (wave function collapse happens at observer level) \cite{wigner},  and
many others (see for example \cite{books4,stat,rod,par,bub,HW} and Ref.s
therein).

 On the other hand, this problem simply does not
exist in Hidden Variable Models since in this case the
specification of the state by using state vectors is insufficient,
there are further parameters (the hidden variables) that we ignore
for characterizing the physical situation. The physical system is
always in a well specified state (corresponding to one of the
quantum mechanical states present in the superposition) univocally
determined by the value of the hidden variables. However, it must
be noticed that for contextual theories one can attribute an
objective state only to those variables which are non-contextual
(we will come back on this point when considering de Broglie-Bohm
model).

\subsection{ The Von Neumann Theorem}

The solution of macro-objectivation and measurement problem
together with EPR argument motivated therefore the search for
hidden variable alternatives to quantum mechanics. Furthermore, an
ulterior argument quoted for supporting HVT was the possibility of
defining quantum probabilities in the frame of relative-frequency
interpretation of probabilities \cite{HW}\footnote{More recently
HVT have also been suggested as a solution of the problems
connected with the definition of a wave function of the universe
in quantum cosmology \cite{thooft}.}. However, historically,
shortly after EPR paper the quest for hidden variable theories
stopped because of von Neumann's claim  \cite{vn} of having
demonstrated a theorem asserting the impossibility of constructing
a hidden variable theory reproducing all the results of QM. So far
the prestige of von Neumann led to an acritical acceptation of
this theorem, but it was then discovered that one of his
hypotheses was too restrictive and thus the program of the
construction of a hidden variable theory was still possible.

In the following we will sketch the von Neumann argument since
this discussion gives some hints on the peculiarity of HVT.

He considered the following situation: the measurement of a
particular quantity of an ensemble is found to yield different
values, although every system is in the same quantum state. We can
give two answers: or the systems are different because of some
"hidden" variable that our theoretical scheme is incapable of distinguishing, either
the systems are really in the same state and the dispersion of
measured values is due to nature itself (which is probabilistic).

The first alternative requires that the total ensemble comprises
as many sub-ensembles as there are different results and every
system in every sub-ensemble is in a dispersion free state,
characterised by a particular value of a hidden variable.

Von Neumann proved that, from assumptions he considered plausible
and  reproducing QM results, no dispersion-free states are
possible.

A hypothesis he introduced was that if  three (or more) operators
$A,B,C$ satisfy the relation $A+B=C$, then the values should
satisfy the relation $v(C)=v(A)+v(B)$, assumption that looks
rather natural for a classical theory. In quantum mechanics of
course the relation for average values $\langle \Psi \vert C \vert
\Psi \rangle =\langle \Psi \vert A+B \vert \Psi \rangle$ is
satisfied, but for non commuting operators one cannot assign
definite values $v(A)$ and $v(B)$ to observables $A,B$ simultaneously. Thus, von
Neumann, by introducing this hypothesis, made impossible the agreement with
QM "ab initio" (see for example \cite{bell66}).

For clarifying this point, let us consider a system of two spin
one half particles. Let $A=\sigma_x$ and $B=\sigma_y$: the
eigenvalues of Pauli matrices are $\pm 1$ and thus  $v(A)$ and
$v(B)$ can assume only the values $\pm 1$. It follows that $
v(A)+v(B)$ can assume the values -2,0 and 2. But $C=A+B$ is just
the operator corresponding to $\sqrt{2}$ times the component of
$\sigma$ along the direction bisecting the x and y axes. As a
result its allowed values are $\pm \sqrt{2}$ in disagreement with
the previous result. It is therefore completely evident that the
hypothesis  $v(C)=v(A)+v(B)$ prevents to satisfy QM results even
in extremely simple examples. This gives also an idea of which
properties, not typical of a classical theory, a HVT must
satisfy.

Probably, if this theorem would have been formulated by someone
less authoritative than von Neumann, it would have been criticised
much earlier (a first critics of the theorem remained completely ignored
\cite{vnJ}) than the Bell confutation of 1966 \cite{bell66} (where also similar arguments of Jauch and Piron \cite{JaPi} were
confuted).
This remains a interesting example of tortuous paths
of science development.

\subsection{ The Bell inequalities}

The subsequent step in discussing possible LHV extensions of QM
was the Bell's finding \cite{bell} that any realistic Local
Hidden Variable theory must satisfy certain inequalities that can
be violated in SQM, thus allowing an experimental test of the
validity of SQM respect to LHV.

More specifically, let us consider a physical system constituted
of two separated subsystems sent to two measurement apparatuses
that measure the expectation value of two dichotomic observables
$A_a,B_b$, $a,b$ being two parameters describing the
setting of measuring apparatus $A$ and $B$ respectively. The two
measurements are performed such to be space-like separated events.
Bell showed \cite{bell} that expectation values of the two observables satisfy
some inequality for every LRT,  which can be violated
in SQM for a specific choice of parameters.

Bell demonstration was based on considering the specific  case
where $A_a,B_b$ are the results ($\pm 1$ in $\hbar / 2$ unities)
of a measurement of spin component along directions $a,b$
respectively for a singlet spin state, Eq. \ref{Singlet}.

Introducing
the expectation value: \be [C(a,b)]_{\psi_0} = \langle \psi_0 |
(\sigma_1 \cdot a) (\sigma_2 \cdot b) | \psi_0  \rangle \ee one
has for parallel analysers \be[C(a,a)]_{\psi_0} = -1 \label{caa} \ee

Furthermore, the locality condition  is requested, stating that measurement of $A$ (B) depends
only upon $a \,(b)$ and $x$
  \be C(a,b) = \int_X A_a(x) B_b(x)
\rho (x) dx \label{loc} \ee
 Therefore, in a LHVT
relation \ref{caa} requires \be A_a(x) = - B_a(x) \label{AB} \ee for every
value of the hidden variable x (the states defined by the hidden
variable x, $\in X$, belong to a space which can be arbitrary. No
specific request on dimensionality or on linearity of operations
with it is required. Only a set of Borel subsets of $X$ is
defined, so that probability measures can be defined upon it).

From Eq. \ref{AB} we can calculate the following function involving
three different orientations of analysers, since $[A_b(x)]^2=1$: \bea | C(a,b) - C(a,c) | =
\left | \int_X [ A_a(x) A_b(x) - A_a(x) A_c(x)] \rho (x) dx \right | = \nonumber \\ \left |  \int_X
A_a(x) A_b(x) [1 - A_b(x) A_c(x) ] \rho (x) dx \right | \leq \int_X [1 -
A_b(x) A_c(x)]  \rho (x) dx \eea

where the last inequality follows from $A,B = \pm 1$.
By using the normalization \be \int_X
\rho (x) dx = 1 \ee follows the Bell inequality \be | C(a,b)
-C(a,c) | \leq 1 + C(b,c) \ee which is always satisfied in every LHVT
but can be violated in SQM, for example for $a,b,c$ coplanar, with
c making an angle of $2 \pi /3$ with a and b making an angle of $
\pi / 3$ with both a,c.

Albeit extremely interesting from a conceptual point of view, this
inequality has been derived with the request $[C(a,a)]_{\psi_0} =
-1 $, which cannot be hold in real experiments due to non unity
efficiency of real set-ups.

This problem was eliminated by Clauser-Horne-Shimony-Holt (CHSH)
who obtained the inequality \cite{CHSH}  \be | C(a,b)
- C(a,c)| + C(b',b)+C(b',c) \leq 2 \label{CHSH} \ee which is one
of the most often used Bell inequalities in experiments.
Incidentally, this inequality was derived for stochastic LHVT, i.e. for theories where hidden variables do
not determine the measurements results completely, so that they still remain random (i.e. one tries to
eliminate any "non-locality", but does not look for a completely deterministic theory).
However, Fine's theorem \cite{fine} stating that
"Necessary and sufficient condition
for the existence of a deterministic factorisable LHVT is the existence of a stochastic factorisable
LHVT for the same experiment" connects the two cases. Furthermore, more recently, it has also been
shown \cite{Perc}  that every probabilistic LHVT can be transformed in a deterministic LHVT by using
additional hidden variables.

Also in this case the demonstration is very simple. Let us consider the inequality:
\bea | C(a,b) - C(a,c) | \leq \int_X | A_a(x) B_b(x) - A_a(x) B_c(x) | \rho (x) dx = \nonumber \\  \int_X
|A_a(x) B_b(x)| [1 - B_b(x) B_c(x) ] \rho (x) dx = \int_X [1 -
B_b(x) B_c(x)]  \rho (x) dx \label{chsh1} \eea
Suppose now that for some $b,b'$ one has $C(b',b) = 1 - \delta$ with $0 \leqq  \delta \leqq 1$,
avoiding in this way the previous condition of perfect correlation (i.e. $\delta = 0$).
By dividing the set X into two regions
$X_{\pm} = \{x| A_{b'}(x) = \pm B_b(x) \}$ one has (being $C(b',b) = 1-\delta =
 \int_X B_b^2(x) \rho(x) dx - 2\int_{X_{-}} B_b^2(x) \rho(x) dx$)
\be
\int_{X_{-}} dx \rho(x) = {\delta / 2 }
\ee
and hence:
\bea
\int_X B_b(x) B_c(x)  \rho (x) dx & \geqslant \int_X A_{b'}(x) B_c(x)  \rho (x) dx -2 \int_{X_{-}} |A_{b'}(x) B_c(x)|  \rho (x) dx
= \nonumber \\ & = C(b',c)- \delta \,\,\,\,\,\,\,\, \,\,\,\,\,\,\,\, \,\,\,\,\,\,\,\,
\eea
From this result and Eq. \ref{chsh1} follows the inequality \ref{CHSH}.

Another example is the inequality proposed by Bell in 1971 \cite{bell71} \be S=|
C(a,b) - C(a,b')| +|C(a',b')+C(a',b)| \leq 2 \label{bellin} \ee
Also this demonstration was produced by using a generalisation of locality suited to
include systems whose evolution is inherently stochastic.

As a further example, let us then consider, still in detail,  the derivation of the
inequality proposed by Clauser-Horne \cite{noen},
which also includes inherently stochastic theories and has been often used in experiments.

 A source emits entangled particles, where
 the first particle goes to detector 1 and the second to detector 2.
 Let us suppose that before the detector $i$ we select a certain property $\theta _{i},$ for example,
 if the particles are entangled in spin, then $\theta _{i}$  is the angle defining the direction
 (respect to the z-axis) along which we are going to measure the spin. The Clauser-Horne sum then reads:

\bigskip

\begin{equation}
CH=P(\theta _{1},\theta _{2})-P(\theta _{1},\theta _{2}^{\prime
})+P(\theta _{1}^{\prime },\theta _{2})+P(\theta _{1}^{\prime
},\theta _{2}^{\prime })-P(\theta _{1}^{\prime })-P(\theta _{2})
\label{eq:CH}
\end{equation}
where $P(\theta _{1},\theta _{2})$ \ represents the joint
probability of observing a particle in 1 with the selection
$\theta _{1}$ and, in coincidence, \ a particle in 2 with the
selection $\theta _{2}$ (apices denote other angles choices).

On the other hand, $P(\theta _{i})$ \ represents the probability
of observing a single particle at $i$ with selection $\theta _{i}$.

Let us now suppose that these probabilities derive by a LHVT,
where $\rho (x)$ \ describes the probability distribution for the
hidden variable $x$. Then we have

\be P(\theta _{i})=\int dx\rho (x) P(\theta _{i},x) \ee and

\be P( \theta _{i},\theta _{j}  )=\int dx\rho (x) P(\theta _{i},
\theta _{j},x) \ee

Again, if the theory is local, then the measurement in 1 does not depend
on the choice of $\theta _{2}$ and viceversa. Thus, we have:

\begin{equation}
P(\theta _{1},\theta _{2},x)=P(\theta _{1},x)\ast P(\theta _{2},x)
\end{equation}
In order to demonstrate the Clauser Horne inequality let us now
consider an algebraic relation for 4 variables: x,x', which lie
between 0 and X ($X \leq 1$) and y,y', which lie between 0 and Y
($Y \leq 1$).

Then, it follows:

\begin{equation}
\  xy-xy' +x' y+x' y' -x' Y-yX\leq 0
\label{che}
\end{equation}

In fact, for $x < x'$, one can rewrite this equation  as \be x
(y-y') +(x'- X) y + (y'-Y) x' \le (x'-X) y + x (y-Y) \ee which is
negative.

On the other hand for $x \ge x'$, one rewrites Eq. \ref{che} as
\be (x-X)y + (y-Y) x' + (x'-x) y' \ee which again is clearly
negative.

By substituting $P(\theta _{1},x)=x$, $P(\theta _{1}^{\prime
},x)=x' $,$P(\theta _{2},x)=y$,$P(\theta _{2}^{\prime
},x)=y'$

and X=1, Y=1, we finally obtain the Clauser-Horne inequality valid for every LHVT

\begin{equation}
P(\theta _{1},\theta _{2})-P(\theta _{1},\theta _{2}^{\prime
})+P(\theta _{1}^{\prime },\theta _{2})+P(\theta _{1}^{\prime
},\theta _{2}^{\prime })-P(\theta _{1}^{\prime })-P(\theta
_{2})\leq 0  \label{ch}
\end{equation}

Beyond the ones already presented here, many different derivations
of Bell inequalities have been proposed,  based on slightly
different initial hypotheses, but all of them are substantially
equivalent \cite{books4} from an experimental point of view: in fact from Fine's
theorem \cite{fine} it follows that all Bell inequalities for two measurements
with two possible outcomes (as the ones discussed here) are equivalent to CHSH, Eq. \ref{CHSH}, in the sense
that states violating one of them violate CHSH as well. Incidentally,
the classification of Bell inequalities for a general number of measurements and outcomes is a very difficult, unsolved,
task (from a computational point of view it is a hard NP problem \cite{pitow}).
Some very recent results for 2,3 measurements/outcomes can be found in Ref.s \cite{colgis,sliw}.

Among various Bell inequalities demonstrations we can quote yet\footnote{We do not enter in details of
these proofs that can be easily find, together with further ones, in earlier reviews
\cite{CS,fp,selleri,MermB} and books \cite{books4}.}:

i) Wigner, Belinfante and Holt proof \cite{belinf,W70,CS}, based
on subdividing the space of states of a two-component system into
subspaces corresponding to various possible values of the
observable of interest and then studying measures on these
subspaces.

ii) Stapp's proof \cite{st1,st2} (see also \cite{st3}), which is
very general since it dispenses with all assumptions about the
state of the system and about probability measures on the space of
states pointing therefore strictly to locality assumption.

iii) Santos' proof \cite{sbell} based on comparison between
classical and quantum logic: for the classical Boolean lattice of
propositions it exists a metric satisfying triangle inequalities
from which one derives quadrilateral inequalities that may be
violated by a non-Boolean lattice of propositions (as the QM one):
these are the Bell inequalities.

Bell inequalities for generic n-level systems were also demonstrated by using
inequalities for Shannon entropy derived in classical information theory
\cite{braucav1,braucav2,schu} (e.g. $H(A|B) \leqslant H(A|B') + H(B'|A') + H(A'|B)$
where $H(A|B)$ is the conditional information for the two observables A,B).

 For other demonstrations (but the list is far
from being complete) the reader can refer to
\cite{desp,mas,sel,gol,af,grass,sm,wubell} and Ref.s therein.

\subsection{Quantum states violating Bell inequalities}

The next point to be clarified  is when a quantum state violates
Bell inequalities \cite{books4}. This point has been largely debated principally
due to its large relevance for quantum information: in the following we will only summarize some of the main results.

 In general two-particles pure entangled states violate some Bell inequality, as generally demonstrated in
Ref. \cite{CFS,gis91,GP92} where it was shown that for
every pure entangled state of two quantum systems is possible to find pairs of observables
whose correlations violate some specific Bell inequality.
For example, the two photons maximally polarisation
- entangled state $ \vert \psi \rangle = {\frac{ \vert H \rangle
\vert H \rangle + \vert V \rangle \vert V \rangle }{\sqrt {2}}}
$ (where H,V denote Horizontal and Vertical polarisations
respectively) maximally violates the former inequality \ref{ch} when the
parameters $\theta_i$, representing the setting of a polarizer
preceding photon detection, are opportunely selected (e.g. $\theta _{1}=67^{o}.5$, $
\theta _{2}=45^{o}$, $\theta _{1}^{\prime }=22^{o}.5$ , $\theta _{2}^{\prime
}=0^{o}$ ). Furthermore, for two particles systems, the amount of violation of the Bell
inequalities has an upper bound \cite{tsir}, e.g. $S \leq 2
\surd 2 $ for Eq. \ref{bellin}.\footnote{Further limits on quantum correlations, which could
eventually be violated by other probabilities sets, can be found in \cite{KT,Land,HY}.
For example \cite{HY} proved that considering two observables $A^a$, $B^b$ such that
$\langle A^a B^{b' } \rangle = 1-\epsilon_1$ and
$\langle A^{a'} B^b \rangle = 1-\epsilon_2$, $ 0 \leqslant \epsilon_1 \, ,
\epsilon_2 \leqslant 2$,
then in QM $| \langle A^a B^{b} \rangle - \langle A^{a'} B^{b' } \rangle | \leqslant \sqrt{2 \epsilon_1} +
\sqrt{2 \epsilon_2} + 2 \sqrt{ \epsilon_1 \epsilon_2}$.}
On the other hand, the situation is not so clear for more than two particles.

In Ref. \cite{popro,GoK} it was proven that no local realistic description is possible for
every pure entangled state of an arbitrary number
of particles, provided additional manipulations are allowed.

Mermin \cite{merm90}, and then Ardehali \cite{ard} and Belinskii and Klyshko \cite{belkly},
have shown  that  N spin-${1\over 2}$
 particles entangled states originate a violation, exponentially growing  with N, of the
  Bell inequality (MABK)

$F \leq 2^{n /2}$,  n even

$F \leq 2^{(n - 1) /2}$,  n odd

where $F$ is the average value of the operator ($\sigma$ are Pauli matrices)
\be
{1 \over 2 i } \left( \prod^N_{j=1} (\sigma_x^j+ i \sigma_y^j) - \prod^N_{j=1}
(\sigma_x^j - i \sigma_y^j) \right)
\ee
which, for example, has the value $F=2^{n}$ for the quantum state
${|\uparrow \uparrow ... \uparrow \rangle + i | \downarrow \downarrow ...\downarrow \rangle \over \sqrt{2}}$.

On the other hand, Scarani and Gisin  \cite{scargis} showed that the pure entangled states (dubbed
generalised GHZ states\footnote{GHZ equality will be described in paragraph 3.6.})
\be
\cos (\alpha) | 0,...,0 \rangle + \sin (\alpha) | 1,...,1 \rangle
\label{genGHZ}
\ee
 do not violate the MABK inequality for $\sin (2 \alpha) \leq 1 / \sqrt{2^{N-1}}$. Other specific cases were discussed
 in Ref. \cite{kz1,kz2,collins}.

Later, it was shown how exist pure
entangled states for $N \geq 2$ that do not violate
any Bell inequality for N particle correlation functions in experiments involving two dichotomic
observables \cite{zuk1,zuk2}. In more detail, Ref. \cite{zuk2} showed that all the states of Eq. \ref{genGHZ} with N odd and
$\sin (2 \alpha) \leq 1 / \sqrt{2^{N-1}}$ satisfy the Bell inequality \cite{zuk1,werwo}
\be
\sum_{s_1,...,s_N=-1,1}| \sum_{k_1,...,k_N=1,2} s_1^{k_1-1}...s_N^{k_N-1} E(k_1,...,k_N)| \leq 2^{N}
\label{zu}
\ee
where $E(k_1,...,k_N)$ is the correlation function that  for the local realistic case,
implying the existence of two numbers $A_j(\vec{n}_1)$ and $A_j(\vec{n}_2)$ of value $\pm 1$ describing
the predetermined result of a measurement by the $j$th observer of the observable defined by
$\vec{n}_1$ and $\vec{n}_2$ respectively, is given by the average over many runs of the experiment:
\be
E(k_1,...,k_N) = \left \langle \prod_{j=1}^N A_j(\vec{n}_{k_j}) \right \rangle_{avg}
\ee
Inequality \ref{zu} (not equivalent to MABK) is equivalent
to the full set of $2^{2^N}$ Bell inequalities for the correlations functions between measurements on $N$ particles
involving two alternative dichotomic observables at each local measurement station \cite{zuk1}.

On the other hand, it has been shown that inequalities involving more than two alternative measurements are stronger
\cite{wuz1,wuz2} and in particular  a general Bell inequality for $N > 2$ and many measurements settings was recently
derived \cite{las}, which is violated by a larger class of states, as for example all the states of Eq. \ref{genGHZ}.

Very recently some further progress was also done \cite{chen} about violation of
Bell inequalities for 3 qubits pure entangled states\footnote{Three qubits pure states form a five parameter family,
whose representation can be found in \cite{ac01}.}, showing that every 3-particle state of the form
\ref{genGHZ} violates some Bell inequality and presenting  numerical results indicating that all pure
3-qubits entangled states violate a Bell inequality.

This for what concerns pure states.
Nevertheless, it was shown that also specific
mixed states maximally violate (CHSH) Bell inequalities \cite{BMR}.

In order to proceed further, let us begin giving a more general definition of entanglement
 valid also for mixed states.
We define a state separable if its density matrix $\rho$ is of the form \be  \rho = \sum_i (\rho^A_i
\bigotimes \rho^B_i) w_i \label{sep} \ee where $w_i > 0$ and
$\rho^A_i$ $\rho^B_i$ are the density matrices of subsystems A and
B. Every non-separable state is called entangled \cite{Werner89}.

A separable state does not violate Bell inequalities. Various criteria for recognising separable states have been
established \cite{peres,horo,horo2,genbel,doh,Ioa}.
Among them, for two-dimensional Hilbert spaces, the Peres-Horodecki one \cite{peres,horo}:
 a state is separable iff the transposition of one of the subsystems
(partial transposition), with respect to any subsystem is positive.
However, for higher dimensional systems this condition is only necessary:
there exist entangled systems whose partial transpose is positive \cite{horo97,ben99}.

A complete characterization of separable states could be given in terms of entanglement witnesses \cite{horo},
i.e. a state $\rho$ is entangled iff there exists a Hermitian operator W (an "entanglement witness")
such that $Tr[W \sigma] \geq 0$ for all separable states $\sigma$, but $Tr[W \rho] < 0$. However, albeit some recent progress (e.g.
see \cite{Lew} and Ref.s therein), the characterization of entanglement witnesses is not known.

Even if the problem whether a general mixed
system violates or not Bell inequalities is still unsolved, specific cases have been discussed
\cite{las,Werner89,horoCHSH,arav,sen,barrett}.
For example, already in 1989 in a seminal paper for these researches Werner has shown \cite{Werner89}
that exist specific non separable
states  (i.e. states not of the form of Eq. \ref{sep}), dubbed Werner states, as
($I$ is the identity matrix)\footnote{In the
following we will use the conventional notation $\vert \Phi_{+,-}
\rangle = {\frac{ \vert 0 \rangle \vert 0 \rangle \pm \vert 1
\rangle \vert 1 \rangle }{\sqrt {(2)}}}$ and $\vert \Psi_{+,-}
\rangle = {\frac{ \vert 0 \rangle \vert 1 \rangle \pm \vert 1
\rangle \vert 0 \rangle }{\sqrt {(2)}}}$ for the so-called Bell
states, forming a basis for entangled states of two particles.}
\be
\rho_W = { 1 - F \over 3 } I + {4 F - 1 \over 3} | \Psi_- \rangle \langle \Psi_- | \,\,\,\,\,\,\,\, \,\,\,\,  F \leqslant 1/2
\ee
 that do not violate any Bell inequality with two projective measurements: a
 result obtained by explicitly building a
LHV model for these states (recently these states have also been
experimentally realized by using a PDC source of entangled photons \cite{WernerExp,WernerExp2}).
This achievement has then been generalised to the case of positive-operator-valued measurements (POVM) by
building a LHV model for POVM on a class of generalised Werner states \cite{barrett}\footnote{Explicitly
showing that a hypothesis suggested in Ref. \cite{teufel} is wrong.}.
 On the other hand, it has been shown that, if several such pairs are tested simultaneously, a
 violation of the CHSH inequality may occur, and no local hidden variable model is compatible with
 the results \cite{per96}. Also, Popescu has shown that by considering local measurements of
 the form $P \bigotimes I$ and $ I \bigotimes P$ on each subsystem  of a Werner state defined in
 a Hilbert space $H = C^d \bigotimes C^d$, where P is a projector on a two-dimensional subspace of $C^d$,
 one gets violation of Bell inequalities for a subensemble when $d \geqslant 5$: non-locality can therefore
 be revealed by sequences of local measurements \cite{pop95}. None the less, since in Popescu's example the observables
  leading to a violation of Bell inequality commute with the local measurements operators, these time
  sequences of measurements can be described by a single observable and Werner's LHV model can be applied,
  but requiring that later measurements influence preceding ones (a "hidden" violation of causality)
  \cite{teufel}.
A general result on the conjecture \cite{teufel} of the equivalence between separability and
the existence  of a "causal" local hidden variable theory, i.e. on the possibility to reveal non-locality
when arbitrary long sequences of general measurements and/or
 measurements on ensemble of states are considered, is still missing.

A very recent discussion of violation of some generalised Bell inequalities by Werner states and mixtures of
 W states (i.e. states of the form $ 1 / \sqrt{N} [|10...0\rangle + |010...0 \rangle + ...+|0...01 \rangle]$) with noise can be found
 in \cite{las,chen}.
Also, a sufficient and necessary condition
for violation of CHSH inequality for a two-dimensional Hilbert space has been presented in Ref.\cite{horoCHSH},
while a condition for a maximal violation of CHSH can be found in Ref. \cite{arav}.

Let us also notice that local actions with classical communication and postselection
(rejection of part of original ensemble) on a mixture that does not violate Bell inequality  can generate  a mixture violating them \cite{gis96,pur1,pur2,pur3,pur4},
 a procedure known as "distillation".
 It has been shown \cite{horo96} that this operation is
possible for every inseparable two level system. On the other hand, it has also been demonstrated \cite{horod} that
there are mixed states, called bound entangled, that are not distillable in spite of being entangled.

It can also briefly be acknowledged that the problem of which states violate Bell inequalities can be related to the one of
"measurement of entanglement" of large relevance in quantum information \cite{PV}.
For a pure state entanglement can be easily quantified in terms of the von Neumann entropy
($E = -Tr{\rho \ln{\rho}}$, where $\rho$ is the density matrix of the system).
The "entropy of entanglement" is defined as the von Neumann entropy of the reduced density matrix operator,
 which for pure states does not depend on which reduced density matrix is used. It satisfies what are reasonable
 requests for a measurement of entanglement, i.e.

i)  to be zero for separable states,

ii) to be invariant for local unitary transformations,

iii) not to increase for local operations, classical communications and subselection,

iv) the value for two factorised states is the sum of the values for these two states.

For mixed states the situation is more complicated and various measures of entanglement, satisfying former
criteria plus the one of reducing to entropy of entanglement for a pure state, have been proposed.
Among them: the relative entropy of the state \cite{REE,REE1}, defined
as the minimum of quantum relative entropy
($S(\rho || \sigma ) = Tr(\rho \ln \rho - \rho \ln \sigma )$ taken over the set $D$ of all separable states
$\sigma$, i.e. \be E(\rho) = \min_{ \sigma \in D} S(\rho || \sigma)\ee
 or the concurrence \cite{conc},
\be C(\rho) = max[0,\lambda_1 - \lambda_2-\lambda_3-\lambda_4]\ee
  $\lambda_i$ being the square roots of eigenvalues
of $\rho (\sigma_y \bigotimes \sigma_y) \rho^T (\sigma_y \bigotimes \sigma_y)$,  where $\sigma_y $ is
the Pauli matrix\footnote{Concurrence is monotonically related to another measurement of entanglement, entanglement of
formation \cite{EF}. For other measures of entanglement see \cite{genbel,neg,ill} as well.}.
The relevant point for our discussion is that it has been shown \cite{VW} how the violation $S$ of
CHSH inequality is, in terms of
concurrence, limited by $ max[ 1,\sqrt{2} C] < S < \sqrt{1 + C^2}$.

Finally, in discussing the states violating Bell inequalities,
 it is interesting to notice that the Quantum Field
Theory vacuum maximally violates Bell inequalities, as pointed out
by Summers and Werner \cite{vacuum1,vacuum2}\footnote{A later related result states that, by considering that
 a fully QFT description requires a q-deformed Hopf algebra of Weyl-Heisenberg one, the bosonic vacuum
 is a, entangled, generalised coherent state of $SU(1,1) \times SU(1,1)$ \cite{Iorio}.}
  in the frame of
algebraic approach to quantum field theory \cite{alg}. This violation vanishes exponentially ($\backsim
e^{-m r}$) with the
spatial separation $r$ of measurements with the length scale determined by the Compton wave length of the
lightest particle (of mass $m$) in the theory (or $\backsim 1 / r^2$ for the massless case). However, in
principle this vacuum entanglement could be 'extracted' to another
physical system \cite{BR} and to be experimentally verified.

\subsection{Theorem about no faster than light transmission by using EPR correlations}

Another theoretical point worth to be discussed before describing experimental tests of quantum non-locality,
is its compatibility with special relativity.

The main motivation of Einstein, Podolsky and Rosen to include the hypothesis
concerning the perfect locality of the system under consideration
derived by the necessity of having no instantaneous transmission
in agreement with special relativity.

However, as we will see, the non-locality of quantum entangled
systems does not allow any transmission of information faster
than light and thus does not raise any problem of compatibility
with relativity, albeit many opposite claims due to many different
authors, also very influent as Popper \cite{popper}.

For the sake of exemplification, before discussing the general theorem, we rapidly examine some
specific examples that allow a clear understanding of why quantum non-locality does
not permit superluminal communication.

Let us consider two observers
(as usual dubbed Alice and Bob) receiving respectively one particle
each of an entangled pair like the one described by the state in
Eq. \ref{Singlet}.

Let us Alice perform a spin test along the z direction: she
obtains a perfectly casual sequence of 1 and -1, each outcome with
probability 1/2. Let us now suppose that also Bob performs the
same test (in principle they could be separated of a space-like
distance and thus in a certain reference frame the Alice's
measurement is before Bob's one, whilst in other reference frames
Bob's measurement precedes Alice's one). Quantum mechanics
predictions tell us that Bob's results are perfectly correlated
with Alice's ones: every time Alice observes a 1 Bob has a -1 and
viceversa. Anyway, if there is no classical communication between
the two, the only result whose Bob disposes is a sequence of 1 and
-1 in a perfectly casual order: he cannot determine in any way if
Alice has performed a measurement or not, his results being a random sequence
 in both the cases. Thus, no information can be
transmitted between Alice and Bob in this way.

However, more complicated schemes can be conceived \cite{ghirdbb}. Let us suppose
for example that Bob could "clone" each particle he receives, i.e.
he put any particle into an apparatus creating 4N photons
exact copies of it.

Now let us also imagine that Alice can decide between performing a
test along the z axis or along a basis at $45^o$ respect to the z
axis. Let us consider the case where she chooses a test along z
and obtains 1 (-1),  1 denoting  spin up and -1 spin down respectively.
Then Bob uses the cloning system on his particle and
sends  N copies to four different apparatuses measuring the
spin along z, -z or the two conjugated directions of the
second basis respectively. He will observe, using z basis, N (0)
particles with -1 and 0 (N) with 1, while on the other basis he will
observe N/2 particles for both 1 and -1 outcomes. Exactly the same result
would be obtained, { \it mutatis mutandis}, if Alice had chosen to
perform the test in the other basis. Namely, in this case if Alice
obtains 1 (-1), Bob would observe, using the second basis, N (0)
photons with -1 and 0 (N) with 1, while on the z basis he will
observe N/2 particles for both outcomes, 1 and -1. Thus by simply looking for
which set-up he observes zero events, Bob would be able to know which
basis Alice has decided to use. Of course, such a knowledge could
be easily used to transmit a signal.

Nevertheless, also this scheme does not work. In fact a general
theorem has been demonstrated stating that it does not exist a way
of performing a general cloning for a quantum state \cite{noclo}.

The demonstration of this theorem is rather simple, and we report
it here for completeness.

Let us suppose to have a cloning machine which acts on an unknown
quantum state $| \Psi_1 \rangle $ and a second "target" state $|
\Psi_0 \rangle$ producing a copy of the first. The action of the
cloning machine can be described by a unitary operator $U$ through
\be | \Psi_1 \rangle \bigotimes | \Psi_0 \rangle \mapsto U (  |
\Psi_1 \rangle \bigotimes | \Psi_0 \rangle ) =  | \Psi_1 \rangle
\bigotimes | \Psi_1 \rangle \label{cloner} \ee

Let us now assume to apply the same cloning machine to an
arbitrary second state $| \Psi_2 \rangle $, the result would be
\be | \Psi_2 \rangle \bigotimes | \Psi_0 \rangle \mapsto U (  |
\Psi_2 \rangle \bigotimes | \Psi_0 \rangle ) = | \Psi_2 \rangle
\bigotimes | \Psi_2 \rangle \label{cloner2} \ee

taking the inner product of the equalities in the two former
equation we obtain:

\be \langle \Psi_1 | \Psi_2 \rangle = (\langle \Psi_1 | \Psi_2
\rangle )^2 \ee

but this relation requires that either $| \Psi_1 \rangle = |
\Psi_2 \rangle $ or $| \Psi_1 \rangle $ is orthogonal to $| \Psi_2
\rangle $, namely the cloning machine can clone only orthogonal
states and therefore a universal cloning machine is impossible.

The former examples give a clear hint of why quantum non-locality
cannot be used for transmitting faster than light communications.
Finally, at the beginning of 80's a general theorem was proved
\cite{GRW80} demonstrating the absolute impossibility of using
quantum non-locality for faster than light transmission, of which
in the following we sketch the demonstration .

Let us consider a composed system $S_1 + S_2$ described by a
statistical operator $W_{12}$. Performing a measurement on $S_1$
corresponds to project on a
specific eigenstate $|s \rangle$; this operation is described by a projection operator $%
P_s^1= |s \rangle \langle s|$. Thus if Alice performs a non
selective (i.e. where one keeps all the outcomes) measurement on
$S_1$, the statistic operator $W_{12} $ transforms according to:

\begin{equation}
W_{12} \rightarrow W^{\prime}_{12}=\sum_s P_s^1 W_{12} P_s^1
\end{equation}

However, all the information on the sub-system $S_2$ is contained in
the reduced statistical operator $W_2$, obtained taking the
partial trace on the Hilbert space $H_1$ (corresponding to subsystem 1):

\begin{equation}
W_{2} =Tr_1[W^{\prime}_{12}]=Tr_1[\sum_k P_k^1 W_{12} P_k^1]=
\sum_k Tr_1[P_k^1 W_{12} P_k^1]
\end{equation}

where we have used the fact that trace is a linear operation.

Then we have, due to the properties of the trace:

\begin{equation}
W_{2} = \sum_k Tr_1[P_k^1 W_{12} P_k^1] = \sum_k Tr_1[P_k^1 W_{12}
] = Tr_1[\sum_k P_k^1 W_{12}] = Tr_1[ W_{12} ]
\end{equation}

which is exactly the same operator we would have obtained without
any measurement on the subsystem $S_1$: thus there is no way to
distinguish where a measurement on system 1 has been made or not, by
performing measurements only on the system 2. The theorem has been
extended  to more general kinds of measurement \cite{GGRW88} and to the case of approximate cloning \cite{BDMS}
as well, guaranteing that no faster-than-light
communication is possible using QM non-locality.

Incidentally, it is interesting to notice that faster than light
communication, on the other hand, would be possible in HVT if one
could know hidden variables values (we will show an explicit example when discussing dBB model).
This fact substantially
forbids to have access to these variables.

\subsection{Toward an experimental test of Local Realism: introductory remarks}

The obvious utmost relevance of Bell inequalities derives from the fact  that they give a
completely general demonstration that every local realistic theory
cannot reproduce all the results of quantum mechanics. An
experimental observation of a violation of these inequalities represents
therefore a conclusive test against these kind of theories.

Nevertheless, unluckily,  to obtain a conclusive experimental
measurement of Bell inequalities is not an easy task.

First of all, referring for example to Eq. \ref{ch}, experimentally one measures
the number of coincidences $%
N(\theta _{1},\theta _{2})$ between two detectors, while in Eq. \ref{ch} the joint probability  $P(\theta
_{1},\theta _{2}) = N(\theta _{1},\theta _{2})/N$ appears, where
$N$ is the total number of pairs emitted by the source, which is
not really measurable because usually a large fraction of the pairs is
lost.

Anyway, by considering the ratio

\begin{equation}
R={[N(\theta _{1},\theta _{2})-N(\theta _{1},\theta _{2}^{\prime
})+N(\theta _{1}^{\prime },\theta _{2})+N(\theta _{1}^{\prime
},\theta _{2}^{\prime })] \over [N(\theta _{1}^{\prime } )+N(\theta
_{2})]}  \label{eq:R}
\end{equation}

$N$ cancels between numerator and denominator and for a LHVT, it is always $%
R\leq 1$, while in SQM it can reach the value $1.207$.

However, the fact that only a subsample of the total number of
produced entangled systems is really detected leads to the
necessity of an additional assumption: we have to ask that the
measured sample is a faithful representation of the whole. In
fact, in principle, this subsample could contain a distribution in
the hidden variables different from the total one, since the
hidden variable values can also be related to the  larger or
smaller probability of the state to be observed. This means that
if the observed sample is not a sufficiently large fraction of the
total one, the experiment is testing Local Realism plus the
additional hypothesis of having an unbiased measured subsample
\cite{santos,decaro,CH1,pea}, a problem known as {\bf detection
loophole}. From an inspection of inequalities \ref{bellin},\ref{CHSH} or
\ref{ch} one can deduce that a detection-loophole free test of LR requires
to observe, for a maximally-entangled state, at least a $82.84 \%$
of the total sample.

This loophole, as we will see, remains the
main unsolved problem for arriving to a conclusive test of LHVT
against SQM: all the experiments performed up to now were unable to solve it (in the few where this
did not happen other strong additional hypotheses were needed).

It must also be noticed that it has been discussed in
many different forms specific for the experiment under
consideration, as, for example, the request for polarisation
entangled photon states that the probability of counts with a
polarizer in place is less than or equal to the probability with
the polarizer removed \cite{noen}, known as no-enhancement
hypothesis.

In order to show explicitly an example of how this loophole manifests itself,
let us consider more in detail the effect of a low detection efficiency for Eq. \ref{eq:R}.
The presence of a low efficiency $\eta$ leads to have much less coincidences ($\propto \eta ^2 N$) than single counts
($ \propto \eta  N$) and therefore to a verification of the inequality $R\leqslant 1$. In order to overcome this problem
one usually substitutes in Eq. \ref{eq:R} single counts with coincidences with no selection on the second channel;
for polarisation entangled photons set-ups this substantially means to introduce the no-enhancement hypothesis \cite{CH1}.
Once this is done, denoting with $\infty$ the absence of selection, Eq. \ref{eq:R}

becomes
\begin{equation}
R={[N(\theta _{1},\theta _{2})-N(\theta _{1},\theta _{2}^{\prime
})+N(\theta _{1}^{\prime },\theta _{2})+N(\theta _{1}^{\prime
},\theta _{2}^{\prime })]\over [N(\theta _{1}^{\prime},\infty  )+N(\infty, \theta
_{2})]} \leqq 1 \label{eq:R2}
\end{equation}
that is the form effectively used in experiments.
Incidentally, in some of them it is used
the simplified form of Eq. \ref{eq:R2} obtained by requiring rotational invariance,
\be
R_{sym}={|N(\pi / 8) - N(3 \pi /8)| \over N(\infty,\infty)} \leqq {1 \over 4}
\label{Rsym}
\ee

The limit of a $82.84 \%$ detection efficiency is rather
difficult to be reached. An important theoretical indication for a
way of overcoming this problem has been obtain by Eberhardt
\cite{eb}, who showed, by a numerical minimization, that this
limit can be lowered to $66.7 \%$ for non-maximally entangled
states (with a smaller violation of the inequality), namely
entangled states where the different components have different weights.
As an example, the region where the Clauser-Horne
inequality \ref{ch} is violated in function of the detection
efficiency $\eta$ and of the degree of entanglement $f$ ($f=1$
for maximally entangled states, $f=0$ non entangled states) for
the state $ \vert \psi \rangle = {\frac{ \vert H \rangle \vert H
\rangle + f \vert V \rangle \vert V \rangle }{\sqrt {(1 + |f|^2)}}} $ is
shown in Fig. 1.

 Another request to be implemented for an
ultimate experiment on local realism is that the two measurements are really space
like separated (locality loophole), a condition that has been
recently well realised (see the following).

Finally, some of other conditions, more easily to be fulfilled from an
experimental point of view (albeit not always met in the experiments
described in the following), for a conclusive test concern the
temporal window of acquisition \cite{santos-t} and the possibility
of not having a background to be subtracted \cite{th}.

\subsection{ First experimental tests of Bell inequalities}

Many different systems have been considered in the literature (as
entangled pairs of ions, $K \bar K$, $\Lambda \bar \Lambda$
etc.) for realizing tests of Bell inequalities, but up to now most
of the experiments have been realized with entangled photons
since these systems present various advantages that we will
discuss in the following.

In the first experiments, performed in 70s and 80s, a polarisation entangled photon pair was
produced using a cascade atom decay or positronium decay.

Denoting with $H$ and $V$ the horizontal and vertical
polarisation, respectively, the state is

\begin{equation}
\vert \Phi_+ \rangle = {\frac{ \vert H \rangle \vert H \rangle +
\vert V \rangle \vert V \rangle }{\sqrt {2}}}  \label{Psi1}
\end{equation}

for the $J=0 \rightarrow 1 \rightarrow 0$ atom decay.

Whilst for positronium decay (as the positronium ground state has
an odd parity) it is:

\begin{equation}
\vert \Psi_- \rangle = {\frac{ \vert H \rangle \vert V \rangle -
\vert V \rangle \vert H \rangle }{\sqrt {2}}} \,\,\, .
\label{Psi2}
\end{equation}

Denoting with $a_{\theta} = a_H cos (\theta) + a_V sin (\theta)$
the annihilation operator with polarisation along a direction
making an angle $\theta$ with horizontal axis (while $a_{H,V}$
denotes annihilation operator for horizontally and vertically
polarized photons respectively), the two corresponding coincidence
probabilities are, remembering that single detection probabilities
are given by $P(\theta)=\langle \psi | a^+_{\theta} a_{\theta} |
\psi \rangle$ and joint probabilities by $P_{12}(\theta_1,
\phi_2)=\langle \psi | a^+_{\phi_2} a^+_{\theta_1}   a_{\theta_1}
a_{\phi_2} | \psi \rangle$,
\begin{equation}
P(\theta _{1},\theta _{2})=1/2\cos ^{2}(\theta _{1}-\theta _{2})
\label{p12c}
\end{equation}

and

\begin{equation}
P(\theta _{1},\theta _{2})=1/2\sin ^{2}(\theta _{1}-\theta _{2})
\label{p12s}
\end{equation}
respectively.

It can be easily shown that with a suitable choice of parameters
probabilities \ref{p12c}, \ref{p12s} allow a violation of Bell
inequalities. For example, the use of the joint probability
\ref{p12c}  in Eq. \ref{eq:R} leads to a maximal
violation of $R=1.207$ for $\theta _{1}=67^{o}.5$ , $\theta _{2}=45^{o}$, $%
\theta _{1}^{\prime }=22^{o}.5$ , $\theta _{2}^{\prime }=0^{o}$.

The main problem concerning positronium decay is the difficulty of
selecting polarisation of high-energy (gamma) photons produced in
the decay. This problem has substantially limited the results
obtained with this source. In short, since no
linear polariser exists for gamma rays, one must analyse the
polarisation by measuring the scattering distribution by means of
the Klein-Nishina formula and introduce the hypothesis that this
result can be correctly related by using QM to the one that would
have been obtained by using linear polarisers. On the other hand
detection efficiency is next to unity for gamma rays.

Experiments with positronium produced in atomic decays (as
$^{64}Co $ or $^{22}Na$) were realised by Kasday, Ullman and Wu
\cite{pos1} and then by others groups \cite{pos2,pos3,pos4}. Most
of them was in good agreement with SQM prediction (with the caution due to the
auxiliary assumptions that we have just discussed), with the
exception of Ref. \cite{pos2}.

More clear results (with less demanding additional assumptions)
have been obtained by using entangled photons produced in atomic
cascade decay, which are in the visible region of the spectrum so
that polarisation can be easily selected. However, in this case,
the atom takes away part of the momentum and thus photon
directions are not well correlated. This leads to the fact that
Eq. \ref{p12c} must be substituted with:

\begin{equation}
P(\theta _{1},\theta_{2})=1/4\eta _{1}\eta
_{2}f_{1}g_{12}(\epsilon _{1}^{+} \epsilon _{2}^{+}+\epsilon
_{1}^{- }\epsilon _{2}^{- }\ast F\ast \cos [2(\theta _{1}-\theta
_{2})]) \label{P12}
\end{equation}
where $\eta _{i}$ is quantum efficiency on channel $i$, $f_{1}$ is
the probability of the photon on branch $1$ to enter the
photo-detector, $g_{12}$ the conditional probability of photon 2
to enter the second detector when photon 1 entered the first, F
measures the degree of correlation in the initial pair and finally
$\epsilon _{1}^{\pm }=\epsilon _{1}^{||}\pm \epsilon _{1}^{\perp
}$, where $\epsilon _{1}^{||}$ and $\epsilon _{1}^{\perp }$ are
the transmission coefficients of the polariser for parallel or
orthogonal polarised light
 to the polariser axis respectively.

Also the single detection probability is modified in the following
way:

\begin{equation}
\ P_i(\theta )=1/2 \eta _{i} f_{i} \epsilon _{i}^{+}  \label{1}
\end{equation}

For atom cascade decay the parameter $F$ in Eq. \ref{P12} is
rather far from 1. Of the whole set of produced photon pairs, only
a small subsample is detected, leading to a large relevance of
detection loophole in this kind of experiments.

As discussed in the previous paragraph, due to low detection efficiency one is bound to
introduce additional hypotheses, as the no-enhancement one.
Because of this loophole, namely of selecting a
subsample of the total number of pairs, one is not simply testing
quantum mechanics against LHVT, but has also introduced a further
hypothesis, which states that the detected sample is a faithful
representation of the initially produced set of pairs.

Even if the detection loophole is rather severe for atomic decay
experiments, beautiful results were obtained using this technique,
giving first clear experimental indications against LHVT.

In general in these experiments an atomic beam emitted by an oven is excited by means of a
laser or an electron beam and then crosses a region between two symmetrically placed photo-detection
apparatuses where entangled photons produced in cascade decays are observed.

Among them one can quote the ones by Freedman and Clauser
\cite{ac1} (based on photon pairs at 551 and 423 nm produced in $4
p^2 \, ^1 S_0 \rightarrow 4p 4s \, ^1P_1 \rightarrow 4 s^2 \,
^1S_0$ cascade in calcium), of Clauser \cite{ac3} and of Fry and Thompson
\cite{ac4} (photons at 436 and 254 nm produced in $7 ^3S_1
\rightarrow 6 ^3 P_1 \rightarrow 6 ^1 S_0$ $^{200}$Hg decay), all
in agreement with SQM and showing a violation of Bell inequalities (having measured $R_{sym} = 0.300 \pm 0.08$,
$R_{sym} = 0.2885 \pm 0.0093$ and $R_{sym} = 0.296 \pm 0.014$ for Eq. \ref{Rsym} respectively).
On the other hand the Holt and Pipkin experiment \cite{ac2} (photons at 567 and,
 405 nm produced in $9 ^1P_1 \rightarrow 7 ^3 S_1 \rightarrow 6 ^3P_0$
$^{198}$Hg decay, where atoms were excited to $9 ^1P_1$ level by a 100 eV electron beam) led
 to results in disagreement with SQM and to no violation of Bell
 inequalities ($R_{sym} = 0.216 \pm 0.013$). However, in this case, a systematic error was
 identified in the form of stresses in the walls of the bulb
 containing the electron gun and mercury vapour \cite{CS} and the experiment of Ref. \cite{ac3}
 substantially reproduced this set-up finding a result in agreement with SQM.

 Typical detection efficiencies reached in this series of experiments were
 less than one per cent (e.g. $0.4 \%$ in Ref. \cite{ac1}).

In this list of 70's experiments one can also quote the
Lamehi-Rachti and Mittig experiment \cite{pp} , based on spin
correlations in proton pairs prepared by low-energy S-wave
scattering. When auxiliary assumptions are introduced, like the ones described in
positronium case, the results agree with SQM
(within limited statistics). A similar experiment, with indications again favourable to SQM, was later
performed by DM2 collaboration \cite{lam}, following a suggestion
of Ref. \cite{torn}, by using $\Lambda \bar \Lambda$ spin
correlated pairs: also in this case the spin was measured
indirectly (through $\Lambda \rightarrow \pi p$ decay).

For a more detailed description of these pioneer experiments see
\cite{CS,fp,selleri}.

\subsection{The Orsay experiment}

The season of cascade atomic decays set-ups was closed by the celebrated
Orsay experiment \cite {asp}, where for the first time the two
measurements were space-like separated with an observed Bell
inequalities violation of more than 5 standard deviations.

In synthesis, in this set-up \cite{asp} (developing a former experiment \cite{asp1})
the polarisation entangled photon pairs with a
wave-length of 422.7 and 551.3 nm were  generated by $ (J=0)
\rightarrow (J=1) \rightarrow (J=0)$ cascade in calcium-40 ($4
p^2 \, ^1 S_0 \rightarrow 4p 4s \, ^1P_1 \rightarrow 4 s^2 \,
^1S_0$  as Ref. \cite{ac1}), which is particularly suited
for coincidence experiments since the life time of the intermediate level is rather short ($ \tau =5$ ns).

The optimum signal to noise ratio for coincidences was obtained by reaching an excitation
rate of about $1 / \tau$ by exciting the atomic beam with a Kripton ion laser (at 406 nm)
and a tunable dye laser (at 581 nm) tuned to resonance for the two-photon process, controlled
by two feedback loops (on the wavelength of tunable laser and on the angle between lasers polarisations).

The entangled photons were then
addressed to detection apparatuses at 6 m from the source, constituted by a large-aperture
aspherical lens, followed by an interference filter, a transport optical system,
an acousto-optic device, polarisers and
photomultiplier tubes.

The  space-like separation between the two detections was obtained
by using rapid acousto-optic switches operating at 50 MHz, which
were selecting different paths for the incident photons in a way
that no communication of the selected basis for the polarisation
analysis was possible between the two different detection systems.
Thus, the two photodetections after polarisation
selection were really two non-causally connected events. Anyway,
some doubts about this statement have been raised since the switch
was not a random one but a periodic one. As we will discuss later,
 also this objection has recently been overcome by  new experiments
\cite{gislong1,gislong2,weihs}, finally eliminating in a uncontroversial way locality
loophole.
The acousto-optic switch was then followed, on each of the two paths,
by a polariser and a photo-multiplier
whose signal was addressed to a fourfold coincidence system.
Typical run time were around 12000 s.

The observed violation of Clauser-Horne inequality, $CH=0.101 \pm 0.020$, was
$83 \%$  of the maximal violation and in good agreement with theoretical SQM
prediction ($CH=0.112$) obtained by including polariser
efficiencies and collection solid angles. Of course, due to low detection efficiency,
this result was obtained by substituting single
counts with coincidences without polarisation selection on the second detector.

It is also worth to notice that in a previous version of this set-up \cite{asp2}
no acousto-optic device was inserted, but polarisation selection was performed by a polarising cube
followed by two photomultipliers in a way that the photon was always (modulus collection and detector
efficiency) observed. Also in this case an evident violation of CHSH inequality, $S= 2.697 \pm 0.015$,
was observed.

In summary, this experiment represented  the final result of the
series of cascade atomic decays ones and allowed to substantially
close the space-like loophole. Nevertheless, collection efficiency
was very low (with coincidences ranging between $0-40 s^{-1} $
against a typical rate of production of pairs of $ 5 \,  10^7
s^{-1}$). This low value, even smaller than the previous ones, was
mainly due to the necessity of reducing the divergence of the
beams in order to get a good switching.
 Thus,  detection loophole was very far from
being eliminated.

Epitomizing, the situation at beginning of 80's was that strong
indications favouring SQM against LHVT had been obtained, mainly in atomic
cascade experiments, arriving also to a substantial closing of
locality loophole. However, detection efficiencies were rather low
leaving a large space for criticisms based on detection loophole.
Furthermore,  limitations of the entangled pair production
techniques were such to leave no hope for relevant progresses in
this sense: since the low detection efficiency was mainly due to
low angular correlation of the pair, even an experiment with perfectly
efficient detectors would not have been resolutive \cite{santosAsp}.

A new kind of sources was therefore needed for progressing toward
a conclusive experiment: this happened in the 90's when
 spontaneous parametric down conversion in non-linear
crystals became largely exploited.

\section{PDC experiments on Local Realism}

\subsection{ The Parametric Down Conversion}

The parametric down-conversion (PDC), or parametric fluorescence,
is a quantum effect without classical counterparts and consists of
a spontaneous decay, inside a non-linear crystal, of one photon
from a pump beam (usually generated by a laser) into a couple of
photons conventionally called signal and idler (for an extensive description of
this phenomenon see \cite{Mandel}). This decay process obeys
(phase matching laws) to energy conservation \be \omega_0 =
\omega_i + \omega _s \ee

 and momentum
conservation \be \vec{k}_0 = \vec{k}_i + \vec{k} _s \ee where
$\omega_0,  \omega_i,  \omega _s$ are the frequencies and
$\vec{k}_0,  \vec{k}_i,  \vec{k} _s$ the wave vectors of pump,
idler and signal photon respectively.
 Furthermore, the two
photons are produced at the same time (within few tens femtoseconds,
as measured with an interference technique).

The probability of a spontaneous decay into a pair of correlated
photons is usually very low, of the order of $10^{-9}$ or lower;
therefore with typical pump power of the order of some milliwatts,
the fluorescence emission lies at the levels of photon counting
regime. Since the photons are produced in pairs and because of the
energy and momentum conservation restrictions, the detection of
one photon in a certain direction and with a given energy
indicates the existence of the pair correlated one, with definite
energy in a well defined direction.

As a first interesting application this property  allows the realisation of a "heralded single photon source",
where the observation of one photon of the pairs "certifies" the presence of the correlated one at a specific
frequency and in a determined direction.

In the type I PDC both photons are produced with the same
polarisation, orthogonal to the pump one.
Photons of equal wave length are emitted on concentric
cones centred on the pump laser
direction (see fig.2), whose diameter depends on the angle between the pump beam and the
optical axis of the crystal, the phase matching angle. When projected into a plane conjugated photons are on on the same diameter
and opposite respect to the
centre of the two concentric cicunferences corresponding to their wave-lengths.

In type II, one produced
photon has equal polarisation to the pump one, the other an
orthogonal polarisation. For an opportune phase matching angle (see later),
they are emitted on intersecting circumferences (see fig.3).

From  a theoretical point of view, in summary, the process of PDC in a crystal with active region of volume $V$
and susceptibility $\chi ^{(2)}$ can be described by the
Hamiltonian (where the sum is over modes allowed by energy and
momentum conservation):

\begin{equation}
H = {1 \over L^3} \sum_{k,s} \sum_{k',s'} E_l \chi^{(2)}_{i,j,l}
(\epsilon^*_{k,s})_i (\epsilon^*_{k',s'})_j \int_V d^3r [ e^{[-i
(k_0-k-k') \cdot r]} a(k,s) a(k',s') + h.c.] \label{HPDC}
\end{equation}

where $k_{0},k,k'$ are the quadrimomenta of the pump, idler and
signal photon respectively, $E_l$ is the vector amplitude of
the pump (strong enough to be treated as a classical field),
whilst $a(k,s)$ and $a(k',s')$ are the annihilation operators for
the produced photons (with polarisation $s,s'$).

Developing to the first order the evolution operator acting on the
vacuum, one obtains a state in the following form

\bea & |\Psi(t) \rangle = \exp{ [ -{i \over \hbar} \int^t_0 H(t')
dt' ]} | 0 \rangle = \nonumber \\
& | 0 \rangle + {1 \over L^3 i
\hbar} \sum_{k,s} \sum_{k',s'} E_l \chi^{(2)}_{i,j,l}
(\epsilon^*_{k,s})_i (\epsilon^*_{k',s'})_j \prod_{m=1}^3
{\sin{[(\vec{k_0} - \vec{k}
-\vec{k'})_m l_m/2]} \over (\vec{k_0} - \vec{k} -\vec{k'})_m/2 } \nonumber \\
&  e^{[i (\omega + \omega ' - \omega _0 ) t/2] }
{ \sin{ \left[ ( \omega + \omega ' - \omega _0 ) t/2 \right] } \over
(\omega + \omega ' - \omega _0 ) /2 }
 | k,s \rangle | k',s' \rangle
 \label{pdc}
\eea

The sum over $k,k'$ does not allow factorisation into a
product of signal and idler states: the state described by Eq.
\ref{pdc} is therefore an entangled state, which can be used for
Bell inequalities tests.

The specific experimental schemes based on this state will be the
argument of next sections; however, we can begin to point out which
is the principal advantage of this technique respect to the
cascade atomic decay one. In PDC the two correlated photons, due
to the conservation of angular momentum, are emitted within very
narrow cones whose specific width can be evaluated smaller than 1
mrad \cite{joo,manpdc}, depending on the kind and dimension of
non-linear  crystal and on the pump width. Due to this high
spatial correlation of entangled photons the collection
efficiency can be in principle very high, at variance with the atomic decay
case.

\subsection{PDC experiments with interferometers}

The type I PDC biphoton state described by Eq. \ref{pdc} presents
a phase and momentum entanglement that can be directly applied for a
Bell inequality measurement by using two separated interferometers
according to the scheme proposed by
Ref. \cite{franson}\footnote{A scheme for a Bell inequalities test by exploiting PDC
entanglement had also previously been  proposed by Klyshko
\cite{klyshko}.} and realized by
\cite{typei1,typei2,typei3,typei4} (some previous experiments based on this scheme
\cite{OZWM,kwprem,rarprem,frprem}
observed interference, but visibility was too low for a Bell inequality test).

The original idea of Franson \cite{franson} consisted in placing
two Mach-Zender interferometers on the path of the two entangled
photons (see fig.4). Let us suppose that the long arm
of the interferometers for the idler and signal photon have a
tunable phase $\phi_i$ and $\phi_s$ respect to the short one, the
final state is of the form \be \Psi_{fr}={1 \over 2} \left [ | s_1
\rangle | s_2 \rangle + | l_1 \rangle | l_2 \rangle e^{i( \phi_1 +
\phi_2)} + e^{i( \phi_1)} | l_1 \rangle | s_2 \rangle + e^{i(
\phi_2)} | s_1 \rangle | l_2 \rangle \right ] \label{frs} \ee
where the subscripts $1,2$ refer to the photon entering the first and
the second interferometer, while $s,l$ denote short and long path,
respectively.

After the interferometers the two photons are sent to detectors.
If both have followed short or long path they arrive in
coincidence, otherwise they are lost for the coincidence window.
Incidentally, this means that, also neglecting every other
inefficiency, only $50 \%$ of the pairs is selected.

The photons arriving in coincidence give a coincidence rate \be
R_c \propto \eta_i \eta_s \langle \Psi_{fr} | a_i^{\dag}
a_s^{\dag} a_i a_s | \Psi_{fr} \rangle = {1 \over 4} \eta_i \eta_s
[ 1 + cos(\phi_1 + \phi_2)] \label{Rcfr} \ee

where $\eta_i \eta_s $ are quantum efficiencies of detectors on
idler and signal path respectively. The striking fact about this
equation for the coincidence rate is that it can be modulated with
$100 \%$ certainty using either of the widely separated phase
plates. This "non-local" effect can be used for testing Bell
inequalities (e.g. as \ref{CHSH},\ref{bellin},\ref{ch}), where now the parameters are the
phases $\phi_{i,s}$.

In the experimental realization of \cite{typei2} a BBO crystal was
pumped by an argon ion laser beam in collinear regime producing photons pairs at 916 nm. A
beam splitter separated the pair in two orthogonal directions (of
course with an efficiency of $50 \%$ that further reduce the total
detection efficiency), addressed to the two interferometers. Detectors were silicon
avalanche photodiodes (cooled at -25$^o$) with a measured efficiency of $16 \%$. A 7
standard deviations violation of Clauser-Horne inequality \ref{ch} was observed. A
similar scheme with a higher violation, 16 standard deviations (only
inferred from a visibility of $80.4 \pm 0.6 \%$),
was also presented in Ref.s \cite{kw,kw2}.

On the other hand, a modified scheme for exploiting momentum -
phase entanglement was used in a previous experiment \cite{typei1}.
In this scheme \cite{HSZ} (see fig.
5) one selects four beams (A,B,C,D) corresponding to
the direction of two conjugated photon pairs. One can therefore
have idler and signal photons (eventually of different
wave-length) emitted either in directions $A,C$ or $B,D$
respectively, originating a state of the form (since one is
working in low rate regime, here and generally in the other experiments discussed in this review,
 the probability of having two pairs
emitted at the same time can be neglected) \be \Psi_{HSZ}={1 \over
\sqrt{2}} \left [ | k_A \rangle | k_C \rangle + | k_D \rangle |
k_B \rangle \right] \label{HSZs}\ee Then both  A,D and B,C are
recombined, after reflection on mirrors, on a beam splitter.
Before recombination a tunable phase shift $\phi_A$ and $\phi_C$
is introduced in paths A,C respectively. On each arm after the beam
splitter is placed a photon-detector and coincidences are measured
among them. Coincidence rate is modulated depending on the phases
$\phi_{A,C}$ in an analogous way to the previous case. The
advantage of this scheme is that in principle one can select all
the produced pairs.

 In the set-up of Ref. \cite{typei1} a kripton-ion
 laser operating at 413.4 nm pumped a KDP crystal. Photon pairs at
 826.8 nm were then selected. The visibility at coincidence level
 was of $82 \%$, leading to a violation of inequality
 \ref{bellin},
 $S=2.21 \pm 0.022$, of 10 standard deviations. The single photon detection rates were around
 $10^5 s^{-1}$ and coincidence rates around $500 s^{-1}$ (with, subtracted,
 accidental estimated to be $100 s^{-1}$),
 leading to an overall detection efficiency of $0.5 \%$  (detectors were silicon
 avalanche photodiodes operating in Geiger mode with quantum
 efficiency larger than $10 \%$). This number is already a little
 larger than the ones obtained in atomic cascade experiments, nevertheless remains
 still very far from what needed for a detection loophole free experiment.
Furthermore, the two measurements were not really space like
separated.

A post-selection free set-up, namely without the problem of eliminating the long-short terms of Franson's scheme,
was then realised \cite{strek} by using type II PDC and polarising beam splitters in the interferometers.
This results in only two different terms: either the horizontally polarised photon in channel 1 passes
through the long path, while
its vertical twin brother passes through the long path in channel two, or both take the short paths.
A $95.0 \pm 1.4 \%$ visibility was observed, but no real test of Bell inequalities was made.

The schemes based on energy-phase entanglement have then found an
interesting development since they are well suited for long
distance fiber transmission of entanglement.

 A first experiment on long distance entanglement transmission was
realised by Tapster et al. \cite{rarlong} by using the Franson
scheme. A pair of entangled photons at 820 nm and 1.3 $\mu m$ was
produced in a lithium iodate crystal pumped by an argon laser at
501.7 nm. The shorter wave-length photon was immediately addressed
to a single mode fiber interferometer, while the other passed
through a 4.3 km single-mode communications fiber before reaching
the interferometer ($1.3 \mu m$ is a well suited wave-length for
propagation in communications fibers).  A $86.9 \%$ visibility was
obtained, sufficient for claiming the possibility of violating
Bell inequalities, even if no results on these was really
presented.

A separation of more than 10 km was then obtained in 1998 \cite{gislong1,gislong2}. This experiment
is particularly interesting since, together with
another based on polarisation entanglement realised at the same time \cite{weihs},
definitively closed locality loophole. In this case a passive\footnote{the use of a passive coupler can
still lead same objections to the effective closing of detection loophole.}
coupler randomly selected which interferometer of the two available
(and therefore which measurement) analyses the photon. The observed CHSH inequality violation was
$S= 2.92 \pm 0.18$ ($2.38 \pm 0.16$ if accidental coincidences are not subtracted).

Since this experiment was bound to realise various
experimental high demanding tasks, it may be interesting to
rapidly describe its scheme. The PDC source consisted in a 655 nm
diode laser with an external cavity (10 mW) and a KNbO$_3$ crystal cut for type I PDC. The analysers were two
temperature controlled Michelson interferometers with Faraday
mirrors (for compensating birefringence in the arms) connected to the source by telecom fibers
and separated of 10.9 km
bee-line. Fibers were equalized at 1 mm level over 18 km (daily thermal expansion of several millimiters
had to be kept into account). Finally, photons were detected by germanium
avalanche photodiodes (with $5 \%$ quantum efficiencies). For this set-up all together the probability for detecting an
emitted photon pair was about $8 \,\,\, 10^{-6}$: a very large separation of detectors was therefore payed by
very low detection efficiency.

A later development of this set-up \cite{hugo}
allowed to realise measurements such that it was possible to
invert the temporal order of them by changing the reference frame ({\it before-before} configuration).
A condition that requires the temporal, $\Delta t$, and spatial
separation, $d$, of the two events  to satisfy the relation  \be | \Delta t
| < { v \over c^2} d \label{bb} \ee (where $v$ is the relative velocity between
the two reference frames) and is stronger that space-like
separation request, verified if \be | \Delta t | < { d \over c} \ee
In this scheme with two interferometers, on each side one of them was kept in the absolute future of the other.
The first detector was therefore the one determining the collapse of the wave function when the photon
 is absorbed in the first layers of photo-detectors.
Since the data could be collected by the second detector only, in the actual
realisation of the experiment (see fig. 6) the "choice device" was a 20 cm diameter black-painted
aluminium disk of 1 cm thickness turning vertically at 10000 rpm. During absorbtion
the rotation provided a good approximation of linear motion defining the moving reference frame.
Observed high visibility $83 \%$ was sufficient for guaranteeing Bell
inequalities violation (even if this were not really evaluated).
This result allowed to exclude some specific models of wave function collapse,
where this happens in the frame of the measuring apparatus (multisimultaneity model)
\cite{SS,ss2}.

The same Geneva group realised later a second experiment \cite{gisbb} further testing this theoretical model
where a high intensity PDC source based on Periodically Poled Lithium Niobate crystal pumped at 657 nm produced
energy-time entangled photons at 1314 nm (with a conversion efficiency 4
orders of magnitude larger than bulk crystals), which were addressed through optical fibers to two
Michelson interferometers, where the beam splitters were substituted by acousto-optic modulators (AOM). Since
the AOM is equivalent to a moving beam splitter it was possible to realise the {\it before-before} configuration of Eq. \ref{bb}.
Again the results were at variance with multisimultaneity predictions \cite{SS,ss2}.

For the sake of completeness, it is worth to notice that also a second kind of entanglement (time-bin)
beyond energy-time was realised with interferometers and PDC in pulsed regime.
The scheme is based on placing on the pump beam
a Mach-Zender interferometer (whose path length difference is large compared to
the pump pulse length) before the non-linear system where a
polarisation entangled pair is generated.
The pump photon can thus follow the short or the long path
originating the superposition \cite{tbent}:
\begin{equation}
\vert \Psi_p \rangle = {\frac{ 1 }{\sqrt{2}}} \left [ \vert s \rangle + e^{i
\phi} \vert l \rangle \right ] \label{ls}
\end{equation}
where $\vert s \rangle $ and $\vert l \rangle $ denote the photon which has
followed the short and the long path, respectively and $\phi$ the phase
difference between the two paths. The PDC process in the non-linear crystal
finally transforms the state of Eq. \ref{ls} in the entangled one
\begin{equation}
 {\frac{ 1 }{\sqrt{2}}} \left [ \vert s \rangle \vert s \rangle + e^{i
\phi} \vert l \rangle \vert l \rangle\right ] \label{tb}
\end{equation}
Even if no direct Bell inequalities test was presented  \cite{tbent}, the observed visibility, $84 \%$,
was sufficiently high for a violation of them, clearly exceeding the upper limit, $71 \%$, for separable states.
Later \cite{tblong} the robustness of this entanglement to decoherence was shown for a 11 km fiber propagation.

In summary, all together this experiments realised very interesting tests of local realism moving toward a conclusive one and
allowing very remote transmission of entanglement.
Nevertheless, in general, total detection efficiency was relatively small.

Before concluding this section is worth to point out how the
propagation of entangled photons to such large distances has found
an obvious application for realising remote quantum key
distribution (QKD) protocols. The main idea of these protocols
\cite{gisRMP} consists in transmitting a cryptographic key by
exploiting the correlation properties of quantum systems. The use
of specific set-ups \cite{QC1,QC2,QC3,Hefei}, analogous to the ones we have
just discussed, allowed QKD up to 150 km in fiber and of some tens kilometers in open air.
This technique is therefore reaching an applicative stage.

\subsection{First tests of local realism by using PDC polarisation entanglement}

A second possibility of using PDC photons for testing Bell inequalities
is to generate polarisation entangled states.

With the purpose of generating such states,  a first scheme is to
use a beam splitter for recombining idler and signal photons produced in type I PDC after
having rotated the polarisation of one of the two photons, see fig.7.

This scheme was the first set-up realised in 1988 by Shih and Alley \cite{SH} and Ou and Mandel \cite{ou}
 for testing local realism with PDC.

 In Shih and Alley set-up, after having  produced degenerate biphoton pairs by pumping a type I non-linear
crystal with a fourth harmonic of a Nd-doped yttrium aluminum
garnet laser (with 100 ps pulses), both photons crossed a $\lambda / 4$ plate,
transforming vertical polarisation into left circular one ($|L \rangle$) and were reflected by a mirror
originating the state
\be
| \Psi \rangle = e^{i \alpha} | R^A \rangle + e^{i \beta} | R^B \rangle
\ee
where $|R \rangle$ denotes right-hand circular polarisation state and $\alpha$ and $\beta$
are the phases associated with path A and B, respectively.

The photons were then superimposed on a beam
splitter,
generating the state: \bea | \Psi_{SH} \rangle = {1 \over 2}
[e^{i(\alpha + \beta)}  | R_1 \rangle | R_2 \rangle - e^{i(\alpha
+ \beta)}  | L_1 \rangle | L_2 \rangle + \nonumber
\\ e^{i(\alpha_1 + \beta_1)} | R_1 \rangle | L_1 \rangle -
e^{i(\alpha_2 + \beta_2)} | R_2 \rangle | L_2 \rangle ] \eea where
suffixes $1,2$ denote the final beam splitter exit and $\alpha_i$,
$\beta_i$ the phase of photon from path A,B to detector
$i=1,2$ ($\alpha_2 + \beta_1 = \alpha_1 + \beta_2 \equiv \alpha + \beta$).
A similar state with linear polarisation was also
obtained by placing a $\lambda / 2$ plate in path A only.

The state post-selected considering the cases when the photons
exit differently from beam splitter can be used for a test of
Bell inequalities. A violation of 3 standard deviations was
obtained ($R_{sym} = 0.34 \pm 0.03$).

A similar set-up was also used by Ou and Mandel \cite{ou},
with a violation of 6 standard deviations of Clauser-Horne inequality ($CH= (11.5 \pm 2.0) /min$).

Incidentally, this technique was recently used \cite{pitent} also for entangling
two photons coming from independent sources (one PDC "heralded" photon and one from an attenuated laser beam),
showing a violation of CHSH inequality, $|S|= 2.44 \pm 0.13$. It must be noticed that entanglement between
 photons, coming from independent sources (and violation of Bell inequalities), was also previously obtained
   \cite{swap} with quantum swapping by using type II polarisation entangled photons,
   as described in next paragraph\footnote{It must
be noticed that in these experiments, where only linear optical elements are used for entangling the two independent photons, only
a limited efficiency is obtained.
A possible way out would be the use of non-linear interactions as Kerr effect \cite{MGkerr}.}.

These set-ups allowed therefore a first realisation of
polarisation entanglement with PDC, nevertheless these kind of
schemes select unavoidably a $50 \%$ of pairs by using a beam
splitter for generating the state and therefore is not well suited
for approaching a detection loophole free test of local realism (a possible solution of this
problem was offered in Ref. \cite{phz}.
However, it requires photon-number resolving detectors: even if some progress toward realisation
of them has been obtained recently \cite{2pd1,2pd2,2pd3,2pd4,2pd5,ra},
reliable efficient photon-number resolving detectors are still far from being available.

\subsection{Bright sources of polarisation entangled photon
states}

In the last years various bright sources  of polarisation
entangled states of two photons (see fig.8) have been produced by
Type II PDC \cite{type2} or by superimposing two type I PDC
emissions, in this case based on two thin adjacent crystals
\cite{whi} or two crystals with an optical condenser between them
\cite{nos} or by inserting them in an interferometer
\cite{shih,shap} (incidentally one can notice that all of them can
be used for generating all four Bell states, $\Phi_{\pm}$ and
$\Psi_{\pm}$).

The schemes with type II PDC  \cite{gartII} are based on the fact
that in this case PDC correlated photons are emitted with
orthogonal polarisations.

In the collinear regime the two degenerate photons exit in two
tangent cones, if one selects the intersection point of them, the
two orthogonally polarised correlated photons exit  in the same
direction and can be separated by a beam splitter, generating the
entangled state of Eq. \ref{Psi2} when one postselects events
where photons have left the beam splitter on different paths
(taking therefore only a $50 \%$ of original pairs). The experimental implementation of this scheme \cite{kiess} led to  a 10 standard
deviations violation of Clauser-Horne inequality \ref{ch}.

In non-collinear regime \cite{typeiith}, if the angle
$\theta_{pm}$, between pump and crystal optical axis, is decreased,
the two cones separate from each other entirely. On the other
hand, if $\theta_{pm}$ is increased they intersect: therefore along
two correlated intersections a,b one superimposes the
probabilities of generating a H (V) or V (H) photon in direction a
(b). Nevertheless, this biphoton state is not yet entangled,
since, due to birefringence in the non-linear crystal, ordinary
and extraordinary photons propagate with different velocities and
different directions inside it. Longitudinal and transverse
walk-offs must therefore be compensated for restoring
indistinguishability between the two polarisations and generating
an entangled state. This is usually done by inserting some
birefringent medium (as quartz) along the optical path of photons.

The first experiment with non-collinear type II PDC has been presented
in Ref. \cite{type2}, where a pump beam at 351 nm (150 mW) pumped a
3mm long BBO crystal. The transverse walk off was estimated to be
negligible compared to coherent pump beam width. On the other
hand, the longitudinal walk off (385 fs) was larger than the coherence
time (determined by interference filters and irises) and  was
compensated by an additional BBO crystal. All four Bell states were generated.
A very strong violation, 102 standard deviations, of CHSH inequality  was reached ($S=-2.6489 \pm 0.0064$
with $\Psi_{+}$) with a
detection efficiency above $10 \%$, representing an
important step toward elimination of detection loophole.

A further important progress realised with type II sources was a
conclusive elimination of detection loophole (together with the experiment described in subsection 3.2
\cite{gislong1,gislong2}). As we have seen, one had already get rid of this loophole in the Orsay experiment,
except for the fact that the selection of measurement was driven
by a periodic (not random) signal. This last problem has been
solved in Ref. \cite{weihs}: polarisation entangled
photons, produced by pumping a type II BBO crystal with a 351 nm
argon-ion-laser, were largely separated by propagating in fibers.
The orientation of the polariser was determined by a random
number generator. The use of an active switch make this experiment
even more indisputable than the interferometer one \cite{gislong1,gislong2}.
Photons were then detected (as in the other
experiments described in this paragraph) by silicon avalanche
photo-diodes with dark counts rate lower than few hundred per
second (and therefore negligible respect to observed signal
counts). A typical observed value of the Bell parameter S was $S =
2.73 \pm 0.02$, i.e. a violation of 30 standard deviations respect
to local realistic predictions. Overall collection efficiency was
of $5 \%$. This relatively small value testify the difficulty of
eliminating at the same  time locality and detection loophole in
order to reach an ultimate experiment.

Concluding the presentation of  type II sources it must be noticed that recently
some very bright sources have been  obtained \cite{kurt,shapII,dig,aspel}.
A first example ($360800 s^{-1}$ coincidence
counts for a 465 mW pump power) was realised in cw regime with a careful maximization of
collection efficiency \cite{kurt}.
The total detection efficiency in the polarisation analysis set-up was $\eta= 0.214$.
A violation of 204 standard deviation of CHSH inequality \ref{CHSH}
was obtained with a measurement time of only 1s per angle setting.
A second example \cite{shapII} pumped at 397 nm a Periodically Poled KTP crystal
generating entangled state by splitting the emitted collinear degenerate photons
reaching a measured coincidence flux of $300 s^{-1}$ for mW of the pump.
The CHSH inequality was violated by $S=2.711 \pm 0.017$.
Finally, by using a type II source producing $20000$ entangled pairs per second (with as a
pump a violet diode laser at 405 nm and 18 mW power) it was possible to transmit
entanglement in open air for more than 600m    with a clear violation of
CHSH inequality, $S = 2.41 \pm 0.10$ \cite{aspel}.
Very recently this result was then extended up to 13km
\cite{Hefei} with a CHSH violation $S = 2.45 \pm 0.09$ (the source being a type II BBO crystal
pumped by a 300mW cw Argon Ion laser at 351 nm).

An alternative to use type II PDC is offered \cite{hardy} by
superimposing the emission of two type I PDC crystals whose
optical axes are rotated of $90^o$ producing two emissions with
orthogonal polarisations. If the optical distance between the two
crystals is smaller than the coherence length of pump laser, one
generates the entangled state:
\begin{equation}
\vert \psi _{NME} \rangle = {\frac{ \vert H \rangle \vert H \rangle + f
\vert V \rangle \vert V \rangle }{\sqrt {(1 + |f|^2)}}}
\label{Psif}
\end{equation}
The explicit value of the parameter $f$ can be fixed according to
the specific choices in the set-up. This is an interesting
property since, as we have discussed before, this non-maximally
entangled states may allow an elimination of detection loophole at
a lower detection efficiency ($66.7 \%$) than the one needed for
maximally entangled ones ($82.8 \%$).

A first scheme used for this purpose is based on superimposing the
emission of two thin adjacent type I crystals. More in details, in
the first example of this kind of source  \cite{whi}, an argon
laser beam at 351 nm, pumped two adjacent BBO crystals 0.59 mm long
with optical axes oriented orthogonally. A rotatable half-wave
plate on the pump beam before the crystals allowed to tune the
laser beam polarisation and therefore the parameter $f$ of the
generated state, Eq. \ref{Psif}. The source was rather bright, giving a
$21.000 s^{-1}$ coincidence rate for 150-mW pump power (an order
of magnitude larger than previous type II sources). A large
violation of Bell inequalities (for maximally-entangled states),
$S= 2.7007 \pm 0.0029$, was observed.\footnote{A further test of
local realism by equalities realised with this scheme by using
non-maximally entangled states will be discussed in a subsequent
paragraph.}

The same year a second source was realized \cite{nos,nos2} by
using an optical condenser for superimposing the two emissions.
This scheme in principle allows a very precise superposition of the whole
parametric fluorescence  even
with long crystals allowing for higher intensities. More in
details, in this scheme (see fig.9) the are two crystals of LiIO$_3$ 250 mm
apart, a distance smaller than the coherence length of the pumping
laser (a 351 nm argon laser beam). This guarantees
indistinguishability in the creation of a couple of photons in the
first or in the second crystal. A couple of planoconvex lenses with 120 mm focal length, centred in between, focalises the spontaneous
emission from the first crystal into the second one, while
maintaining exactly, in principle, the angular spread. A hole of 4
mm diameter is drilled into the centre of the lenses to allow
transmission of the pump radiation without absorption and, even
more important, without adding stray-light, because of
fluorescence and diffusion of the UV radiation. The pumping beam
at the exit of the first crystal is displaced from its input
direction by birefringence: a small quartz plate (5 x5 x5 mm) in
front of the first lens of the condenser compensates this
displacement, so that the input conditions are prepared to be the
same for the two crystals. Finally, a half-wavelength plate
immediately after the condenser rotates the polarisation of the
Argon laser beam that excites in the second crystal a spontaneous
emission which is cross-polarised with respect to the first one.

The coincidence rate was analogous to the previous scheme, while a
test of Bell inequalities by using non-maximally entangled states
with $f \simeq 0.4$ led to a violation of Eq. \ref{eq:R2} $R=1.082 \pm
0.006$ when the set of polarisers specific for maximising the
violation for this non-maximally entangled state was chosen
($\theta_1 =72^o.24$, $\theta_2=45^o$, $\theta_1 ^{\prime}=
17^o.76$ and $\theta_2 ^{\prime}= 0^o$).

In the last years, many studies were devoted to produce sources in pulsed regime
(which are particularly useful for quantum information for timing reasons).
When the pump pulses are very short (typically hundreds of femtoseconds), amplitudes for
photon pairs produced at different depth inside the crystal become distinguishable, reducing
two-photon interference visibility \cite{kr}. This problem required either to use thin ($\approx 100 \mu m$)
non-linear crystals \cite{sergthin} or narrow band spectral filters (for increasing coherence length) in front of detectors
\cite{gr1,gr2,digiu}. However, these solutions significantly reduce available flux of entangled photon pairs.

Quite recently, a bright source  \cite{shih} was obtained by pumping
with a femtosecond mode locked laser two type I BBO crystals
inserted in a Mach-Zender interferometer.

A bright source (always in femtosecond pulsed regime) was also obtained
by addressing back, both rotated by a $\lambda / 4$ wave plate,
the PDC emission and the pump beam to the same type I crystal by
means of a spherical mirror \cite{roma,mat}. A 213 $\sigma$ violation of Bell inequality was observed.

A scheme based on an interferometer was
also realised for cw regime \cite{shap}.
The scheme consists in a Mach-Zender interferometer
where the pump enters through a first beam splitter pumping two identical type II crystals
inserted each in a different arm (A,B) originating the biphoton state
\be {\frac{ \vert H_A (\omega_s) \rangle \vert V_A(\omega_i) \rangle +
\vert H_B (\omega_s) \rangle \vert V_B(\omega_i) \rangle }{\sqrt {2}}} \ee
After the polarisation on one of the arms has been rotated by a half-wave plate the two emissions are
recombined on a polarizing beam splitter producing the entangled state ${\frac{ \vert H_1 (\omega_s) \rangle \vert V_2(\omega_i) \rangle +
\vert V_1 (\omega_s) \rangle \vert H_2(\omega_i) \rangle }{\sqrt {2}}}$, where 1,2 refer to the two PBS ports.
This source (whose practical implementation was based on a single crystal with counterpropagating pump beams)
reached a flux of entangled photons of 12000 pairs $s^{-1} $ mW$^{-1} $ and
100 $\sigma$ violation of CHSH inequality. Detection efficiency was $\sim 18 \%$.

Finally, again in cw regime, in the set-up of Ref.\cite{Bu} two collinear type I PDC emissions
(after having rotated one of them by a half-wave plate) were superimposed in a modified
Mach-Zender scheme.

Altogether the realisation of these high efficiency collection
very bright sources of polarisation entangled photons points out a possible way for
reaching an ultimate test of local realism. At the moment the
principal problem remains the detection loophole: the highest
detection efficiencies reached so-far are some tens per cent,
whilst they should arrive, as we have seen, to $82.84 \%$ for maximally-entangled
states or $66.7 \%$ for non-maximally entangled ones. The more
stringent limitation for reaching highest efficiency is in the
quantum efficiency of detectors (at the moment around $70 \%$ for
commercial avalanche photo-diodes detectors in visible spectrum).
Nevertheless, technological progresses, in particular in
superconductor based detectors (see for example \cite{ra,ros,y}), could led in the next years to a large
improvement of these limits.

\subsection{Applications of bright sources of entangled photon pairs}

For the sake of completeness, it is worth to point out that the
bright sources of entangled photon pairs described in the two former subsections
have found very important applications in the developing field of Quantum Information and related
areas of research (as Quantum Imaging and Quantum Metrology).

Quantum Information is an exponentially growing area of physics with promising technological applications
addressed to study codification, elaboration and transmission of information by using specific properties of Quantum
States.

A description of this field is largely beyond the purposes of this review and can be found, for example,
in \cite{Qcrip,Qinf,Qinf2,Qinf3,Qinf4}. Here, we would like only to mention that entanglement
and related quantum non-locality are the main resources exploited by these applications.

Among the main results of these studies we may quote the discovery that a quantum computer could efficiently
solve problems that do not have efficient algorithms on a classical one (as factorisation in
prime numbers), the ideation and practical implementation of absolutely secure protocols of communication
(as already mentioned), the realisation of quantum communication schemes without a classical equivalent
as teleportation (i.e. remote reconstruction of an unknown quantum state by sharing an entangled
state and classical transmission of information of a measurement on the state to be teleported),
dense coding (encoding two bits of information by manipulating only one subsystem of a shared entangled
state) and quantum swapping (teleportation of entanglement), under shot-noise measurements in interferometers
 by using entangles states, etcetera\footnote{Of course these results are obtained in SQM, some of them
 should be revised in HVT framework and in particular if hidden variables were (eventually partially)
 accessible.}.

In particular, applications of sources of entangled photons to this field range from quantum
cryptography \cite{QI1,QI2,QI3,QI4,QI5,QI9},
teleportation protocols \cite{QI6,QI7,QI8,QI10,swap}, quantum
imaging \cite{QI11,QI12,QI13}, linear quantum optical gates
\cite{QI14,QI15,QI16,QI17,QI18} (which are a fundamental element for building a quantum optical computer),
quantum metrology \cite{Qmet2,Qmet3,Qmet4},
entanglement manipulation \cite{QI20,QI21,QI22,QI23,QI24,QI24b},
quantum tomography \cite{QI25,QI26}, etc.

\subsection{Tests of local realism by equalities}

Beyond Bell inequalities Local Realism can also be tested by
measuring some specific product of observables for entangled
states for which the results of SQM and LHVT are different.

A first example was given by Greenberger-Horne-Zeilinger
\cite{GHZ,GHSZ}.

Let us consider a three photon entangled state
\be \Psi_{GHZ}= {1 \over \sqrt{2}} ( \vert H \rangle \vert H
\rangle \vert H \rangle + \vert V \rangle \vert V \rangle \vert V \rangle ) \label{ghz} \ee

Rewriting the state by using the bases \bea & \vert 45
\rangle = {\frac{ \vert H \rangle + \vert V \rangle }{\sqrt {2}} }
\, \,\,\,\,\,\, \vert -45 \rangle = {\frac{ \vert H \rangle -
\vert V
\rangle }{\sqrt {2}} }  \\
 & \vert R \rangle = {\frac{ \vert H \rangle + i \vert V \rangle
}{\sqrt {2}} } \,\,\,\,\,\,\,  \vert L \rangle = {\frac{ \vert H
\rangle - i \vert V \rangle }{\sqrt {2}} } \label{RL} \eea one has
\bea \Psi_{GHZ}= {1 \over 2} ( \vert R \rangle \vert L
\rangle \vert 45 \rangle + \vert L \rangle \vert R \rangle \vert
45 \rangle + \nonumber \\ \vert R \rangle \vert R \rangle \vert
-45 \rangle + \vert L \rangle \vert L \rangle \vert -45 \rangle )
\label{ghz2} \eea or \bea \Psi_{GHZ}= {1 \over 2} ( \vert
45 \rangle \vert 45 \rangle \vert 45 \rangle + \vert 45 \rangle
\vert -45 \rangle \vert -45 \rangle + \nonumber \\ \vert -45
\rangle \vert 45 \rangle \vert -45 \rangle + \vert -45 \rangle
\vert -45 \rangle \vert 45 \rangle ) \label{ghz3} \eea This state
has some significant properties. First of all any individual or
two-photon joint measurement is maximally random. Secondly, if one
attributes the value $+1$ to $R,45$ measurements and the value
$-1$ to $L,-45$, then the state is such that the product of three
measurements is always $-1$. Thus once two
measurements are known, the third can be inferred with certainty
without performing it: it is therefore an element of reality
according to EPR definition.

Let us then consider a measurement on the basis $45,-45$ for all
the photons in the framework of a local realistic model. From Eq.
\ref{ghz2} derives that whenever the result $45$ ($-45$) is
obtained for one photon, the other two must carry opposite
(identical) circular polarisations. Let us then consider the
specific example where one measures a $-45$ polarisation both for
photons 2 and 3. Since photon 3 has $-45$ polarisation, in a hidden variable framework photons 2
and 1 must have equal circular polarisations. On the other hand,
since photon 2 has $-45$ polarisation, photons 1 and 3 must have
equal circular polarisations as well. If circular polarisations
are element of reality fixed by some hidden variable, then all the
three photon must have identical circular polarisations, but if
photons 2 and 3 have identical circular polarisations it follows
that photon 1 has linear polarisation $-45$, thus one can
simultaneously measure the outcomes $-45,-45,-45$: but this is
at variance with SQM result deriving from Eq. \ref{ghz3}.
Similar results are obtained for other outcomes as well. There is
therefore a sharp difference between SQM and LHVT.

This result implies that if one makes three space-like separated
measurements on a suitable entangled state on an opportune basis he
obtains a completely different result according if SQM or LHVT are valid.
Nevertheless, also in this case one cannot obtain a conclusive
test if detection efficiency is not sufficiently high, i.e.
detection loophole appears here as well \cite{GHSZ,lar}. The
presence of detection loophole is substantially due to the fact
that some specific subset of hidden variables can simply
correspond to undetected events in presence of a certain
measurement. Thus, if detection efficiency is not sufficiently high,
the set of hidden variables which would give results at variance
with SQM could simply correspond to  undetected events. This limit is
rather stringent requiring detector efficiency above $90.8 \%$
\cite{GHSZ} if emission rate of particle triples is known or a
ratio between triple and double coincidence rate above $75 \%$ in
a general case \cite{lar}.

Even if an experimental realization of GHZ test presents
these problems concerning a conclusive test of local realism,
it remains very interesting. In the last years GHZ
entanglement has been realised\footnote{An indication of GHZ effect
was previously observed in a NMR experiment \cite{GHZNMR}, however
in this case the "experiment was performed on thermal states,
[thus] though the spins in the experiment mimic the effects of
entanglement, they are not in fact entangled."}, both for the
original version of three entangled particles \cite{GHZexp,GHZexp2} and
for extensions to four photons \cite{zhao}. Both the experiments
gave results in agreement with SQM and at variance with LHVT
within experimental uncertainties (see fig. 10).

Let us sketch how GHZ polarisation entangled state of Eq.
\ref{ghz} were generated \cite{GHZexp,GHZexp2}. The scheme
consisted in transforming two pairs of polarisation entangled
photons produced simultaneously in a type II crystal pumped by a
high intensity UV 200 fs pulse into three entangled photons by
using postselection \cite{GHZp}. More in details, in some rare
event two entangled pairs ${\frac{ \vert H \rangle \vert V \rangle
- \vert V \rangle \vert H \rangle }{\sqrt{2}}} $ were produced by
the same pulse. The selection of the desired state was then
obtained by inspecting a posteriori the four-fold coincidence
recording obtained by the apparatus in fig. 11: the photon
registered at detector T is always horizontally (H) polarized and
thus its partner in b must be vertically (V) polarised. The photon
reflected at the polarising beam splitter in arm $a$ is always V,
being turned into equal superposition of V and H by a $\lambda /2$
wave-plate, and its partner in arm b must be H. Thus if all four
detectors click at the same time, the two photons at detectors
$D_1$ and $D_2$ must either both have been VV
 or HH. The photon at $D_3$ was therefore H or V, respectively.
 The indistinguishability of both cases was obtained by using
 narrow band filters (4 nm) to increase coherence time to about
 500 fs. The observed outcomes (see fig. 10) agree very well with SQM predictions.

Extension of this scheme were then realised for entangling 4
\cite{zhao,4en} or 5 photons \cite{zhao5}. Even if the experimental
difficulty in realising this state limits the real possibility of
observing a violation of local realistic predictions, nevertheless
these achievements are interesting since it has been shown (as hinted before)
that violations of local realism
become stronger with the increasing number N of entangled particles
\cite{merm90,Massar2,colhyb}, in the sense that SQM can violate specific Bell inequalities by
an amount growing with N and with a lower detection efficiency limit for a detection loophole free experiment.

Finally, let us notice that tests of local realism based on equalities were also
proposed by Hardy (and others) for two-particle entangled
states \cite{hareq,jordan,tbm,goldeq} (a previous demonstration was given for the six dimensional
space of spin 1 particles \cite{HeRe}).
In synthesis, this result can be obtained by considering a polarisation entangled state of the form
$\alpha |H \rangle  |H \rangle  - \beta |V \rangle |V \rangle $. On photon 1 (2) one then performs polarisation
measurements along one of the n+1 possible directions $A_i$ ($B_i$) with $0 \leq i \leq n$,
corresponding to project on the states $|A_i \rangle$
($|B_i \rangle$).
Hardy's theorem states that the propositions about joint probabilities,
where $A_i = 1$ ($A_i = 0$) means that $A_i$ has been measured with outcome
$A_i$ (its ortoghonal):

\bea
P_i= P(A_i =1, B_i=1) \neq 0 &  \cr
 P(A_i =1, B_{i-1} = 0) = 0 & \,\,\,\,\,\, \,\,\,\,\,\, \,\,\,\,\,\, \,\,\,\,\,\, \,\,\,\,\,\,  1 \leq i \leq n \cr
 P(A_{i-1} =0 , B_i=1) = 0  & \,\,\,\,\,\, \,\,\,\,\,\, \,\,\,\,\,\, \,\,\,\,\,\, \,\,\,\,\,\,    1 \leq i \leq n \cr
 P(A_0 =1, B_0=1) = 0 &
\label{Heq}
\eea
for some specific choice of $A_i,B_i$ lead to contradiction with local realism, but can be verified in SQM.

The experimental realisation of this
scheme has been made by Rochester \cite{roceq,roceq2} and Rome \cite{dem2,dem3,dem4}
groups by using polarisation entangled
photons with results in agreement with SQM.

In little more detail, in Ref. \cite{roceq,roceq2}
equalities \ref{Heq} are directly checked. Since, due to experimental imperfections, one does not measure
exactly zero results where expected, an estimate of the probability $P_i$ is obtained from these data by
using EPR arguments, showing how this estimate is fourteen time smaller than the measured one,
contradicting local realism of about 45 standard deviations
($0.0070 \pm 0.0005$ respect to $0.099 \pm 0.002$). A similar comparison is performed in Ref. \cite{dem2}
as well (with  14 standard deviations from the local realistic prediction); whilst in Ref. \cite{dem3,dem4}, to avoid problems associated with a {\it nullum experiment}, is tested
an inequality on $P_i$ obtained by associating
Hardy theorem with Clauser-Horne inequality. Results are at variance with LR predictions (30,37,26 and 21 standard deviations
for $i=4,5,10,20$, respectively \cite{dem4}).

Nevertheless, also for Hardy's equalities
the detection loophole reappears \cite{Gar,sroc}, requiring a $82.84 \%$
collection efficiency for a conclusive test with maximally
entangled states (the same as for Bell inequalities). The
experimental efficiency was of $\approx 10 \%$ in Ref. \cite{dem2}.
A test of these equalities, always in agreement with SQM, was also realised with non-maximally
entangled states \cite{kweq}.

In synthesis, experiments about local realism based on equalities represent a very sharp test of SQM against LHVT. None the less also
in this case detection loophole appear as the strongest limitation toward a conclusive experiment. This limit is more
difficult to overcome for GHZ scheme, whilst a resolutive test of LR by Hardy's two particle scheme is in
a situation analogous to Bell inequalities case.

\subsection{LHVT built for surviving PDC experiments}

As we have seen in the former paragraphs, the most recent PDC
experiments have posed very strong constraints on the existence of
LHVT. Nevertheless, some space for LHVT is still left by the detection
loophole.

Let us briefly comment on some specific models that have been
built in order to show explicitly how one can still build a LHVT within the
limits of quantum optical tests of local realism\footnote{For the
sake of completeness, it is worth to notice that recent attempts
to build a LHVT non-violating Bell inequalities in general
\cite{acc1,acc2,acc3,HP} have been shown to be incorrect
\cite{g1,z,app}.}.

A first example of this kind of models appeared in Ref. \cite{pea}
and was then excluded by more recent experiments. However, it gave a general
scheme for building a LHVT exploiting detection loophole. The main idea is to
consider not only 2 outcomes ($\pm 1$) of the measured variables, but three ($\pm 1$,
undetected) and to build a hidden variable distribution able to reproduce SQM predictions
for the events where both the particles have been observed.

This scheme has then be applied, more recently, for building
new LHVT models  \cite{GG,l,M,d},  with the specific
purpose of giving examples of LHVT not yet excluded by present
experiments.

In order to give an idea of how they are built, let us present a very simple one,
built for a singlet state of two spin $1/2$ particles \cite{GG}.
Each particle is characterised by a hidden variable $\vec{\lambda}$ with a uniform a priori
probability distribution and by its quantum state $\rho$. For the singlet state the vectors $\vec{\lambda}$
are opposite for the two particles.
If the spin is measured along a direction $\vec{a}$, the outcome $\pm 1$
is determined by the sign of the scalar product $ (\langle \vec{\sigma} \rangle _{\rho} - \vec{\lambda}) \cdot \vec{a} $
where $\langle \vec{\sigma} \rangle _{\rho}$ denotes the expectation value of the Pauli matrix.
Furthermore, one assumes that at one of the measurement apparatuses ($A$) an outcome is produced only with probability
$| \vec{\lambda_A} \cdot \vec{a}|$. The resulting correlation function is
\be
E(\vec{a},\vec{b}) = \int d\vec{\lambda} \rho(\vec{\lambda} | \, outcome \,\, produced) sign(\vec{\lambda} \cdot \vec{a} )
sign(-\vec{\lambda} \cdot \vec{b} ) = - \vec{b} \cdot \vec{a}
\ee

as in SQM. Single particle distributions are also correctly reproduced.
The model can then be  modified \cite{GG} for being symmetric for the two measurement apparatuses.
Its validity requires an effective efficiency of less than $75 \%$ for both of them.

 A more ambitious program was started by a British-Spaniard group with the
 hope to build a real alternative to Quantum Optics \cite{s0,s1,s2,s3,s4,s5,Santos2}.
The main idea was that the probability distribution for the hidden
variable is given by the Wigner function, which is positive for
photons experiments. Furthermore a model of photodetection, which
departs from quantum theory,  is built in order to reproduce
available experimental results.

A great merit of this model is that it gives a number of
constraints, which do not follow from the quantum theory and are
experimentally testable.

In particular, there is a minimal light signal level that may be
reliably detected: a difference from quantum theory is predicted
at low detection rates, namely when the single detection rate $R_S
$ is lower than
\begin{equation}
R_S < { \eta F^2 R_c^2 \over 2 L d^2 \lambda \sqrt{ \tau T} }
\label{rate}
\end{equation}
where $\eta$ is the detection quantum efficiency, $F$ is the focal
distance of the lens in front of detectors, $R_c$ is the radius of
the active area of the non-linear medium where entangled photons
are generated, $\tau$ is the coherence time of incident photons, d
is the distance between the non-linear medium and the
photo-detectors, $\lambda$ the average wavelength of detected
photons. $L$ and $T$ are two free parameters which are less well
determined by the theory: $L$ can be interpreted as
the active depth of the detector, while $T$ is the time needed for
the photon to be absorbed and should be approximately less than 10
ns, being, in a first approximation, the length of
the wave packet divided for the velocity of light.

This prediction has recently been tested by an experiment measuring Clauser - Horne inequality with
polarisation entangled photons with a
strongly negative result for this model \cite{nos2}. Further recent negative
tests of this model, based on other specific predictions, can also be found elsewhere \cite{nosW,nosww}.

Finally, for the sake of completeness, it can be mentioned that very
recently, one of the authors of the previous model presented a new
LHV model \cite{ss3}, which does not have the same degree of
development of the former one, but in its simplicity allows one to
reproduce all Bell inequalities tests performed with polarisation
entangled photons. In the same paper it is suggested that a test
of the model can be performed by comparing the visibility:

\be V_a= { N(0) - N(\pi /2) \over N(0) + N(\pi /2)} \ee

with

\be V_b = \sqrt{2} { N(\pi /8) - N(\pi /8) \over N(\pi /8) + N(3
\pi /8)} \ee

where $N(\theta)$ are the coincidence counts measured on a
polarisation maximally entangled photon state when the two
polarizers are set to two angles differing of $\theta$. In fact, in the model of Ref. \cite{ss3}
\be V_b / V_a > 1+ \cos^2
(\pi \eta / 2) \left [ V_b - { sin^2 ( \pi \eta / 2) \over (\pi
\eta / 2)^2} \right] \label{ins} \ee is expected, result that can
be violated in SQM. This
prediction could probably be tested in a near future\footnote{Some first tests, based on previous data, did not
produce a conclusive answer \cite{ss3,nosMinsk}.}.

\subsection{ Test of local realism in Hilbert spaces with
dimension larger than 2}

In the last years it emerged that the use of higher dimension
Hilbert spaces ($d>2$, where states are dubbed qudits in analogy
to the quantum information word qubit denoting two level systems),
instead of the traditional $d=2$ ones, can lead to a larger
violation of Bell-like inequalities
\cite{kz1,kz2,collins,Massar2,Massar1,Massar3,Kas2,Kas3,Gisen,CHnos}.
 This result is related \cite{Acin} to the the discovery that
 quantum communication based on qudits presents a higher
security than the traditional qubit schemes
\cite{cq1,cq2,cq3a,cq3,cq4,cq5}.

More in details, concerning  Bell inequalities, various studies
were addressed to understand the limit quantum efficiency for a
loophole-free test of local realism (LR) and the resistance to
noise. For example, in Ref. \cite{Massar1} Bell inequalities with
enhanced resistance to detector inefficiency were investigated.
This is of particular interest since, as we have seen, the
loophole due to low detection efficiency $\eta$ of the detection
apparatuses is the last unsolved problem for  a conclusive test of
local realism. The result was that the limit for the smallest
detection efficiency $\eta^*$ necessary for a loophole free test
of LR, decreases for $d>2$ maximally entangled states of a $1-2
\%$ respect to the value $\eta^* = 82.84 \%$ for $d=2$ maximally
entangled states with 2x2 number of settings of the detection
apparatuses.

Later it was then shown \cite{Massar2} that for a specific
hidden variable model differences between Quantum Mechanics (QM)
and Local Realistic Theories (LRT) are observable up to $\eta^* >
{ M_A + M_B -2 \over M_A M_B -1}$, where $M_A$ and $M_B$ are the
number of measurements available to the two experimenters sharing two subsystems of a general
entangled state. An asymptotic result for large $d$ was obtained
in Ref. \cite{Massar3}.

On the other hand, the resistance to noise of  some specific Bell
inequalities tested by using maximally entangled states generated
by multiport beam splitters  was investigated as well \cite{kz1,kz2}, showing how it increases with $d$. More in
details, it was shown how considering a mixed state ($0 \le F \le
1$)
 \be \rho = (1-F) | \Psi \rangle \langle \Psi | + F
\rho_{noise} \label{noise}\ee where $\rho_{noise} $ is a diagonal
matrix with entries equal to $1/9$, the threshold value of $F$ for
violating a Clauser-Horne inequality grows from 0.2929 for qubits,
to 0.30385 for qutrits up to 0.3223 for states in a $d=16$ Hilbert
space.

  In Ref. \cite{kz2} it was also demonstrated that,
for maximally entangled states, the limit detection efficiency
decreases from $0.8285$ for $d=2$ up to $0.8080$ for $d=16$ (being
$0.8209$ for qutrits, $d=3$). A specific
Clauser-Horne like inequality was then proposed and investigated for
the previous maximally entangled system \cite{Kas3} (inequality that includes
also the ones presented in \cite{collins}). Similar results
concerning the resistance to noise of LR tests performed with
qudits were obtained in Ref. \cite{Gisen} as well.

 A further contribution came by showing \cite{CHnos}, performing a
numerical study of a generalized Bell inequality \cite{Kas3} on
two specific examples (qutrits generated by tritter, three arms interferometer, or biphotons as qutrits), how also in the case of qutrits the use of non-maximally
entangled states allows a reduction of the detection efficiency
for a conclusive test of local realism respect to maximally
entangled ones. Also a stronger reduction to noise was found
($F_{th} = 0.3216$ referring to Eq. \ref{noise}). Nevertheless,
this reduction of the requested state detection efficiency is
smaller (from 0.8209 to 0.8139 for tritters and from 0.8505 to
0.7413 for biphotons) than what obtained for qubits.

Finally, it must also be acknowledged that recently some papers were addressed to the
study of local realism with 3-4 qutrits. In Ref. \cite{kazu,cema} GHZ paradox was generalized to 3-4 qutrits,
in Ref. \cite{ac3qut} a Bell inequality for 3 qutrits was presented and in Ref. \cite{kas34} a numerical study
on violation of local realism for 3-4 qutrits was performed (indicating a stronger violation than for 3 qubits case).

All these results, beyond the large conceptual interest, have also
stimulated new experimental tests based on qutrit  photons
entangled states.

A first realisation of qutrits was proposed and realised by Moscow
group \cite{moscow1,moscow2,moscow3} by exploiting the superposition of
three biphoton states produced in PDC ($HH,HV,VV$), but in this
case no test of local realism was performed.

On the other hand, a first test of local realism was made by
Bouwmeester and collaborators \cite{bou} by using the
rotationally-invariant four photon state \be {1 \over \sqrt{3}} (
| HH, VV \rangle - | HV,VH \rangle + | VV, HH \rangle ) \ee
 produced in type II PDC for violating Bell inequality for
"spin 1" systems. The results $S= 2.27 \pm 0.02$ was obtained, in
evident disagreement with LR $S<2$ prediction (here and in the following always
apart from detection loophole).

Another test of LHVT was then presented \cite{Vaz} based
on the use of orbital angular momentum ($0, \hbar,-\hbar $)
entangled photons generated by sending both correlated
down-converted beams through holographic modules consisting of two
displaced holograms, which project photons onto a
specific superposition of Laguerre-Gaussian modes (describing
specific orbital angular momentum states).  Detection apparatuses
are then made of holograms (suited for projecting into a specific LG
mode) preceding photo-detectors. An extension of
Clauser-Horne-Shimony-Holt inequality to qutrits was observed to
be violated of 18 standard deviations.

Finally, Geneva group \cite{gisqun} obtained a 24 standard
deviations violation of local realistic predictions by measuring
violation of an inequality proposed by Collins and others
\cite{collins}. In this set-up  bin-time entangled
photon pairs were used (see fig. 12). They were created via a periodically poled waveguide and two
tritters, i.e.  balanced interferometers with three arms (a generalisation
of the time-bin entanglement of Eq. \ref{tb}). Tritters were then also used for analysing
the entangled qutrits.

Some progresses toward realization of qudits in higher dimensional spaces
(up to now without tests of local realism) can be found in Ref.s \cite{ried,nev}.

\subsection{Tests of non-contextuality}

Few years after the proposal
of Bell inequalities, Bell \cite{bell66} and Kochen-Specker
\cite{KS} posed further limits to HVT showing that every
non-contextual HVT cannot reproduce all the results of SQM, where
non-contextuality is defined as the request that each observable
has a value in an individual system that would give the result of
a measurement regardless of which sets of mutually commuting
observables we choose to measure it with.

The difference between  Bell-Kochen-Specker theorem and Bell inequalities is that the first one rules out
the assignment of non-contextual values to an arbitrary observable, whilst Bell inequalities rules out it
even when it is restricted to cases in which it can be
justified on the basis of locality. Thus, the Bell-Kochen-Specker theorem permits to eliminate non-contextual
hidden variable theories (NCHVT) which form a subset of local realistic hidden variable  theories,
tested by Bell inequalities (a LHVT requires non-contextuality between observables
when the measurements are space like separated,
but in general can be contextual when they are not causally disconnected).

More in details Bell and Kochen-Specker considered a physical
system of spin 1 and an arbitrary choice of three orthogonal
directions $a,b,c$. The eigenvalues of the square spin components
are $0,1$. Furthermore, the sum of them satisfy the following equation \be
S_a^2+S_b^2+S_c^2 = s (s+1) =2 \label{s1}\ee since we are dealing
with particle of spin 1 (s=1).

Let us then consider a set of
directions containing many different orthogonal triads. The three
observables consisting of the squared spin components along
orthogonal triads commute and therefore they can be measured
simultaneously. The values of such measurements (0 or 1) must
satisfy the same constraint \ref{s1} as the observables
themselves. Therefore, two of the values must be 1 and the other
0. The no hidden variable theorem is based on finding a quantum
mechanical state for which the statistics for the results of
measuring any three observables associated with orthogonal triads
could not be realized by any distribution of assignments of 0 or 1
to any direction of the set, consistent with the constraint.

Bell gave a general demonstration \cite{bell66}, based on Gleason theorem \cite{Gleason},
of the impossibility of satisfying this request. Independently, a year later
 Kochen and Specker \cite{KS} explicitly showed a finite set of
directions (117) which do not satisfy it.

Therefore, in principle an experimental test of Bell-Kochen-Specker theorem could
exclude every non-contextual HVT, requiring at least
some observables of the theory to be context-dependent (however,
it must be emphasized that for every HVT not all the observables
are contextual).
For the sake of completeness, it must be noticed that the experimental relevance of Bell-Kochen-Specker theorem
was questioned \cite{meyer,kent} due to real imperfect laboratory experiments. A claim later confuted in Ref. \cite{mermconf,simon,cabconf}.
However, a direct experimental realization of Bell-Kochen-Specker test of NCHVT disclosed to be rather difficult.

Tests of NCHVT were obtained \cite{MWZ} with polarisation entangled photons produced by parametric down conversion
either by using a version of Greenberger-Horne-Zeilinger for NCHVT \cite{zukpla} (that
reduces to only two particles test)
and an "event ready" test of Bell inequality for only one particle (see related theoretical works in
Ref. \cite{roycon,cabcon}).
Both tests showed large violation of non-contextuality (of more than 300 and 170 standard deviations respectively).
However, the relatively low collection efficiency, about $8 \%$, imposed to invoke fair sampling assumption
for these experiments as well.

Another test, directly based on the original theoretical proposal of
\cite{SZWZ1,SZWZ2}, has been recently realised \cite{KSexp} by using polarisation and
path of a single photon (a "heralded photon" produced by PDC) to form a two qubits system.
Also here the data agree largely more with SQM ($80 \%$) than with NCHVT
($20 \%$ of data sample). None the less, again detection loophole is not eliminated.

\subsection{Other Quantum Optical experiments connected with local realism and quantum non-locality}

Before concluding this section, we would like just to list some other
recent quantum optical experiments, whose results are connected with the studies of local realism.

A first interesting possibility is to realise a state that substantially reproduces the original EPR one.
A first scheme was realized by Ou et al. \cite{ouEPR1,ouEPR2}, following the theoretical proposal of \cite{RD,RD1},
 by employing a subthreshold non-degenerate optical parametric
oscillator to generate correlated amplitudes for signal and idler beams of light.
The role of position and momentum variables is played by quadrature-phase amplitudes:
the amplitudes of signal beams ($X_s,Y_s$) can be inferred from measurements of amplitudes ($X_i,Y_i$) of the spatially
separated idler beam. The observed values of variances of inferred observables $\Delta^2_{inf} X_s$ and $\Delta^2_{inf}
Y_s$ give
$\Delta^2_{inf} X_s \Delta^2_{inf} Y_s = 0.70 \pm 0.01$ in agreement with EPR paradox
$\Delta^2_{inf} X_s \Delta^2_{inf} Y_s < 1$, namely showing an apparent violation of
Heisenberg uncertainty principle.

Recently, an EPR state was also realised \cite{boyd} by producing position (x) -momentum (p) photon entangled states
by means of type II collinear PDC. Observed product of variances of inferred $x_S$ and $p_S$,
$\Delta^2_{inf} x_S \Delta^2_{inf} p_S = 0.01$, dramatically violates
EPR criterium.

Concerning EPR states it can also be mentioned that, after a paper \cite{ban} showing how the EPR state can violate Bell inequalities even if its
Wigner function is positive (at variance with previous claims \cite{bEPR,sEPR}), an experiment \cite{manEPR}
was addressed, following the theoretical scheme of \cite{gEPR}, to test Bell inequalities by homodyne measurements on
states produced by a pulsed nondegenerate optical parametric amplifier. A violation larger than 7
standard deviation was observed (of course apart from the various loopholes present in this experiment as well \cite{manEPR}).

Another interesting experiment recently realized \cite{bjork} concerns the demonstration of
quantum non-locality at single particle level.

The original idea \cite{tan} consisted in generating by a beam splitter an entangled state between a
single photon state $|1 \rangle$ and the vacuum $| 0 \rangle$
\be
| \Psi_{BS} \rangle = { |1 \rangle_1 |0 \rangle_2 + i |0 \rangle_1 |1 \rangle_2 \over \sqrt{2} }
\ee
where subscripts $1,2$ denote the two exits of the beam splitter, both addressed to a second beam splitter where they are
combined with a local oscillator (LO) for performing a homodyne detection, whose phase $\vartheta_{1,2}$
is the local parameter. The exits $c_{1,2},d_{1,2}$
of the beam splitters on beam $1,2$ are then all measured by photo-detectors.
Ref. \cite{tan} showed that
correlation functions among intensities $I$
\be
E(\vartheta_1,\vartheta_2) = {\langle (I_{d_1} -I_{c_1}) (I_{d_2} -I_{c_2}) \rangle \over \langle
(I_{d_1} + I_{c_1}) (I_{d_2}+ I_{c_2}) \rangle }
\ee
violates Bell inequalities.

After the discussion of Ref.s \cite{sanT,tan2,granT} pointed out that this original proposal required
additional assumptions, a new version of a test of local realism at single particle level, overcoming these problems,
 was proposed \cite{HarT}.

The experimental set-up of the Russian-Swedish group \cite{bjork} substantially implements the scheme of \cite{tan},
but in a version where single photon and LO co-propagate after the beam splitter, eliminating some problem of
the original scheme.

In detail, the single photon was produced by type-I PDC in a LBO crystal pumped with a
femtosecond-pulsed mode-locked titanium-sapphire laser after frequency doubling at 390 nm.
A fraction of the initial laser beam at 780 nm was used as LO.
The observed visibility ($91 \pm 3 \%$ after background subtraction) is
 sufficient for Bell inequalities violation.

As a third example, let us consider a recent experiment \cite{bouwhard} that realised
a photon version of the proposal of Ref. \cite{hareq1} (originally based on electrons and positrons).

The original Hardy's proposal considers two interferometers, one for electrons and the other for positrons,
arranged in such a way that two of their arms intersect. If both particles are at this intersection at the same time
they annihilate: this implies in a local realistic model, where particles have defined trajectories,
that a certain output of the two interferometers can never appear.
On the other hand, this output has non-vanishing probability in SQM.

The scheme realised experimentally \cite{bouwhard} uses indistinguishable photons produced in PDC
as substitute of electron and positron and photon bunching at a beam splitter as the annihilating interaction.
An inequality relating output probabilities valid for LHVT is then tested.
 Since this inequality is violated of 12 standard deviations, always apart from detection loophole (that appears here as well),
the experiment disagree with LHVT prediction.

Finally, another interesting possibility, recently investigated, is the one of entangling
a relative high number of photons, generating in this way a superposition of two "quasi-classical states"
(a "Schr\"odinger kitten") \cite{MGG1,tomb,dak,demG,lvG,bouG} with the hope of better understanding the macro-objectivation process.
Entangled states of few photons have been effectively obtained \cite{demG,lvG,bouG,dem04}.
However, recently entanglement of more clearly "macroscopic" systems, as SQUID, has been obtained as well \cite{squid}.
Here it is also worth to mention recent works where photon entanglement has been transferred to plasmons
(involving approximately $10^{10}$ electrons) and then back to photons \cite{alt,fas} that have still shown a Bell
inequalities violation \cite{alt}.

Concluding this section, we would like to address the interested reader to the papers \cite{spop,short,chiao,ssr}
for further experiments, and comments on them, somehow connected with quantum non-locality and local realism.

\section{Test of local realism with other physical systems than
photons}

As we have discussed in the previous paragraphs, the largest part
of tests of local realism performed up to now have been realised by
using photons, since entangled photons pairs are relatively easy to
produce, simple measurement schemes are available and photons can
be easily propagated for long distances. However, a conclusive
experiment has not yet been realised mainly due to insufficient
quantum efficiency of single-photon detectors, even if relevant
progresses in this sense have been obtained in the last years.

Therefore, it remains the interest for investigating the
possibilities of using other physical systems. Among them we will
consider in the next paragraphs the ones more discussed in the literature
and most interesting
according to our opinion: mesons and ions. Nevertheless, it is
worth to mention that relevant results were obtained also with
other systems as:

i) Polarisation correlation of $^1 S_0$ proton pairs produced in
nuclear reactions \cite{pp} (discussed in subsection 2.8).

ii) Neutrons interferometry \cite{neu}, with $S= 2.051 \pm 0.019$.
Detectors had high efficiencies, but one had large losses in
interferometers that made impossible to eliminate the detection loophole. Furthermore,
entanglement was generated between two different degrees of freedom in a single particle
(spatial and spinor part of the wave function): thus, obviously, the two measurements were not separated.

iii) Atoms entangled in a superposition involving two circular Rydberg states
produced by single photon exchange in high Q
cavity \cite{har}. Purity ($\thickapprox 0.63$) was too small for a test of local realism: further experimental
improvements are needed before this scheme could really be interesting for this purpose.

iv) Single atom entangled with single photon \cite{mo}. Where
atomic, hyperfine levels of a trapped $^{111}$Cd$^{+}$ ion, and photonic, polarisation, degrees of
freedom are probabilistically entangled following a spontaneous
emission of a photon from an atomic excited state. A value $S=
2.203 \pm 0.028$ was observed. Since photon detection takes place approximately 1.1 meters away from
the atom and detection of atomic degree of freedom takes $125 \mu s$,
locality loophole is far from being solved. Also detection efficiency is rather low due to small acceptance angle and
 transmission loss ($\thicksim 1 \%$), low quantum efficiency of detectors ($\thicksim 20 \%$) and restriction
  of the excitation probability to $\thicksim 10 \%$ for suppressing multiple excitations.

\subsection{Tests of Bell inequalities with mesons}

In the last years many papers
\cite{uch,BGH,BF1,BF2,nosk4,BF,gh,six,BK1,BK2,BK3,BK4,BK5,BK6,Bra,DG,BK7,BK8,BK9,BK10} have
been devoted to study the possibility of making local realism
tests by the use of pseudoscalar meson pairs as $K \bar{K}$ or $B
\bar{B}$\footnote{Properties of other neutral pseudoscalar mesons (as D, $D_s$ and $B_s$)
make them less interesting for this kind of studies.}.
In fact, if the pair is produced by the decay of a
particle at rest in the laboratory frame (as the $\phi$ at $Da
\phi ne$), the two particles can be easily separated to a
relatively large distance allowing a space-like separation of the
two subsystems and permitting an easy elimination of the
locality loophole. Furthermore, very efficient particle
detectors are available leading to the hope of easily eliminate
detection loophole . Finally, a very low noise is expected as
well.

These proposals are based on the use of entangled states of the
form : \be |\Psi \rangle = { | K^0 \rangle | \bar K^0 \rangle  -
| \bar K^0 \rangle | K^0 \rangle \over \sqrt{2} }  = { |
K_L \rangle | K_S \rangle  - |  K_S \rangle | K_L \rangle \over
\sqrt{2} }  \label{psik} \ee

where $| K^0 \rangle$ and $| \bar K^0 \rangle$ are the particle
and antiparticle related by charge conjugation and composed by a
quark of flavour $d$ with an anti-strange $\bar s$ and a $\bar d $ with a $s$
respectively. Whilst mass eigenstates are
\be | K_L \rangle = {p | K^0 \rangle  + q | \bar K^0
\rangle \over \sqrt{|p|^2 + |q|^2}} \ee and \be | K_S \rangle = { p | K^0 \rangle
- q |  \bar K^0 \rangle \over \sqrt{|p|^2 + |q|^2
} } \ee where $p = 1+ \varepsilon$ and $q = 1 - \varepsilon$ in terms of the (small)
 electroweek CP-violation parameter $\varepsilon$ ($|\varepsilon|= (2.26 \pm 0.02) 10^{-3}$).
The $K_L$ is the long living state,
corresponding for $\varepsilon=0$ to CP=-1 eigenstate ($|K^0_- \rangle$) for which 2 pions decay is forbidden,
and the $K_S$ is the short living state, corresponding for $\varepsilon=0$ to CP=+1 eigenstate ($|K^0_+ \rangle$), for which
2 pions decay is allowed\footnote{see, for example, Ref. \cite{PrLe} for more details on $K_0$ phenomenology.}.

In general the small violation of CP symmetry can be neglected in considerations about local realism, except for
a class of proposals that suggested to test local
realism by measuring CP-violation parameters \cite{uch,BGH,BF1,BF2}.
The original idea \cite{uch} was to build the Bell inequality
\be P(K_S^0,\bar K^0) \leq P(K_S^0,\bar K^0_+) + P(K_+^0,\bar K^0)
\ee
based on joint measurement probabilities of
a $K_S^0$, $\bar K^0$ and the (unphysical) $K^0_+$  with the state of Eq. \ref{psik} and (with some additional hypothesis on phases) to transform it in an inequality on the parameter $\varepsilon$,
\be
Re \{\varepsilon \} \leq |\varepsilon|^2
\ee
One later work \cite{BGH} obtained more stringent bounds independent of any phase convention,
that are violated by measured data \cite{PDB}.
Similarly, an inequality on the CP-violation parameter $\varepsilon '$ (see \cite{PDB,PrLe}
for a definition) was obtained \cite{BF1,BF2}.
Nevertheless, it must be noticed that all these results require additional assumptions \cite{nosk4}, such
as the validity of relations derived in SQM in LHVT as well.
Therefore, albeit giving an interesting connection between local realism and specific properties (CP violation)
of electroweak lagrangian, they do not really represent  a  conclusive test of local realism.

Concerning other proposals, first of all it must be noticed that
a simple hypothesis \cite{furry,six}
where the state (\ref{psik}) collapses shortly after its
production in two factorised states \be |  K_S \rangle | K_L
\rangle , \, | K_L \rangle | K_S \rangle \ee (or similarly for other pseudoscalar mesons) is already excluded, see for example results of
Ref. \cite{albr}.

On the other hand, a simple test of local realism
based on a correlation function defined such that it
takes the value 1 when two or none $\bar K^0$ are identified and
-1 otherwise, would not lead to a violation of Bell inequalities
due to the specific values of $ K^0 \bar K^0 $ mixing parameters
 \cite{gh}.

Anyway, other Bell inequalities and hidden variable schemes can be
considered. Nevertheless, the statement about high efficiency in
detection of pseudoscalar mesons does not survive to a deeper
analysis \cite{nosK,nosK2,nosK3}. The main concern about this
statement derives from the fact that in most experimental tests
proposed up to now, one must tag the $P$ or $\bar{P}$ (where $P$
denotes a pseudoscalar particle and $\bar{P}$ its antiparticle)
through its decay. This requires the selection of $\Delta S =
\Delta Q$ semileptonic decays (i.e. the ones where the strangeness
and charge changes of the hadrons are the same), which represent
only a (small) fraction of the total possible decays of the meson.
For example, one has \cite{PDB} the following branching ratios\footnote{ $B$ denotes the
pseudoscalar meson analogous to $K$
where the quark $s$ is substituted by the heavier quark $b$.}:
\bea
 BR(K^0_S \rightarrow \pi^{+} e^{-} \nu_e) = (3.5 \pm 0.2) 10^{-4} \cr
BR(K^0_L \rightarrow \pi^{+} e^{-} \nu_e) = 0.1939 \pm 0.0014  \cr
BR(K^0_L \rightarrow \pi^{+} \mu^{-} \nu_{\mu}) = 0.1359 \pm
0.0013 \cr BR(B^0 \rightarrow  l^{+} \nu_{l} X) = 0.105 \pm 0.008
\cr BR(B^0 \rightarrow  l^{+} \nu_{l} \rho ^-) = (2.6 \pm 0.7)
10^{-4} \cr BR(B^0 \rightarrow  l^{+} \nu_{l} \pi ^-) = (1.33 \pm
0.22) 10^{-4} \label{BR} \eea where $X$ means anything, $l$ denotes
a generic lepton (as $e ^-$ electron, $\mu ^-$ muon) and $\nu_{l}
$ its related neutrino, $\pi$ denotes the pseudoscalar meson
composed of $u$ and/or $d$ quarks and antiquarks.

Besides this problem, one has to consider experimental cuts on the
energies of the decay products, which  inevitably further reduce
this fraction. Moreover, an additional part of the pairs is lost by
decays occurring before the region of observation. Finally, most
of these proposals involve the regeneration phenomenon\footnote{i.e. the possibility
of regenerating $K_S$ from a $K_L$ beam by explotig different interaction amplitudes of $K^0$
and $\bar{K^0}$ with matter \cite{PrLe}.}, which
introduces further strong losses; on the other hand if no choice
of the measurement set up (e.g. the presence of the regeneration
slab) is introduced, the space-like loophole cannot be really
eliminated.

The result of these considerations is that one is unavoidably led
to subselect a fraction of the total events. As one cannot exclude
a priori  hidden variables related to the decay properties of the
meson and losses, one cannot exclude the sample to be biased and
thus the detection loophole appears here too. This is in complete
analogy with polarisation photons experiments, where the detection
loophole derives by the fact that one can envisage losses
related to the values of hidden variables that determine if the
photon passes or not a polarisation (or another) selection.
Namely, in a local realistic model the properties of a particle
are completely specified by the hidden variables. Also for mesons decays and losses, in
a LRT, can happen according to the values of the
hidden variables (both in a deterministic or in a probabilistic
way). States with different hidden variables can decay in
different channels, with the condition that the branching ratios
{\it averaged } over the hidden variables distribution reproduce
the quantum mechanics predictions.

From this discussion follows that, for what concerns the experiments based on Bell inequalities
\cite{BF,BK1,BK2,BK3,BK4,BK5,BK6,DG}, the same limits for the
total efficiency previously discussed remain valid. As the total
branching ratio in $\Delta S = \Delta Q$ semileptonic decays, Eq.s \ref{BR}, is
much smaller than 0.8284 (and the same happens for any other
selection, as far as we know), this unavoidably implies that a
loophole free test of Bell inequalities cannot be  performed in
this class of experiments. The eventual use of non-maximally
entangled states,  lowering the efficiency threshold to 0.67, does
not substantially change the situation.

This problem does not appear in Ref.
\cite{BF}, however other additional hypotheses are needed (see Eq.
15 and discussion after Eq. 18 of \cite{BF}), and thus this
proposal does not allow a general test of LRT as well.

Finally, it must also be noticed  that the only observation of
interference between the two terms of the entangled wave function,
Eq. \ref{psik}, as in Ref \cite{CLEO}, does not exclude  general
LHVT, for this feature can be reproduced in a general class of
local realistic theories.

In summary, we can conclude that the proposed Bell inequalities measurement on
pseudoscalar mesons  pairs can not allow a conclusive test of
local realism.

Nevertheless, they represent a prominent example of
studying local realism with other physical systems than photons.

In this sense the recent experimental test of Bell inequalities
\cite{belle} with a pair of $B^0 \bar{B}^0$ mesons is very
interesting. Briefly, an entangled state of the form \ref{psik}
has been produced from $\Upsilon (4S)$ decay, the flavour has then
been identified by reconstructing a semileptonic decay for one
meson and from lepton tagging (a multidimensional likelihood
method) for the other. After having normalised the correlation
function to the undecayed pairs, an evident violation of CHSH
inequality has been obtained, $S = 2.725 \pm 0.167_{stat} \pm
0.092_{syst}$ \footnote{For a specific discussion of loopholes of
this experiment, next to general arguments of Ref.s
\cite{nosK,nosK2}, see Ref. \cite{Bra}.}.

\subsection{Other tests of local realism with mesons}

Let us then consider other proposals for testing local realism
with mesons, not based on a  Bell inequalities measurement. Two
proposals of this kind have been recently advanced by F. Selleri
and others concerning a  $K^0 \bar{K}^0$ \cite{BK7} (very lately
further developed in Ref. \cite{DG}) or  a $B^0 \bar{B}^0$ \cite{BK8}
system respectively.

In Ref. \cite{BK7,Selnew} a very general model is proposed, where the $K^0 \bar{K}^0$ pair is
local-realistically described by means of two hidden variables.
One ($\lambda_1$) determines a well defined CP value,
another $\lambda _3$ determines the times when  a sudden jump
between defined valued of strangeness $S$ (i.e. between a $K^0$ and a
$\bar K^0$) happens. These jumps are necessary for explaining the observed $K^0$-$\bar K^0$
oscillations (in the model a further parameter $\lambda_2$ driven by $\lambda _3$
and determining $S$ is introduced as well).

Denoted by $K_1$ the state
with CP=1, S=1, $K_2$ the state with CP=1, S=-1, $K_3$ the state
with CP=-1, S=1 and $K_4$ the state with CP=-1, S=-1, the initial
state can be, with probability $1/4$, in anyone of the states $CP=
\pm 1$, $S=\pm 1$. Each of these pairs give, in the
local-realistic model (LRM), a certain probability of observing a
$\bar K^0 \bar K^0$ pair at proper times $t_a$ and $t_b$ ($\ne
t_a$) of the two particles, which are \cite{BK8}:
\bea
P_1[t_a,t_b]= [E_S(t_a) Q_-(t_a) - \rho(t_a)] \cdot E_L(t_a)
p_{43}(t_b | t_a) & \nonumber \\
P_2[t_a,t_b]=[E_S(t_a) Q_+(t_a) +
\rho(t_a)] \cdot E_L(t_a) p_{43}(t_b | t_a) & \nonumber \\
P_3[t_a,t_b]=[E_L(t_a) Q_-(t_a) + \rho(t_a)] \cdot E_S(t_a)
p_{21}(t_b | t_a) &  \label{P} \\ P_4[t_a,t_b]=[E_L(t_a) Q_+(t_a)
- \rho(t_a)] \cdot E_S(t_a) p_{21}(t_b | t_a) \, \, \nonumber
 \eea corresponding to an initial state with $K_1$ on
the left and $K_4$ on the right, $K_2$ on the left and $K_3$ on
the right, $K_3$ on the left and $K_2$ on the right and $K_4$ on
the left and $K_1$ on the right, respectively.

In Eq. \ref{P}, we have introduced $ E_S(t) = exp(- \gamma_S t)$
and $ E_L(t) = exp(- \gamma_L t)$, where (in units $c=\hbar=1$)
$\gamma_{S}=(1.1163\pm 0.0007) 10^{10} s^{-1}$ and
$\gamma_{L}=(1.9305 \pm 0.0058) 10^{7} s^{-1}$ denote the decay
rate of $K_S$ and $K_L$ \cite{PDB}.

Furthermore, the function $Q_{\pm}$ are defined through: \be
Q_{\pm} ={ 1 \over 2}  \left[ 1 \pm {2 \sqrt{E_L E_S} \over E_L +
E_S} \cos( \Delta m t) \right] \ee where $\Delta m = (0.5292 \pm
0.0010) 10^{10} s^{-1}$ is the mass difference $ M_{K_L}
-M_{K_S}$. We have also introduced the symbol $p_{ij}(t_a | t_b)$
for denoting the probability of having a $K_j$ at time $t_b$
conditioned to have had the state $K_i$ at time $t_a$. From Ref.
\cite{BK7,BK8} one has: \be
 p_{21}(t_b | t_a) = E_S^{-1}(t_a) [p_{21}(t_b | 0) - p_{21}( t_a | 0)  \cdot E_S(t_b-t_a)]
\ee and \be p_{43}(t_b | t_a) = E_L^{-1}(t_a) [p_{43}(t_b | 0) -
p_{43}( t_a | 0)  \cdot E_L(t_b-t_a)] \ee

where \be
 p_{21}(t | 0) = E_S(t) Q_-(t) - \rho(t)
\ee and \be
 p_{43}(t | 0) = E_L(t) Q_-(t) + \rho(t).
\ee Finally, $\rho(t)$ is a function not perfectly determined in
the model (see discussion in Ref. \cite{BK7,BK8}), but which is
limited  by \bea -E_S Q_+ \le \rho \le E_S Q_- \cr -E_L Q_- \le
\rho \le E_L Q_+ \cr \label{ro} \eea

The LRM probability of observing a $\bar K^0 \bar K^0$ pair is
given by the sum of the four probabilities of Eq. \ref{P}
multiplied by $1/4$. Since it is rather different from the quantum
mechanical prediction, \bea P_{QM}[\bar{K^0}(t_a),
\bar{K^0}(t_b)]= {1 \over 8} [ e^{[-(\gamma_S t_a + \gamma _L
t_b)]} +
 e^{[-(\gamma_L t_a + \gamma _S t_b)]} - \nonumber \\ 2 e^{[-(1/2) (\gamma_S + \gamma_L) (t_a + t_b)] }
 \cos (\Delta m (t_a - t_b)) ]
\eea
 an experimental measurement of this quantity could represent a conclusive
test of local realism \cite{BK7,BK8}.

However, it has been  shown \cite{nosK3} how also in this case detection loophole manifests itself.
 When the total detection efficiency is lower than 1, the different
probabilities can contribute in different ways since the hidden
variables, which determine the passing or not the test, could also
be related to the decay properties of the meson pair and losses.
As for the cases previously discussed, the hidden
variables values completely characterise the state, and thus, in
principle, even its decay properties. If this is the case,
different coefficients $a_i$   can multiply the four
probabilities. One has therefore: \bea P[\bar{K^0}(t_a),
\bar{K^0}(t_b)]= 1 / 4 \cdot [ a_1 P_1 [t_a,t_b] + a_2 P_2
[t_a,t_b] + \nonumber \\ a_3 P_3 [t_a,t_b] + a_4 P_4 [t_a,t_b]
] \eea

The freedom of the choice of these parameters permits therefore to
reproduce the quantum mechanical prediction (the same happens for B mesons \cite{nosK2}),
 as shown by an explicit numerical calculation in Ref.  \cite{nosK}.
Thus also this scheme, albeit interesting,  is not suitable for a conclusive test
of local realism.

Finally, let us consider a recent proposal \cite{Garb}, which
seems to overcome the objections of Ref. \cite{nosK,nosK2}. It is based
on generating a non-maximally entangled state by placing a
regenerator slab on the path of kaons pairs produced in $\phi$
decays, where $r$ is the regeneration parameter, or by considering
kaons produced in $p \bar p$ annihilation at rest, where $r$
measures the relative strength of p to s wave channels and on
selecting the $K_S$ surviving after $T = 10 \tau_S$

\be |\Phi \rangle = { R | K^0  \rangle | K^0  \rangle  + R | \bar
K^0 \rangle | \bar K^0\rangle + (2-R) | \bar K^0  \rangle |  K^0
\rangle  - (2+R) | K^0 \rangle | \bar K^0 \rangle \over 2
\sqrt{2+|R|^2} } \label{psi3} \ee where $R=-r exp[-( i \Delta m +
(\Gamma_S - \Gamma_ L)/2) T]$ and $R'=-r^2/R$, $\Delta m$ is the
difference between $K_L$ and $K_S$ masses, whilst $\Gamma_S$,
$\Gamma _L$ their respective decay widths.

Then one selects, with an appropriate choice of parameters, the
case $R=-1$, for which SQM predicts the probabilities of joint
detection:

\bea P_{QM} (K^0 , \bar K^0) = \eta \eta'/12 & \nonumber \\ P_{QM} (K^0 ,
K_L) = 0 & \nonumber \\ P_{QM} (K_L , \bar K^0) = 0 & \\ P_{QM} (K_S ,  K_S)
= 0 \nonumber \label{PQM} \eea where $\eta$ and $\eta '$ are the
detection efficiencies for identifying $K^0$ and $\bar K^0$,
respectively.

In a hidden variable model with distribution $\rho(a)$ of the
hidden variable (or variables set) $a$ the probability of
observing a $K^0$ to the left and $\bar K^0$ to the right is: \be
P_{LR}(K^0,\bar K^0) = \int da \rho(a) p_l(K^0|a) p_r(\bar K^0|a)
= \eta \eta'/12 \leq \int_{A_{0,\bar 0}} da \rho(a) \ee where $
p_l(K^0|a), p_r(\bar K^0|a)$ are the single kaon probabilities of
detecting a $K^0$ to the left and $\bar K^0$ to the right
respectively and $ A_{0,\bar 0}$ is the set of hidden variables
corresponding to a $K^0$ to the left and $\bar K^0$ to the right.

In a LHVT the necessity of reproducing Eq. \ref{PQM} requires that
if a $K^0$ ($\bar K^0$ ) is observed to the left (rigth) a $K_S$
propagates  to the right (left). Thus one has $ p_l(K_S|a) =1,
p_r( K_S|a)=1 $ if $a $ belongs to $ A_{0,\bar 0}$. This is at
variance with SQM predictions \ref{PQM}, since:

\be P_{LR}(K_S,K_S) = \int da \rho(a) p_l(K_S|a) p_r( K_S|a)  \geq
\int_{A_{0,\bar 0}} da \rho(a) \label{PLR} \ee

This result seems therefore to show that even if the detection
efficiency of strangeness eigenstates is small, nevertheless LHVT
can be tested without any additional hypothesis if the $K_S$,
$K_L$ can be determined with perfect efficiency. Experimentally,
this determination is realised by looking to the decays between
time $T_0 =10 \tau_S$ ($\tau_S = 89.35 \pm 0.08 $ ps is the mean life of $K_S$),
where the state \ref{psi3} is produced, and
time $T_1$ such that a negligible $K_L$ contribution to decays
used for tagging $K_S$ is still expected in the interval.

However, also for this scheme detection loophole reappears
\cite{nosK3}, in fact the former discussion does not consider that
in a deterministic theory hidden variables could also fix the
channel of decay and the precise time of decay.

Therefore, Eq. \ref{PLR} becomes: \be P_{LR}(K_S, K_S) =\sum_C
\int_{T_0}^{T_1} dt \sum_{C'} \int_{T_0}^{T_1} dt' \int da \,
\rho(a) p_l(K_S|a) p_r( K_S|a) p_{l,C}(t|a) p_{r,C'}(t'|a)
\label{PLR2} \ee where $C$ and $C'$ run over the different decay
channels (allowing an identification of $K_S$) and $ p_{i,C}(t|a)$
gives the probability of the $i=l,r$ (left, right) meson to decay
into the channel $C$ at time t.

Let us now consider how this modifies the discussion concerning
Eq. \ref{PLR}.
 If the efficiencies $\eta$ and $\eta '$ of
$K^0$ and $\bar K^0$ detection were high, the situation would not
substantially change . However, unluckily, they are very small.
The method for this detection consists \cite{CLEO} in looking to
distinct interaction of $K^0$ and $\bar K^0$ with matter
(interaction and therefore identification that in principle could
depend on the hidden variables value): this led in Ref.
\cite{CLEO} to the identification of 70 unlike-strangeness events
and 19 like-strangeness events over $8 \cdot 10^7$ analysed
events! Thus, the few $K^0$-$\bar K^0$ identified events could
easily correspond to $K_S$ which would  not have decayed in the
temporal window that allows their identification and thus could
not contribute to the integral \ref{PLR2} (on the other hand if
$\eta$ and $\eta '$ were sufficiently large this would not be
possible). Furthermore, it must also be considered that $| K_S
\rangle$ and $| K_L \rangle $ are not perfectly orthogonal, for
$\langle K_S | K_L \rangle = 3.3 \cdot 10^{-3}$ \cite{PDB}. This
means that a fraction of $K_S$ in the LHVT belongs to a hidden
variable set corresponding to decays characteristic of $K_L$ (and
vice versa), giving  a further fraction of states not contributing
to $ P_{LR}(K_S,K_S) $ as defined in Eq. \ref{PLR2}. Albeit very
small, this contribution cannot be neglected due to the small value
of $P_{LR}(K^0,\bar K^0) = \eta \eta'/12$.

Summing up, the small fraction of simultaneously identified left
$K^0$ and right $\bar K^0$ could be easily accounted for, in a
LHVT, by a fraction of $K_S$ that does not decay in an
identifiable form, since they decay outside of the temporal window
$T_0 < T < T_1$ or in an allowed channel for $K_L$, as three pions
or pion, lepton, neutrino.

The numerical results of Ref. \cite{nosK3} show that  a loophole free test of local realism in this scheme
 requires $\eta \approx \eta ' > 9 \%$ that seems very difficult to be obtained
experimentally. Therefore, also with this scheme an ultimate test
of Local Realism  will be hardly obtained.

In summary, from the discussion of the two last subparagraphs, we can therefore reach the conclusion that,
even if representing an interesting possibility to study local realism with other physical systems beyond photons,
at the moment the suggested LHVT tests with pseudoscalar mesons cannot lead to a conclusive result
due to specific forms in which detection loophole reappears here as well.

\subsection{Bell inequalities experiments with ions}

Finally, let us consider experiments based on entangled ions, which are rather interesting since
very high detection efficiencies can be reached.
On the other hand, as we will see, in this case the elimination of
locality loophole is hard.

For the sake of exemplification, we will describe in the following
a recent experiment performed at NIST \cite{nist}\footnote{Incidentally, also
this technique allowed entanglement
of more particles (4) \cite{nist2}. Anyway no test of local
realism was performed in this case.}.

In synthesis, the set-up consisted in generating an entangled state
of the form \be | \Psi_2 \rangle ={ |0 \rangle |0 \rangle - |1
\rangle |1 \rangle \over \sqrt{2}} \ee by coupling two levels of
$2S_{1/2}$ ground state of  $^9$Be$^+$ ions ($ |0 \rangle = |
F=1,m_f=-1 \rangle$, $ |1 \rangle = | F=2 ,m_f=-2 \rangle$) by a
coherent stimulated Raman transition. The two laser beams used to
drive the transition had a wavelength of 313 nm and a difference
frequency near the hyperfine splitting of the states, $\omega_0
\cong 2 \pi \cdot 1.25$ GHz. The fidelity $F= \langle \Psi_2 |
\rho_{exp} | \Psi_2 \rangle$ of the generated state (described by
a density matrix $\rho_{exp} $) respect to the theoretical one $|
\Psi_2 \rangle$ was measured to be of $88 \%$.

After having produced the state $| \Psi_2 \rangle$  a Raman pulse of short
duration ($\sim 400$ ns) was applied again transforming the state
of each ion $i$ as \bea | 0 \rangle _i \rightarrow {( | 0 \rangle
_i- i e^{i \phi_i} | 1 \rangle _i) \over \sqrt{2}} \nonumber \\
| 1 \rangle _i \rightarrow {( | 1 \rangle _i- i e^{i \phi_i} | 0
\rangle _i) \over \sqrt{2}} \eea where the phase $\phi_i$ is the
phase of the field driving the Raman transitions and represents
the parameter used in the Bell inequality test. In the experiment this phase was set
either by varying the phase of the radio-frequency synthesizer that determines
the Raman difference frequency (controlling the total phase) or by motion $\Delta x_j$ of a
ion along the trap axis, which gives a phase change $\Delta k \Delta x_j$ where $\Delta k $
is the difference wave vector (controlling the differential phase).

Finally, the state of an ion was probed with circularly polarized
light from a 'detection laser beam'. During this detection pulse,
ions in the state $| 1 \rangle$ scatter many photons, whilst ions
in the state $| 0 \rangle$ scatter very few photons. For two ions
one can have three cases: zero ions bright, one ion bright, two
ions bright. In the one-ion-bright case Bell's measurement
requires only knowledge that the states of two ions are different
and not which one is bright.

The measured CHSH inequality violation was $S= 2.25 \pm 0.03$,
in agreement with SQM predictions once imperfections of the experiments are kept into account.

The total detection efficiency was $\approx 98 \%$, where the $2 \%$
decrease was mainly due to misidentification of a bright ion due to
imperfect circular polarisation of detection light that can cause
$| 1 \rangle \rightarrow | 0 \rangle$ transitions. No fair
sampling hypothesis was therefore needed.

Nevertheless, in this set-up the measurements on two ions not only
are not space-like separated, but even one has a common measurement
on the two ions. Therefore, one cannot absolutely speak of
non-locality in this case.

In principle the two ions could be separated, but this looks a
very difficult experimental task.

In conclusion,  experiments with ions allow a very high detection
efficiency, but the realisation of an ultimate experiment with
this technique looks not easy because of the difficulty in having
space-like separated measurements.

\section{Some conclusions about Local Hidden Variable Theories}

In summary, in the previous paragraphs we discussed how EPR
argument originated the debate about  the possible existence of  a
local realistic theory, where physical predictions are
deterministic and quantum probabilities derive by our ignorance of
some hidden variable fixing all the properties of a physical
system. The great beauty of Bell theorem resides in allowing a
general test of the possible existence of whatever local hidden
variable theory against standard quantum mechanics. In 70's
various experiments based on cascade atomic decays gave strong
indications against LHVT, culminating in the Orsay experiment
closing also (with some minor caveats) locality-loophole (i.e. the
request of having space-like separated measurements). However,
these experiments had very low detection efficiency leaving open
the possibility that the selected subsample was not a faithful
representation of the whole one (detection loophole).

The use of entangled photons produced by parametric fluorescence
allowed in the 90's to approach a detection-loophole free
test of LHVT and to close without any objection the locality one.
Furthermore, also other interesting experiments about (or connected with) local
realism were performed with this source of entangled states.

Even if, on account to these experiments, little space remains for LHVT,
nevertheless, due to the very fundamental relevance of a
conclusive test, the work for realising new experiments on Bell
inequalities is still going on.

Progresses in photon-detector could eventually allow one to close every loophole
with PDC entangled photons. On the other hand, entangled ions are
detected with very high efficiency, but to obtain separated
measurements is hard. Entangled mesons have suscitated a certain
interest as well, but detection loophole reappears in a form hard
to be eliminated. Furthermore, other physical systems have also been used for
experimental realisations of tests of LRT, but for the moment, with less success.

Finally, before concluding this part, for the sake of completeness
we would like also to quote some recent theoretical proposals
\cite{aug,matP,fry,garc,cin,sam,zan} for eliminating the detection
loophole that have not yet found an experimental implementation.

Altogether, the hope to reach in the next years a conclusive answer about local realistic
alternatives to standard quantum mechanics seems to be rather reasonable.

\section{ Non-local Realistic Theories}

As we have seen, experiments on Bell inequalities indicate that, a
part from the detection loophole, no Local Realistic Theory can
represent a valid alternative to SQM. However, it is not excluded
the possibility of considering non-local (or eventually lacking
other classical properties, e.g. counterfactual definiteness
\cite{deB}) Realistic Theories (NLRT), namely theories where the
action on a subsystem influences superluminally (eventually
instantaneously) the whole system. Of course, in order to preserve
compatibility with special relativity, this influence must be
built such not to introduce faster than light communication.

In particular two non-local hidden variable theories (NLHVT) have
suscitated a large interest: the Nelson stochastic \cite{nelson}
mechanics and, even more,  the de Broglie-Bohm model \cite{debroglie,bohm52}.
In the following subsections, we will describe dBB model and then summarize
the main ideas of Nelson's one.

For the sake of completeness, it can be noticed that also the model of Ref.
\cite{SS,ss2} (quoted in sub-section 4.2),
 and the Eberhard one \cite{ebnlhvt} are non-local in the sense that they require some superluminal hidden communication.
 However, it has been shown that for them arises the problem of possibility of faster than light communication
\cite{SG,SG1}. In more generality it has been demonstrated
that this problem appears for every model where one has a finite speed
superluminal hidden communication in a preferred frame without local hidden variables \cite{SG1}.

  Also Bohm-Bub \cite{bub} model
can be interpreted \cite{mattu,durt} as a non-local hidden
 variables theory  where there are both local hidden variables and a super-luminal connection (mixed model),
 but with finite velocity\footnote{For some generalisations of Bhom-Bub model see \cite{belinf,durt,tut,chri}.}.

 The problem of faster than light communication arises in this case \cite{durt} as well.
None the less, since it has attracted a certain interest,
in the following lines  we rapidly sketch, for the sake of exemplification,
the way how this model is built for a two-dimensional Hilbert
 space.

 To a general wave function of the form
 \be
 |\psi \rangle = \psi_1 | s_1 \rangle + \psi_2 |s_2 \rangle
 \ee
 where $| s_1 \rangle $, $|s_2 \rangle$ are eigenstates of  a certain observable S representing  a basis of the
 two-dimensional Hilbert space, one associates a vector in  the dual space
 \be
 \langle \xi | = \xi_1 \langle s_1| + \xi_2 \langle s_2 |
 \ee
The components of $\langle \xi |$ are the hidden variables. They are supposed to be randomly distributed on the
 hypersphere of unit radius in Hilbert space defined by $\sum_i |\xi_i|^2=1$ (assumption that allows us to reproduce usual
 QM average for all observables). Furthermore, in order to describe wave function collapse, one postulates that,
  in addition to Schr\"odinger equation, during a measurement process of the observable S the following evolution equations
  operate ($R_i = {|\psi_i|^2 \over |\xi_i|^2}$) :
  \bea
  {d \psi_1 \over dt} = \gamma (R_1 -R_2) \psi_1 |\psi_2|^2 \\
 {d \psi_2 \over dt} = \gamma (R_2 -R_1) \psi_2 |\psi_1|^2 \eea
which lead to $|\psi \rangle \rightarrow e^{i \phi_1} |s_1 \rangle $ if $R_1 > R_2$ and
to $|\psi \rangle \rightarrow e^{i \phi_2} |s_2 \rangle $ if $R_2 > R_1$ (i.e. to the wave function collapse).
Some indications of possible tests of the model against SQM in repeated measurements have also been given in \cite{bub,durt}
(see \cite{pap} for an earliest experiment and \cite{durt,tut} for criticisms on it).

Before concluding this introduction to NLRT, it is still worth to shortly mention studies concerning what
superluminal classical communication could reproduce quantum correlations.
Whilst an unlimited instantaneous communication between subsystems allows one to reproduce SQM results,
a communication among a limited number of subsystems (the so called hybrid models) does not generally  allow
this result. For example, Svetlichny
has shown that for the three subsystems case superluminal communication between arbitrary pairs cannot reproduce the results of
measurements performed on quantum states \cite{svet}. This result was then generalised to an arbitrary number of subsystems
\cite{colhyb}. Finally, in Ref. \cite{JLM} this last outcome was extended to a larger class of
 classical communication (in the Graphs theory language of \cite{JLM} to partially paired graphs respect to
 separable graphs considered in \cite{svet,colhyb}).
On the other hand, concerning bipartite systems, efforts concentrated on studying the number of bits of the superluminal
communication needed for reproducing the quantum correlations \cite{BCT,stei,MBCC},
reaching the result that one bit suffices for maximally entangled states \cite{ton}, whilst this resource is
insufficient for non-maximally entangled ones \cite{BGS}.

\subsection{ The de Broglie - Bohm theory}

A particularly interesting example of NLHVT is the theory of de
Broglie - Bohm \cite{debroglie,bohm52}\footnote{we do not discuss
here the history of this model and which were the precise
positions of de Broglie and of Bohm etc., but we adopt, as became
conventional, the name of de Broglie-Bohm for the version
presented in the following.} for the , where  the hidden variables
are the positions of the particles, which evolve according to a
Hamilton-Jacobi equation that includes also a quantum potential
term that, as we will see, is derived from Schr\"odinger equation
and allows one to reproduce (for an ensemble of particles)
probabilistic predictions of SQM.

In order to derive the explicit form of this equation of motion,
let us consider the usual Schr\"{o}dinger equation
\begin{equation} i\hbar \frac{\partial \Psi (x,t)}{\partial
t}= \left[-\frac{\hbar ^{2}}{2m}\Delta +V(x) \right]\Psi (x,t) \label{Sc}
\end{equation} and write the wave function as:

\begin{equation}
\Psi =R(x,t)\cdot exp[iS(x,t)/ \hbar]
\end{equation}

separating real and imaginary part one obtains the "Hamilton-Jacobi" equation
\bigskip
\begin{equation}
\frac{\partial S}{\partial t}+\frac{(\nabla S)^{2}}{2m}+V+Q=0
\end{equation}

and the continuity equation:

\begin{equation}
\frac{\partial \rho }{\partial t}+\nabla \cdot (v\rho )=0
\end{equation}

where

\begin{equation}
\rho(x,t)= |\Psi |^2=R(x,t)^2
\end{equation}

and

\begin{equation}
v={\nabla S \over m} \label{geq}
\end{equation}

is the velocity of the particle given by the "guidance equation",
which constitutes the real novelty of the dBB model and has no
equivalent in SQM.

\begin{equation}
Q=-\frac{\hbar ^{2}}{2m}\frac{\Delta R}{R}
\end{equation}
can be considered as a ''quantum'' potential to be added to the
usual one, V, in the Hamilton-Jacobi equation. Incidentally, let us notice that,
since $R$ appears both in numerator and denominator, $Q$ is unchanged when
the wave function is multiplied by an arbitrary constant, i.e. it does not depend on the {\it strength}
of $\Psi$, but only on its {\it form}: this has relevant consequences for the {\it interpretation} of the
{\it field} $\Psi$ and of the quantum potential. We do not discuss this point (see \cite{books1,holland}),
but simply notice that in \cite{books1} the quantum potential is interpreted as a form
of information ({\it active information}).

The particle can thus be thought to move according to the combined
effect of the quantum and external potential:

\begin{equation}
m\frac{d^{2}x}{dt^{2}}=-\nabla (V+Q)
\end{equation}

where the position $x$ is therefore the hidden variable, which, by
assumption, cannot be determined experimentally (without inducing
a collapse of the wave function) and therefore remains always
unknown (as we will detail more precisely at the end of this
subsection). The positions of an ensemble of particles are
distributed according to $\rho(x,t)$.

Let us notice the ingeniousity of this construction: the quantum
potential is such that if one assumes that the particles are
initially statistically distributed in $x$ according  to $\rho (x,0) = | \Psi
(x,0)|^2$ (the "quantum equilibrium" postulate), then the evolution implies that at the time $t$ they
exactly reproduce the statistical distribution predicted by
quantum mechanics $ \rho (x,t) = | \Psi (x,t)|^2$.

Measuring the properties of a statistical ensemble of particles no
difference appears respect to SQM. However, now every single
particle has a well defined  position and follows a well defined
trajectory. The non-epistemic nature of SQM probabilities becomes
epistemic and completely related to our ignorance of hidden
variable values.

It must be emphasized that one cannot even in principle know the
hidden variables values (the coordinates of the single particles):
if this would be possible one can show that this would lead to
some different prediction from SQM. Furthermore, in this case, one
could transmit faster than light signals. In order to understand
how, let us consider a singlet state as (\ref{Singlet}) sent to
two Stern-Gerlach apparatuses. In the dBB theory a correct
description of spin (generalizable to many body case as well) can
be derived by considering Dirac equation and its non relativistic
limit. The precise description of this formulation is however
rather ample and beyond the purposes of this review: the
interested reader  can find it in Ref.s \cite{books1,holland}.
However, the main result of this formulation is that once the
Pauli equation is derived as non-relativistic limit of Dirac one,
the guidance equation \ref{geq} must be substituted by
\begin{equation}
v={ j \over \rho} \label{geqP}
\end{equation}
where $j= \Psi ^\dag \alpha \Psi $ is the Dirac current and $\rho$
the probability density. By using this guidance equation follows
that the spin can be interpreted as depending on a context
dependent circulatory motion of the trajectories. A particle with
spin is still regarded as a simple point particle whose only
intrinsic property is its position and that has a velocity
determined by the guidance condition \ref{geqP}, which in the
non-relativistic approach includes a circulatory contribution
determining the spin. In the interaction with a Stern-Gerlach
apparatus a wave packets splits according to this circulatory
contribution; the result in each individual case depends on the
precise initial conditions, i.e. on the value of the hidden
variable $x$.

The former discussion rapidly summarises the correct description
of spin in dBB model. Nevertheless, in order to have some clear
hint about why is even in principle impossible to know the hidden
variable, in the following we present the simplified argument of
Ref.\cite{ghirdbb} for a singlet state of two spin one half
particles each addressed to a Stern-Gerlach measurement device.
Simplifying a bit, the context dependent spin of a particle
crossing a Stern-Gerlach apparatus is determined by the hidden
variable $x$ and by the direction of the magnetic field B: the
particle cannot cross the horizontal plane, if the position of the
particle is in the upper (lower) semiplane it is deviated toward
the high (low) and one attributes to the particle a spin up
(down). If the field is reversed one would arrive to the opposite
conclusions concerning the spin. The value of the spin depends
therefore on the general context in which the measurement is
performed (i.e. dBB theory is a contextual one). Nevertheless, the
dBB reproduces, by construction, the results of QM: thus if one
particle is found with spin up the other must have spin down. Let
us now imagine that Alice can have access to the hidden variable
before performing the Stern-Gerlach test, with her measurement
preceding the Bob's one. If she observes the particle in the upper
semiplane she leaves the magnetic field as it is and the particle
(deviated toward the high) is attributed to have spin up, on the
other hand if she sees the particle in the lower semiplane, she
reverses the gradient of the magnetic field and the particle
(deviated toward the low) is again attributed to have spin up. On
the other side Bob leaves his apparatus untouched. But now he will
receive only spin down particles which will all be deviated in the
same semiplane: he immediately knows that Alice is performing a
measurement and thus receive an instantaneous information from
her. The access to the hidden variable value would therefore allow
a super-luminal transmission: thus, compatibility with special
relativity requires that hidden variables can not be known even in
principle.

 The contextuality \footnote{in the more refined description of spin hinted before,
the two spin one half EPR particles initially do not have any
circular motion corresponding to spin \cite{books1}, it emerges,
for both, only when one of the particles enters a Stern-Gerlach
device: a clear example of contextuality and non-locality of dBB
theory.} we have discussed in the former example constitutes the
main problem of the dBB model (but not for his supporters): the
value of some observables (as the spin) depends on the global
context in which the measure is performed. The quantum potential
depends on all the particles included in the wave function. Even
an action on an extreme far particle can lead to relevant changes
on the measured properties of a particle entangled with this. Let
us notice that we can still define isolated systems grouping the
particle in sets, such that the members of different sets are
non-interacting and non-entangled.

In general, the result of contextuality is that not all the
observables are objective, namely completely characterised by the
hidden variable of the theory, but most of them depend on the
whole context in which the measurement is performed. However, one
can always identify a complete system of observables, which is
non-contextual: in the case of dBB, this is given by the
positions (on the other hand we have seen as spin is contextual).

In order to give some other explicit example of how dBB theory works, let us now
consider the double slit Young experiment in the framework of this theory.

Each particle follows a well defined trajectory and crosses one of the slits. The shape of its trajectory
varies according to its initial position in the slit (see fig. 13). The trajectories are
determined by the quantum potential and give origin, when a
statistical ensemble of particles is considered, to the SQM
interference figure. Let us notice that the closing or opening of
the second slit immediately affects the particle at the first
slit: here is another example of the contextuality and non-locality of the theory.
Of course, also in SQM the wave function is affected by the
opening or closing of the slit, but in this case no trajectory is
defined and one cannot assert the particle to cross one or the
other of the two slits.

It is worth to remind that contextuality is a common property to
every HVT, which completely reproduces the statistical result of
QM. As we have seen this assertion is contained
in the Bell and Kochen-Specker theorems.

Finally, even without entering into the details, it must also be
noticed that for dBB theory Lorentz invariance is not valid for the
individual event (even if any violation would not be detectable
experimentally, where a statistical sample of events would be
necessary) \cite{books1}.  This is related to the existence of the
instantaneous non-local  effect of the quantum potential on the
particle in a well defined position on its trajectory.

In summary, dBB model has been built for reproducing the results of SQM
for a statistical ensemble of particles, but
attributing to every single particle trajectories evolving
according a Hamilton-Jacobi equation, containing beyond the
classical potential also a second (the "quantum") one.
This theory has been deeply investigated in various aspects (as many-body systems,
interpretation of transition processes, tunnel phenomena, etc.) and large efforts have
been devoted to built a relativistic version of it. For a detailed
description of all these studies we address to specific books
\cite{books1,holland} and to some recent works (see \cite{Duerr1,Duerr2,Hol,Hom}
and references therein).
Here, we would like only to hint at a recent relevant progress
concerning bosonic fields. Whilst,
for fermionic fields, dBB interpretation is of fermionic particles
"guided" by fermionic fields, for bosonic fields an analogous
interpretation was difficult to be realised and usually they were
treated differently from fermions assuming directly bosonic fields
as basic elements of the theory \cite{books1}. However, recently a
coherent theory of bosonic particles guided by bosonic fields has
been developed \cite{ghose1}, based on Kemmer-Duffin formalism
\cite{Kemmer}, reestablishing parallelism between fermions and
bosons. Let us also mention that some variants of dBB model have been presented
where the wave function  describes a real
physical wave propagating in space-time (substantially dating back to the original de
Broglie proposal \cite{debroglie,croca}),
or with "exotic" density distributions \cite{valentini},
or with modified guidance equation (within constraints of relativistic covariance)
\cite{Hol2}.

\subsection{Nelson Stochastic Model}

As a further example of NLHVT let us consider  the  stochastic mechanics
introduced by Edward Nelson in 1966 \cite{nelson}. The foundation
of this model is based on two basic hypotheses. The first assumes
that dynamical systems follow trajectories (also here positions are
therefore  the hidden variables) perturbed by
an underlying Brownian motion. The second is a particular form of
the second principle of dynamics, where the classical acceleration
is replaced by a suitable form of stochastic acceleration.

It can be shown that the basic equation of stochastic mechanics
can be derived from variational principles, in complete analogy
with classical mechanics, based on the same classical action, but
exploiting stochastically perturbed trajectories as trial
trajectories. With this purpose let us consider (following the
presentation of Ref. \cite{guerra}) a classical Lagrangian \be
L(q,{d q \over dt}) = {1 \over 2} m \left ({d q \over dt} \right)^2- V(q) \ee and
introduce a diffusion process, with density $\rho(x,t)$, satisfying the
forward Ito stochastic equation (for details on stochastic equations see for example the book
\cite{Qnoise}) \be dq(t) = v_{(+)}(q(t),t) dt + d w(t) \ee Let us
also introduce a diffusion constant $\nu$, pertaining the Brownian motion $d w(t)$, and forward ($v_{(+)}$),
backward ($v_{(-)}$) and osmotic velocities ($u$) defined by \be 2
m \nu = \hbar \ee and \be v={1 \over 2} (v_{(+)} + v_{(-)}) \ee \be
u={1 \over 2} (v_{(+)} - v_{(-)}) = \nu \nabla log \rho \ee

The density $\rho$ satisfies the continuity equation: \be
\partial_t \rho(x,t) = - \nabla \cdot (\rho v)
\ee

Once the average stochastic action ($E$ denoting an overall
average) \be A(t_0,t_1;q) = \int_{t_0}^{t_1} E \left[{1 \over 2} m
\left({\Delta q \over \Delta t}\right)^2- V(q)\right] dt \ee is introduced, it is
found that stationarity of it under arbitrary small variations
$\delta v_{(+)}$, with the constraint $\delta \rho(.,t_1)=0$,
requires the current velocity field to satisfy the Hamilton-Jacobi
equation \be m v = \nabla S\ee where
\begin{equation}
\frac{\partial S}{\partial t}+\frac{(\nabla S)^{2}}{2m}+V+Q=0
\label{NelsD}
\end{equation}
and \be Q= { - \hbar^2 \over 2m} {\triangle \sqrt{\rho} \over
\sqrt{\rho}} \label{QNels}\ee

a form equivalent to Schr\"odinger equation \ref{Sc} (similarly to
what described in the former subsection) when the wave function is
rewritten as
\begin{equation}
\Psi =\sqrt{\rho(x,t)} \cdot exp[iS(x,t)/ \hbar]
\end{equation}

Furthermore, introducing the forward and backward transport operators
\be
D_{\pm} = \partial_t + v_{\pm} \cdot \nabla \pm \nu \triangle
\ee
and the Nelson acceleration
\be
a(x,t) = {1 \over 2} ( D_{+} v_{-} +  D_{-} v_{+})
\ee
one can rewrite the dynamical equation \ref{NelsD} in the Newton form
\be
m a = - \nabla V .
\ee

In summary, Nelson's model is a scheme where particles have well
defined trajectories (as in classical mechanics), but with a
stochastic component of motion. When this last part is built
suitably the equations of motion become equivalent to
Schr\"odinger equation. Thus, in principle, SQM and Nelson's model
are equivalent from a predictive point of view, differing only in the interpretation, since in
Nelson's model one attributes to a single particle a trajectory
and therefore an hidden variable (the position). Finally, it must
be emphasized that, due to the form of the term \ref{QNels}, the
transition probabilities for a subsystem can depend on the
properties of a far away (space-like separated) different
subsystem \cite{nels2}, showing the non-locality of this model.

Extension to quantum field theory have been elaborated \cite{GuRu,GuLo,lim}
(a similar approach can also be found in Ref. \cite{PaWu,DaHu}).

The interested reader can find specific reviews of this model in Ref.s \cite{GuNe,BoHi}.
Other HVT where particles follow definite trajectories, as in dBB and Nelson's models,
can be found in section 4 of Ref. \cite{HW} (where in general various other HVT models are quoted).

\subsection{Experimental tests of NLHVT against SQM}

As we discussed in subsection 2.4, Bell inequalities are based on the locality hypothesis
and thus do not concern NLHVT. One can therefore pose the question of how to compare these
theories with SQM or whether, at least for some of them, there is a perfect
equivalence from a predictive point of view with SQM.

A first experimental test of non-local HVT was proposed
\cite{croca} and realised \cite{cexp1,cexp2} for the variant of
dBB theory where the wave function is not only assumed to give
"guidance condition" by quantum potential, but it is a real
physical wave propagating in space-time. This means that if the
wave function splits on different paths, even after detection of
the particle on one path, the waves on the other paths ("empty
waves") can manifest physical effects (such as interference).

This theoretical proposal \cite{croca}, following and developing some former ones \cite{selb,tarb,crocab},
 considered a source
producing a pair of photons by PDC that are addressed to a
modified Mach-Zender interferometer (see fig. 14).

In little more detail, two identical photons produced at the same time (for example in PDC) are addressed
to two different beam splitters part of the same arm of a Mach-Zender interferometer.
If one assumes that the photon is composed of a localized particle
and a real wave propagating according with d'Alembert equation and
that detection of a particle in one of possible channels does not
induce collapse of the wave on other possible channels, then for
lossless beam splitters with transmission and reflection
coefficients $t,r$ respectively,  the outgoing waves $\Psi_1$ and
$\Psi_2$ from the other two beam splitters of the interferometer are,
in terms of waves $\Phi_1$ and $\Phi_2$, associated
to signal and idler photon, respectively \bea \Psi_1 = t^2 \Phi_1 +
r^2 t \Phi_2 + tr^2 e^{i \alpha} \Phi_2 \,\, , \nonumber \\
\Psi_2 = t^2 e^{i \alpha} \Phi_2 \eea where $\alpha$ is the phase
difference between the two paths.

The coincidence probability is then given by: \be P(D_1,D_2)
\propto |\Psi_1|^2 |\Psi_2|^2 = |t|^4 |\Phi_2|^2 [ |t|^4
|\Phi_1|^2 + 2 |r|^4 |t|^2 (1 + \cos{\alpha}) |\Phi_2|^2] \ee

which depends on the phase difference between the two optical
lengths of the interferometer: an effect essentially due to
overlapping in the path BS$_3$-D$_1$ of signal wave function with
empty wave generated by the idler photon going through BS$_1$ and
BS$_4$.

On the other hand, the usual quantum optical results, easily
calculated by combining annihilation operators in the
interferometer, does not present any dependence on the phase
difference between the two optical lengths of the interferometer.

This experiment has been realised by Rochester group \cite{cexp1}
with an equivalent scheme showing perfect agreement with Quantum
Optics result.

Since some doubts was arisen \cite{croca2} about the real
superposition of wave packets in the interferometer, a second
version of the experiment was then realised answering to these
objections \cite{cexp2,cexp3}. The perfect agreement with SQM
results represented a conclusive negative test of empty wave models.

Here we can also hint at the proposal of  a possible experimental test for the "wavelet" version of
de Broglie model compared to SQM \cite{wavelet}.

Moreover, before concluding this section, we would like to mention possible
tests of the conventional dBB model against SQM. As previously described de Broglie-Bohm model
was built to be completely equivalent to Standard Quantum Mechanics.
Nevertheless, some authors have suggested that the constraints due to the existence of trajectories
could lead to differences between the
two, which could eventually be investigated by an experimental
test (see for example \cite{Fel,Red,ghose2,ghose0,ghose3,pla,Golshani}): namely dBB and SQM are not simply
different interpretations of the same theory, but really different theories\footnote{We do not enter here
in the complex debate of when two theories can be considered different, of what means different
interpretations of the same theory, etc. \cite{books4,books1,books2,pop,lak,fey,kuhn}.}.

In particular, a recent proposal of two teams
\cite{ghose0,ghose3,pla,Golshani} of a variant of double
slit experiment gave rise to a certain interest. In little more detail, this scheme considers two identical particles crossing at the same time a double slit each at a specific slit: the calculation of trajectories show that they never cross the double slit symmetry axis and therefore no coincidences are expected in the same semiplane at variance with SQM result.
Even if the debate about the validity of this
theoretical prediction or its specificity for some variant of dBB model (or an application of this scheme to
Nelson Stochastic model as well) is not settled yet
\cite{ghose3,mar1,mar2,mar3,st,gol2,intr}, it is worth to mention that a recent
experimental realisation of this scheme \cite{ds1,ds2} has obtained results in perfect agreement with SQM
 (see Fig. 15 for the scheme of this set-up and experimental data).

Finally, discussing possible experimental evidences that could be obtained against dBB model, it is also worth to
 quote some results \cite{surr}
showing that in specific cases (as interferometers with which way detectors)
Bohm trajectories can be macroscopically at variance with the recorded track
(in this sense they are called "surrealistic" in this paper)\footnote{For later developments of this idea,
as "weak" (i.e. such to change minimally the measured state) and "protective" (weak and adiabatic)
measurements, see \cite{AES} and Ref.s therein.}.
Nevertheless, this achievement is not considered
really compelling from dBB model supporters \cite{DFGZ,DHS,bar2}, who argue that in these cases a position detector
does not really measure the real position of the particle (on the Bohm trajectory) but
is affected by the non-local quantum potential.

\section{Determinism at Planck scale?}

Before concluding this review, it is interesting to mention that
in the last years a further revival of interest for hidden variables theories has arisen starting from a paper of the Nobel laureate
't Hooft where it was suggested that determinism could reappear at Planck scale \cite{thooft}.
A point of large relevance
into the program of reaching a completely unified theory including also gravity.

The main idea of this proposal is that at Planck scale physical systems (including gravity) are described by a
deterministic theory, but at larger scales we have loss of information (for dissipation):  "quantum states"
are equivalence classes of the deterministic states, the loss information are the hidden variables.

In order to have an idea of how this scheme works, let us present a simple example from Ref. \cite{thooft}.

Let us consider a discrete system with four states $e_1, e_2, e_3, e_4$ whose deterministic evolution,
after every time step, is
\be
e_{1} \rightarrow e_{2}, \, e_{2} \rightarrow e_{1}, \, e_{3} \rightarrow e_{3}, \,  e_{4}
 \rightarrow e_{1}
 , \,
\ee
Even if evolution is deterministic, it can be useful to introduce a Hilbert space in order to handle it
probabilistically. This evolution is described by the (non-unitary) operator:
\be U=
\left(%
\begin{array}{cccc}
  0 & 1 & 0 & 1 \\
  1 & 0 & 0 & 0 \\
  0 & 0 & 1 & 0 \\
  0 & 0 & 0 & 0 \\
\end{array}%
\right)
\label{U}
\ee
However, after a short lapse of time only the states $e_1, e_2, e_3$ survive. Thus one can simply erase the state $e_4$
and considering $e_1, e_2, e_3$ as the "quantum" system with a unitary evolution described
by the upper 3x3 part of $U$, Eq. \ref{U}.

This system may therefore be described in three equivalence classes:
\be
E_1=\{ e_1 \}, \, E_2=\{ e_2,e_4 \}, \, E_3=\{ e_3 \}, \,
\ee
with unitary evolution operator (H is a Hamiltonian operator)
\be
U'= e^{-i H}=
\left(%
\begin{array}{ccc}
  0 & 1 & 0 \\
  1 & 0 & 0 \\
  0 & 0 & 1 \\
\end{array}%
\right)
\ee

This simple model shows how, if information is allowed to dissipate, one has to define quantum states as equivalence
classes of states, where two states are equivalent iff, some time in the future, they evolve into one and the same state.
Equivalence classes that form a smaller set of the complete set of primordial
states  that one starts off with. A  continuum model is then presented \cite{thooft} as well.

Summarizing, the main idea presented in Ref. \cite{thooft} is therefore that a quantum state is defined as an equivalence
 class of states all having the same future.
These equivalence classes are described by observables, that in
quantum terminology correspond to a complete set of commuting
operators at every time ("beables" following Bell terminology
\cite{bEPR}). A beable describes the information available on a
system at Planck scale. All other quantum operators are
"changeables" (they do not commute with all beables). A physical
system can evolve deterministically at Planck scale, but a
probabilistic theory can derive at larger (spatial) scales due to
loss of information. If this is the case, Bell inequalities
experiments with photons, electrons etc. would not be resolutive
for testing this deterministic theory since photons, electrons
etcetera do not represent true degrees of freedom of it (i.e.
correspond to "changeables" and not to "beables").

A further  indication of how this mechanism could work was given later \cite{blas}, showing how
 a quantum harmonic oscillator can emerge from a pair of classical oscillators with dissipation.

On the same line in Ref. \cite{BMM} it was shown how quantum field theory in (3+1)- dimensional
Minkowsky space could emerge as low energy limit of a (4+1)-dimensional classical gauge theory. Here the fifth
dimension would be the hidden variable. Incidentally, let us notice that the local dynamical theory in more dimensions would generate
fundamentally non-local effects in lower dimensional space: in order to examine such a model a
Bell inequalities test should be performed at the scale
of the transition between classical dissipative dynamics to quantum dynamics.

Later developments of these proposals and related arguments can be found in
Ref.s \cite{bru,bla2,bla3,thooft2,thooft3,bm03,elz,kato,wet}
(see also the very recent paper \cite{chen2} where hidden variables are two additional time dimensions).

Finally, it can also be mentioned a further work \cite{had}, where it is shown that a orthomodular lattice of
propositions characteristic of quantum logic can be constructed for manifolds in Einstein's theory of general relativity,
where both state preparation
and measurement apparatus constrain results of experiments (future observations represent hidden variables).

Altogether these highly speculative proposals arose a new interest for the search of a deterministic theory
beyond quantum mechanics. Only further studies will be able to show if this ideas could have interesting
developments.

\section{Conclusions}

In this review, after a general introduction to the researches on local realistic alternatives
to standard quantum mechanics, we have presented
the most recent results (with a larger emphasis to experimental ones) about these studies.

This problem is of the utmost importance for our understanding of the nature, namely for clarifying
if nature is intrinsically probabilistic or
quantum mechanical probabilities derive by our ignorance of some hidden parameters and therefore an underlying
deterministic theory is conceivable.

The transition from the XIX century point of view of a perfectly
deterministic nature described by classical mechanics to the
actual quantum mechanical point of view of a probabilistic world
has been difficult and largely debated and still many points at
the very foundations of quantum mechanics need a clarification. It
should also be noticed that whilst this new paradigm has been
largely accepted by physicists community, its assimilation in
diffused culture is still rather limited.

Furthermore, as we have seen, a conclusive experiment falsifying in an absolutely uncontroversial way
local realism is still missing.

More in details, for what concerns {\bf local} hidden variable theories, since Bell theorem it is known that
 a general answer about their validity can be given by an experiment. In the last 40 years various experiments have addressed this problem:
strong indications favouring standard quantum mechanics have been obtained, but no conclusive experiment has
yet been performed, mainly due to low detection efficiencies that demand for additional assumptions.
Nevertheless, relevant progresses toward this goal have been made in the last ten years and in my opinion
an ultimate experiment could not be far in the future.

However, we have to acknowledge that this personal opinion is not generally shared: on one side some authors deem that
the large amount of experimental data disfavouring Local Hidden Variable Theories is already largely sufficient
for excluding them, on the other side other authors (see for example \cite{santult}) claim that the lack of a conclusive
 experiment after 40 years and in particular the "resistance" of detection loophole to be eliminated could point out a practical
impossibility of falsifying local realism.
These discussion largely involve methodological questions \cite{pop,lak,fey,kuhn} which are amply
beyond the purposes of this paper.

Even if Local Realistic Theorem will be excluded by an ultimate Bell inequalities experiment,
{\bf non-local} hidden variable theory will still remain a possible alternative to standard quantum mechanics.
Following the discussion of last sections, in our opinion
 a large space still remains for relevant contributions to study  this possibility
  both from a theoretical and an experimental point of view.

In general interesting developments in this area can be expected in the next years,
also in connection with the related developing field of quantum information.

{\bf Acknowledgements}

I would like to thank E. Cagliero, V. Carabelli and F. Piacentini
for help in revising the manuscript.

\newpage
{\bf FIGURE CAPTIONS}

Fig.1 Contour plot of the quantity $CH/N$ (see Eq. \ref{eq:CH}; N is the total number of detections)
in the plane with $f$ (non maximally entanglement parameter, $\vert \psi \rangle = {\frac{ \vert H \rangle \vert H \rangle + f
\vert V \rangle \vert V \rangle }{\sqrt {(1 + |f|^2)}}}$ ) as x-axis and detection efficiency $\eta$ as y-axis.
The leftmost region corresponds to the region where no detection loophole free test of Bell
inequalities can be performed. The contour lines are at 0, 0.025, 0.06, 0.1, 0.15, 0.2.

Fig.2 Scheme of type I PDC. The two circumferences (continuous and dashed) correspond to two different wave
 lengths. The spots indicate
the directions of emission of two entangled photons.

Fig.3 Scheme of type II PDC emission. Two circumferences, where are emitted degenerate photons (continuous line) and
correlated photons of different wave lengths (dashed lines), are shown. H, V denote horizontal and vertical
polarisation, respectively.

Fig.4 Franson scheme for Bell inequality test from Ref. \cite{franson}. A source emits two energy-time entangled photons
that after having crossed a Mach-Zender interferometer are detected by $D_1$ or $D_1'$ and $D_2$ and $ D_2'$ respectively.

Fig. 5 Set-up of Ref. \cite{HSZ} for a Bell inequality test. A source S emits two particles (1,2) in four beams A,B,C,D.
An entangled state is realised by superposition on beam splitters.

Fig.6 Scheme of the experiment (from Ref.\cite{hugo}) for a Bell inequality test on energy-time entangled photons
by two remote measurement devices.

Fig. 7 Scheme for generating polarisation entangled photon states by superimposing on a beam splitter two correlated
photons produced in type I PDC after having rotated polarisation of one of the two (from Ref. \cite{SH}).

Fig. 8 Typical set-up for PDC entangled photon Bell experiments. On the left one can recognise the pump laser,
 a titanium-sapphire mode locked laser pumped by a diode laser and with second harmonic generation.
 At the centre of the picture the non-linear crystal, followed by a filter for eliminating the UV pump and
 fibre couplers (preceded by interferential filters) collecting the photons to be addressed to the detectors
 (one of them is on the background).

Fig. 9 Sketch of a bright source of polarisation entangled photons realised by superimposing two type I PDC emissions
(from Ref. \cite{nos}). CR1 and CR2 are two $LiIO_3$ crystals cut at
 the phase-matching angle of $51^o$. L1 and L2 are two identical piano-convex lenses with a hole of 4 mm in the centre.
 P is a 5 x 5 x 5 mm quartz plate for birefringence compensation and $\lambda / 2$ is a first order half wave-length plate at 351 nm.
U.V. identifies the pumping radiation at 351 nm. The infrared beam (I.R.)
is used for system alignment (dashed line identifies the correlated emission).

Fig. 10 Outcomes of GHZ test from Ref. \cite{GHZexp}. These data show that terms predicted by SQM (tall bars)
appear in a fraction $0.85 \pm 0.04$
 of all cases against a fraction $0.15 \pm 0.02$ of other terms.

Fig. 11 Experimental scheme for generating a GHZ pair from Ref. \cite{GHZexp}.  Pairs of polarisation-entangled photons
 are generated by a short pulse of ultraviolet light (200 fs, $\lambda$= 394 nm) pumping a BBO crystal.
Observation of the desired GHZ correlations requires fourfold coincidence and therefore simultaneous emission of two pairs.
The photon registered at T is always horizontally polarised (H) and thus its partner in b must be vertically polarised (V).
The photon reflected at the polarizing beam-splitter (PBS) in arm a is always V,
being turned into equal superposition of V and H by the $\lambda /2$ plate, and its partner in arm b must be H.
Thus if all four detectors register at the same time, the two photons in D1 and D2 must either both have been VV and
reflected by the last PBS or HH and transmitted. The photon at D3 was therefore H or V, respectively.
Both possibilities are made indistinguishable by having equal path lengths via a and b to D1 (D2) and by using narrow
bandwidth filters to stretch the coherence time to about 500 fs, substantially larger than the pulse length.
Polarizers oriented at 45° and $\lambda /4$ plates in front of the detectors allow measurement of linear 45$^o$ (-45$^o$)
 (circular R/L) polarisation.

Fig. 12 Experimental apparatus for tritter generation of qutrits from Ref. \cite{gisqun}.
 Different paths combination originate 5 peaks  in arrival time histogram (shown on the right).
Coincidences for central peak (shown on the left as a function of Alice's and Bob's phase vectors)
project onto one of three orthogonal qutrit states.

Fig. 13. Bohm trajectories calculated for a particle crossing a double slit (from Ref. \cite{pla}).

Fig. 14 Outline of the experiment for testing empty waves hypothesis (from Ref. \cite{croca}).
Two identical photons (produced by PDC) enter through two different beam splitters (BS$_{1,2}$) a Mach-Zender
interferometer. Coincidences are measured between the two photo-detectors (D$_{1,2}$)
at the exits of the second pairs of beam splitters (BS$_{3,4}$).

Fig. 15 The double slit experiment of Ref. \cite{ds2}. In the upper window the set-up scheme.
A pump laser at 351 nm generates type I parametric down conversion in a lithium-iodate crystal.
Conjugated photons at 702
nm are sent to a double-slit (two slits of $10 \mu m$ separated of $100\mu m$) by a system of two piano-convex
lenses in a way that each photon of the pair crosses a well
defined slit. A first photodetector is placed at  1.21 m  a second
one at 1.5 m from the slit. Both the single photon detectors (D)
are preceded by an interferential filter at 702 nm (IF) and a lens
(L) of 6 mm diameter and 25.4 mm focal length. Signals from
detectors are sent to a Time Amplitude Converter and then to the
acquisition system (multi- channel analyser and counters).
In the lower window the experimental coincidences data are compared with quantum mechanics predictions.
 On the x-axis we report the position of the first detector respect to the median symmetry axis of the double slit.
 The second detector is kept fixed at -0.055 m (the region without data around this point is due to the superposition of the two detectors).
The x errors bars represent the width of the lens before the
detector. Coincidences are clearly observed in the same semiplane at variance with \cite{ghose0,ghose3,pla,Golshani} result.
\end{document}